\documentclass[%
 superscriptaddress,
 twocolumn,
 showpacs,
preprintnumbers,
nofootinbib,
 amsmath,amssymb,
 prd,
balancelastpage,
nofootinbib,
floatfix
]{revtex4-1}

\usepackage{graphicx}
\usepackage{dcolumn}
\usepackage{bm}
\usepackage{amsmath}
\usepackage{diagbox}
\usepackage{rotating}
\usepackage{multirow}
\usepackage{xcolor}
\usepackage{hepunits}

\usepackage{hyperref}
\usepackage[mathlines]{lineno}
\newcommand{\be}{ \begin{eqnarray} }
\newcommand{\ee}{ \end{eqnarray} }
\newcommand{\sx}{$\sigma_x$}
\newcommand{\ev}{eV$_{\mathrm{ee}}$}

\begin{document}

\preprint{FERMILAB-PUB-21-498-AE-E-QIS}

\title{Characterization of the background spectrum in DAMIC at SNOLAB}

\author{A.~Aguilar-Arevalo} 
\affiliation{Universidad Nacional Aut{\'o}noma de M{\'e}xico, Mexico City, Mexico} 

\author{D.~Amidei}
\affiliation{Department of Physics, University of Michigan, Ann Arbor, Michigan, United States}  

\author{I.~Arnquist}
\affiliation{Pacific Northwest National Laboratory (PNNL), Richland, Washington, United States} 

\author{D.~Baxter}
\affiliation{Fermi National Accelerator Laboratory, Batavia, Illinois, United States}
\affiliation{Kavli Institute for Cosmological Physics and The Enrico Fermi Institute, The University of Chicago, Chicago, Illinois, United States}

\author{G.~Cancelo}
\affiliation{Fermi National Accelerator Laboratory, Batavia, Illinois, United States}

\author{B.A.~Cervantes Vergara}
\affiliation{Universidad Nacional Aut{\'o}noma de M{\'e}xico, Mexico City, Mexico} 

\author{A.E.~Chavarria}
\affiliation{Center for Experimental Nuclear Physics and Astrophysics, University of Washington, Seattle, Washington, United States}

\author{N.~Corso}
\affiliation{Kavli Institute for Cosmological Physics and The Enrico Fermi Institute, The University of Chicago, Chicago, Illinois, United States}

\author{E.~Darragh-Ford}
\affiliation{Kavli Institute for Cosmological Physics and The Enrico Fermi Institute, The University of Chicago, Chicago, Illinois, United States}

\author{M.L.~Di~Vacri}
\affiliation{Pacific Northwest National Laboratory (PNNL), Richland, Washington, United States} 

\author{J.C.~D'Olivo}
\affiliation{Universidad Nacional Aut{\'o}noma de M{\'e}xico, Mexico City, Mexico} 

\author{J.~Estrada}
\affiliation{Fermi National Accelerator Laboratory, Batavia, Illinois, United States}

\author{F.~Favela-Perez}
\affiliation{Universidad Nacional Aut{\'o}noma de M{\'e}xico, Mexico City, Mexico} 

\author{R.~Ga\"ior}
\affiliation{Laboratoire de Physique Nucl\'eaire et des Hautes \'Energies (LPNHE), Sorbonne Universit\'e, Universit\'e de Paris, CNRS-IN2P3, Paris, France}

\author{Y.~Guardincerri}
\thanks{Deceased January 2017}
\affiliation{Fermi National Accelerator Laboratory, Batavia, Illinois, United States}

\author{ T.W.~Hossbach}
\affiliation{Pacific Northwest National Laboratory (PNNL), Richland, Washington, United States} 

\author{B.~Kilminster}
\affiliation{Universit{\"a}t Z{\"u}rich Physik Institut, Zurich, Switzerland }

\author{I.~Lawson}
\affiliation{SNOLAB, Lively, Ontario, Canada }

\author{S.J.~Lee}
\affiliation{Universit{\"a}t Z{\"u}rich Physik Institut, Zurich, Switzerland }

\author{A.~Letessier-Selvon}
\affiliation{Laboratoire de Physique Nucl\'eaire et des Hautes \'Energies (LPNHE), Sorbonne Universit\'e, Universit\'e de Paris, CNRS-IN2P3, Paris, France}

\author{A.~Matalon}
\affiliation{Kavli Institute for Cosmological Physics and The Enrico Fermi Institute, The University of Chicago, Chicago, Illinois, United States}
\affiliation{Laboratoire de Physique Nucl\'eaire et des Hautes \'Energies (LPNHE), Sorbonne Universit\'e, Universit\'e de Paris, CNRS-IN2P3, Paris, France}

\author{P.~Mitra}
\affiliation{Center for Experimental Nuclear Physics and Astrophysics, University of Washington, Seattle, Washington, United States}

\author{A.~Piers}
\affiliation{Center for Experimental Nuclear Physics and Astrophysics, University of Washington, Seattle, Washington, United States}

\author{P.~Privitera}
\affiliation{Kavli Institute for Cosmological Physics and The Enrico Fermi Institute, The University of Chicago, Chicago, Illinois, United States}
\affiliation{Laboratoire de Physique Nucl\'eaire et des Hautes \'Energies (LPNHE), Sorbonne Universit\'e, Universit\'e de Paris, CNRS-IN2P3, Paris, France}

\author{K.~Ramanathan}
\affiliation{Kavli Institute for Cosmological Physics and The Enrico Fermi Institute, The University of Chicago, Chicago, Illinois, United States}

\author{J.~Da~Rocha}
\affiliation{Laboratoire de Physique Nucl\'eaire et des Hautes \'Energies (LPNHE), Sorbonne Universit\'e, Universit\'e de Paris, CNRS-IN2P3, Paris, France}


\author{M.~Settimo}
\affiliation{SUBATECH, Universit\'e de Nantes, IMT Atlantique, CNRS-IN2P3, Nantes, France}

\author{R.~Smida}
\affiliation{Kavli Institute for Cosmological Physics and The Enrico Fermi Institute, The University of Chicago, Chicago, Illinois, United States}

\author{R.~Thomas}
\affiliation{Kavli Institute for Cosmological Physics and The Enrico Fermi Institute, The University of Chicago, Chicago, Illinois, United States}

\author{J.~Tiffenberg}
\affiliation{Fermi National Accelerator Laboratory, Batavia, Illinois, United States}

\author{D.~Torres Machado}
\affiliation{Universidade Federal do Rio de Janeiro, Instituto de  F\'{\i}sica, Rio de Janeiro, Brazil}

\author{M.~Traina}
\affiliation{Laboratoire de Physique Nucl\'eaire et des Hautes \'Energies (LPNHE), Sorbonne Universit\'e, Universit\'e de Paris, CNRS-IN2P3, Paris, France}

\author{R.~Vilar}
\affiliation{Instituto de F\'isica de Cantabria (IFCA), CSIC--Universidad de Cantabria, Santander, Spain}

\author{A.L.~Virto}
\affiliation{Instituto de F\'isica de Cantabria (IFCA), CSIC--Universidad de Cantabria, Santander, Spain}

\collaboration{DAMIC Collaboration}
\noaffiliation

\date{\today}

\begin{abstract}
We construct the first comprehensive radioactive background model for a dark matter search with charge-coupled devices (CCDs). We leverage the well-characterized depth and energy resolution of the DAMIC at SNOLAB detector and a detailed {\tt GEANT4}-based particle-transport simulation to model both bulk and surface backgrounds from natural radioactivity down to 50~eV$_{\text{ee}}$. We fit to the energy and depth distributions of the observed ionization events to differentiate and constrain possible background sources, for example, bulk $^{3}$H from silicon cosmogenic activation and surface $^{210}$Pb from radon plate-out. 
We observe the bulk background rate of the DAMIC at SNOLAB CCDs to be as low as $3.1 \pm 0.6$ counts kg$^{-1}$ day$^{-1}$ keV$_{\text{ee}}^{-1}$, making it the most sensitive silicon dark matter detector. Finally, we discuss the properties of a statistically significant excess of events over the background model with energies below 200~eV$_{\text{ee}}$.
\end{abstract}


\maketitle

\section{Introduction}

The particle nature of dark matter is one of the most elusive mysteries in physics~\cite{Kolb:1990vq,review_Hooper}. After decades of searching for ``heavy'' weakly interacting massive particles (WIMPs) with masses 10--$10^3$ GeV/$c^2$~\cite{goodman,lewin_review_1996}, all potential
signals have so far been excluded or remain unverified. 
This effort has required unprecedented understanding of radioactive background sources down to the keV energy scale~\cite{Schumann:2019eaa}. In the absence of a detection of the heavy WIMP, many experiments are refocusing on the possibility of lower mass dark matter particles that would produce energy signals below 1~keV~\cite{CosmicVisions}. 
Such dark matter models are theoretically well-motivated but largely unexplored hypotheses that include low-mass WIMPs~\cite{Zurek:2013wia}, hidden-sector particles~\cite{hiddenPhoton, subGeV_essig}, axions and axion-like particles~\cite{Graham_2015}, and strongly interacting massive particles (SIMPs)~\cite{SIMP,SIMP_collar}, among many others. A review of simplified models for low-mass dark matter can be found in Ref.~\cite{Lin:2019uvt}.

A dominant method of searching for lower mass dark matter is to use detectors that measure the small ionization signals produced from a dark matter particle recoiling off of either the nuclei or electrons in a material. This technique has been demonstrated by cryogenic detectors~\cite{cdmsHVeV,edelweissHV}, noble liquid time-projection chambers~\cite{darksideS2,xenonS2}, and gaseous detectors~\cite{newsG}, to name a few. Here, we will focus on the use of charge-coupled devices (CCDs) by the Dark Matter In CCDs (DAMIC) experiment at SNOLAB to search for ionization produced by dark matter scattering in silicon. More specifically, we will detail the construction of the first comprehensive radioactive background model for a CCD detector down to a threshold of 50~eV$_{\text{ee}}$ (electron-equivalent energy available as ionization), which was recently used to set limits on GeV-scale WIMPs coupling to nuclei~\cite{damic2020}. In Ref.~\cite{damic2020}, we also reported an unexpected excess of events above the background model with energies $<200$\,\ev . This paper revisits the same data and provides supplementary information for a robust exploration of the reported event excess.

This paper is organized as follows. 
In Section~\ref{S:Setup}, we summarize our experimental setup, including a detailed description of the DAMIC at SNOLAB detector, CCD sensor operation, and our data analysis. 
In Section~\ref{S:Backgrounds}, we expand on this discussion with a focus on backgrounds, as from radiocontaminants in detector materials and on their surfaces. We detail our application of the {\tt GEANT4} simulation package~\cite{geant4} to these background sources and our modeling of the partial charge collection region discovered near the back surface of the CCDs. 
In Section~\ref{S:BackModel}, we present the construction of our radioactive background model from a fit to data above 6~k\ev\ with templates constructed from simulated events, as well as an independent cross-check on the activity of surface $^{210}$Pb and the extrapolation of this model into our WIMP-search region.
In Section~\ref{S:likelihood}, we revisit our ``WIMP search,'' explore the event excess over the background model, and discuss its significance relative to our background model uncertainties.
Finally, in Section~\ref{S:Discussion}, we summarize the results of this analysis and discuss future improvements and applications. 

\section{Experimental Setup}\label{S:Setup}

DAMIC at SNOLAB is the first dark matter detector to employ a multi-CCD array. Following its original deployment~\cite{damic2016}, it was upgraded for lower backgrounds, more CCD detectors, and longer exposure. The results of the latest detector installation include the search for hidden-sector dark matter particles from its interactions with electrons~\cite{damicDMe} and the most sensitive search for silicon nuclear recoils from the scattering of WIMPs with masses below 10~GeV~\cite{damic2020}. Here, we describe in detail the DAMIC at SNOLAB detector, including CCD operation and data analysis.

\subsection{DAMIC Detector}\label{S:Detector}

The DAMIC detector at SNOLAB consists of an array of eight CCD modules in a tower-like configuration that was installed in January 2017.
The topmost CCD module (CCD~1) was fabricated with ultra-low radioactivity copper and is shielded from the others by ultra-low radioactivity lead.
One of the other eight CCDs was disconnected soon after installation due to luminescence from one of the amplifiers, which produced unwanted charge throughout the CCD array.
Of the remaining devices (CCDs 1--7), one device (CCD 2) was initially not operational but unexpectedly came back online after a temperature cycle and electronics restart caused by an unplanned power outage at SNOLAB in April 2017.

The DAMIC at SNOLAB CCDs were packaged by Fermi National Accelerator Laboratory (Fermilab).
The CCD package consists of the CCD sensor and Kapton flex cable glued onto a silicon support frame cut from high-resistivity silicon wafers from Topsil (of the same origin as those used for CCD fabrication).
The Kapton flex, fabricated by Cordova Printed Circuits Inc., was first glued using pressure-sensitive adhesive ARclad IS-7876, after which the sensor was glued using epoxy Epotek 301-2 and cured for two days in a laminar flow cabinet.
A fully automatic Fine Wire Wedge Bonder was used to connect with aluminum wires the pads on the CCD to the corresponding pads on the flex cable.
The flex cable has a miniature AirBorn connector installed on its end, which carries the signals to drive and read the CCD.

Each CCD module consists of a CCD package installed in a copper support frame. The modules slide into slots of a copper box~\cite{Chavarria:2020hju} with wall thickness of 6.35~mm, which, in addition to mechanical support, acts as a cold IR (infrared radiation) shield during operation.
Copper trays also slide into slots of the box to make the shelves that hold two 2.5~cm thick ancient lead (smelted $> 300$ years ago) bricks above and below CCD~1.
The copper frame for CCD~1, the copper tray that holds the top ancient lead brick, and an additional $\sim 1$~mm thick copper plate placed on top of the lower ancient lead brick were made from high-purity copper electroformed by Pacific Northwest National Laboratory (PNNL)~\cite{efcu}.

All other copper parts of the CCD modules and box were machined using propylene glycol as lubricant at The University of Chicago from oxygen-free high conductivity (OFHC) Copper Alloy 10100 procured from Southern Copper \& Supply Company.
At least 3~mm of copper was removed from all surfaces of the stock plates when machining to minimize possible contamination introduced in the copper when it was rolled into plates.
The machined copper parts and brass fasteners used for their assembly were cleaned and passivated with ultra-pure (Fisher Scientific Optima grade or equivalent) water and acids following the procedure from Ref.~\cite{HOPPE2007486}.
The packaged CCDs and modules were transported to SNOLAB by car and taken underground to be installed in the detector.

The copper box is suspended below an 18~cm high lead shield inside a 20~cm diameter, 79~cm long cylindrical copper cryostat.
The cryostat is held at pressures of 10$^{-6}$~mbar (10$^{-4}$~Pa) by a HiCube 80 Eco turbomolecular pump.
The internal cylindrical lead shield has a central 3.15~cm diameter hole to accommodate a concentric copper cold finger that makes thermal contact between the top of the copper box and a Cryomech AL63 cold head above the shield.
A temperature sensor and a heater installed at the interface between the cold head and the cold finger are connected to a LakeShore 335 unit to control the temperature of the system.
The Kapton flex cables run along the side of the lead shield through an outer channel before they connect to a second-stage Kapton flex extension that includes an amplifier for the CCD output signal.
The flex extensions connect to the vacuum-side of a vacuum interface board (VIB) that acts as the electronics feedthrough.

\begin{figure}[t!]
	\centering
	\includegraphics[width=\linewidth]{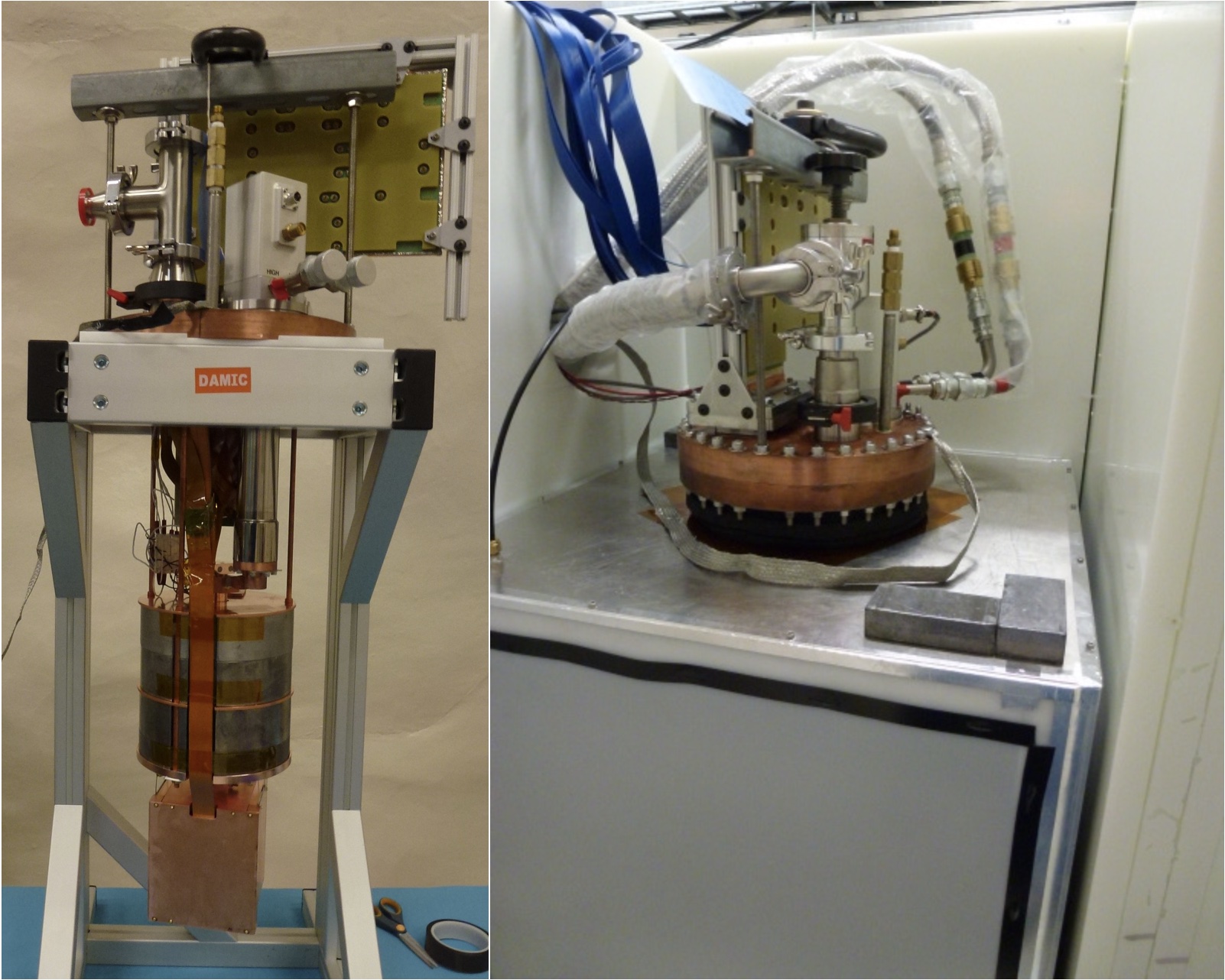}
	\caption{Photographs of the DAMIC detector at SNOLAB. \emph{Left:} cryostat insert showing the Kapton flex cables running from the CCD box to the VIB along the channel in the internal lead shield. The cold head is also visible. \emph{Right:} sealed copper cryostat inside its radiation shield, with electronics and service lines connected to the feedthrough flange.}
	\label{fig:setup}
\end{figure}

\begin{figure}[t]
	\centering
	\includegraphics[width=0.45\textwidth]{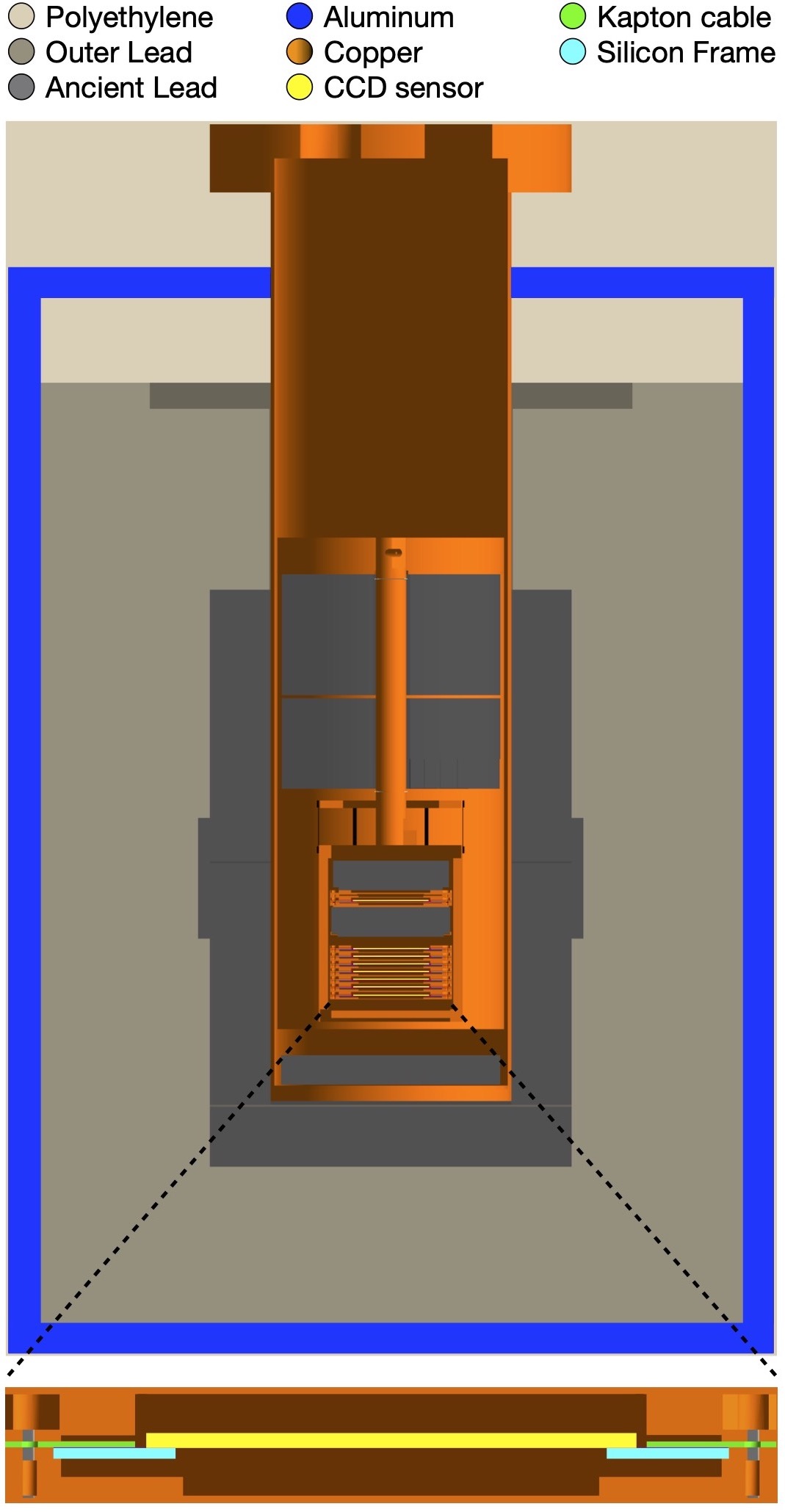}
	\caption{Cross-section of the \texttt{GEANT4} geometry of the DAMIC at SNOLAB detector, with each detector part considered in this analysis colored according to the top legend. For the copper parts, we include shadows from the 3D geometry to allow different parts to be visible. A zoomed cross-sectional view of a CCD module with modified aspect ratio is shown at the bottom.}
	\label{fig:geometry}
\end{figure}

The copper cryostat is lowered into a cylindrical hole in a rectangular lead castle with its flange remaining above the lead so that the ports are accessible.
In this configuration, the CCDs are shielded by at least 20~cm of lead in all directions, with the innermost 5~cm being ancient lead, and the outer lead being low-radioactivity lead from the Doe Run Company~\cite{exo}.
All ancient lead surfaces were cleaned in an ultra-pure dilute nitric acid bath~\cite{ABGRALL201622}.
A hermetic box covers the lead castle and seals around the neck of the cryostat.
Nitrogen gas with a flow rate of 2~L/min is used to keep the volume inside this box (around the cryostat) filled with pure nitrogen slightly above atmospheric pressure and free from radon.
Photographs of the detector are shown in Fig.~\ref{fig:setup} and a diagram of a cross-section of the geometry is shown in Fig.~\ref{fig:geometry}.

Around the lead castle and above the cryostat flange, 42~cm of high-density polyethylene shields against external neutrons.
A horizontal access hole through the polyethylene feeds through the pumping line, the helium lines from the outside compressor to the cold head, the boiloff nitrogen from a liquid-nitrogen dewar, and electrical cables. 
Blue coaxial cables connect the air-side of the VIB to the CCD controller, the Monsoon system developed for the Dark Energy Camera~\cite{Mclean:2012pka,Flaugher:2015}, that is located outside the shield. 
The Monsoon sends the clock signals and biases to the CCDs, and processes the analog CCD signal.
The Monsoon is programmed by a DAQ computer that receives the measured pixel values via optical fiber.

\subsection{Charge-Coupled Devices}\label{S:CCD}

DAMIC CCDs were developed by Lawrence Berkeley National Laboratory (LBNL) MicroSystems Lab~\cite{1185186} and consist of a $\sim 675~\mu$m-thick substrate of n-type high-resistivity ($>$10~k$\Omega$~cm) silicon with a buried p-channel and an array of $4116 \times 4128$ pixels on the front surface for charge collection and transfer. Each pixel is 15~$\times$~15~$\mu$m$^2$ in area and consists of a three-phase polysilicon gate structure. The CCDs feature a 1~$\mu$m thick in-situ doped polysilicon (ISDP) backside gettering layer that absorbs heavy metals and other impurities from the silicon substrate during manufacturing and acts as the backside contact to fully deplete the device during operation~\cite{HOLLAND1989537,heavyMetals}. The overall thickness of a CCD is estimated from the fabrication process flow to be $674 \pm 3$~$\mu$m, with an active thickness of $669 \pm 3$~$\mu$m and $\sim 2~\mu$m-thick dead layers on the front and back surfaces. Considering the pixel array area of 62~$\times$~62~mm$^2$, the total active mass of each CCD is 6.0~g.

A substrate bias of 70~V creates an electric field across the fully-depleted silicon bulk, in what we will refer to as the $\hat{z}$-direction. A nuclear or electronic recoil in the silicon will ionize silicon atoms and produce electron-hole (e-h) pairs over the band gap of 1.12~eV~\cite{bandgap}. Charges are drifted by the electric field, and holes are collected in the potential minimum below the polysilicon gates, where they are stored for the duration of an exposure. Clocking the three-phase gate potentials allows efficient transfer of charges across the CCD in the $\hat{y}$-direction and into the serial register (last row). Channel stops formed by ion implantation prevent movement of charge in the $\hat{x}$-direction (across columns) in the main pixel array. Clocking the gate potentials in the serial register shifts charges in the $\hat{x}$-direction and into the ``sense'' node, where the pixel charge is measured. Charge transfer inefficiencies (i.e., fraction of the charge left behind after every pixel transfer) in LBNL CCDs are typically $10^{-6}$~\cite{1185186}.

Two MOSFET readout amplifiers are located at opposite ends of the serial register, one of which is used to read out the transferred charge and the other to simultaneously read empty mirror pixels (as charges are always clocked in the same $\hat{x}$-direction).
The pixel charge is estimated with the correlated double sampling (CDS) technique~\cite{janesick2001scientific}.
First, any residual charge in the low-capacitance sense node is cleared with a reset pulse.
A reference value for the sense node is then obtained by integrating its potential over 40~$\mu$s.
The pixel charge is then transferred to the sense node and its potential measured again over the same integration time.
The difference between the two measured values is then proportional to the pixel charge.
After the two measurements, the pixel charge is discarded with a reset pulse and the procedure is repeated with the next pixel. 
The CDS technique has been previously demonstrated in DAMIC CCDs to have an uncertainty in the measurement of the pixel charge as low as 1.6~$e^-$~\cite{damicDMe} once its integration time is optimized to suppress high-frequency noise.
In the DAMIC electronics, the CDS operation is performed with an analog circuit whose output is sampled once per pixel by a 16-bit digitizer.
From the sequence of digitized values, we construct an image whose pixel values above the digitizer baseline are proportional to the charge collected in the CCD pixel array.

For this analysis, we used a subset of DAMIC data where a variable in the readout software was modified so that 100 rows of the CCD pixel array were transferred into the serial register before the serial register was clocked and the pixel charge measured (referred to as 1x100 readout mode). This downsampling procedure at the readout stage results in an effective pixel size of 15~$\times$~1500~$\mu$m$^2$. This procedure reduces the position resolution in the $\hat{y}$-direction but allows the charge from an ionization event (spread out over multiple rows) to be read out in a smaller number of measurements, substantially reducing the uncertainty in the total ionization charge (energy) from an interaction. With 1x100 readout, only a few percent of low-energy events populate more than one row in an image, which simplifies the analysis to one dimension (along rows), while retaining some coarse resolution in $\hat{y}$ to study the spatial distribution of observed events.

\begin{figure}[t]
	\centering
	\includegraphics[width=0.49\textwidth]{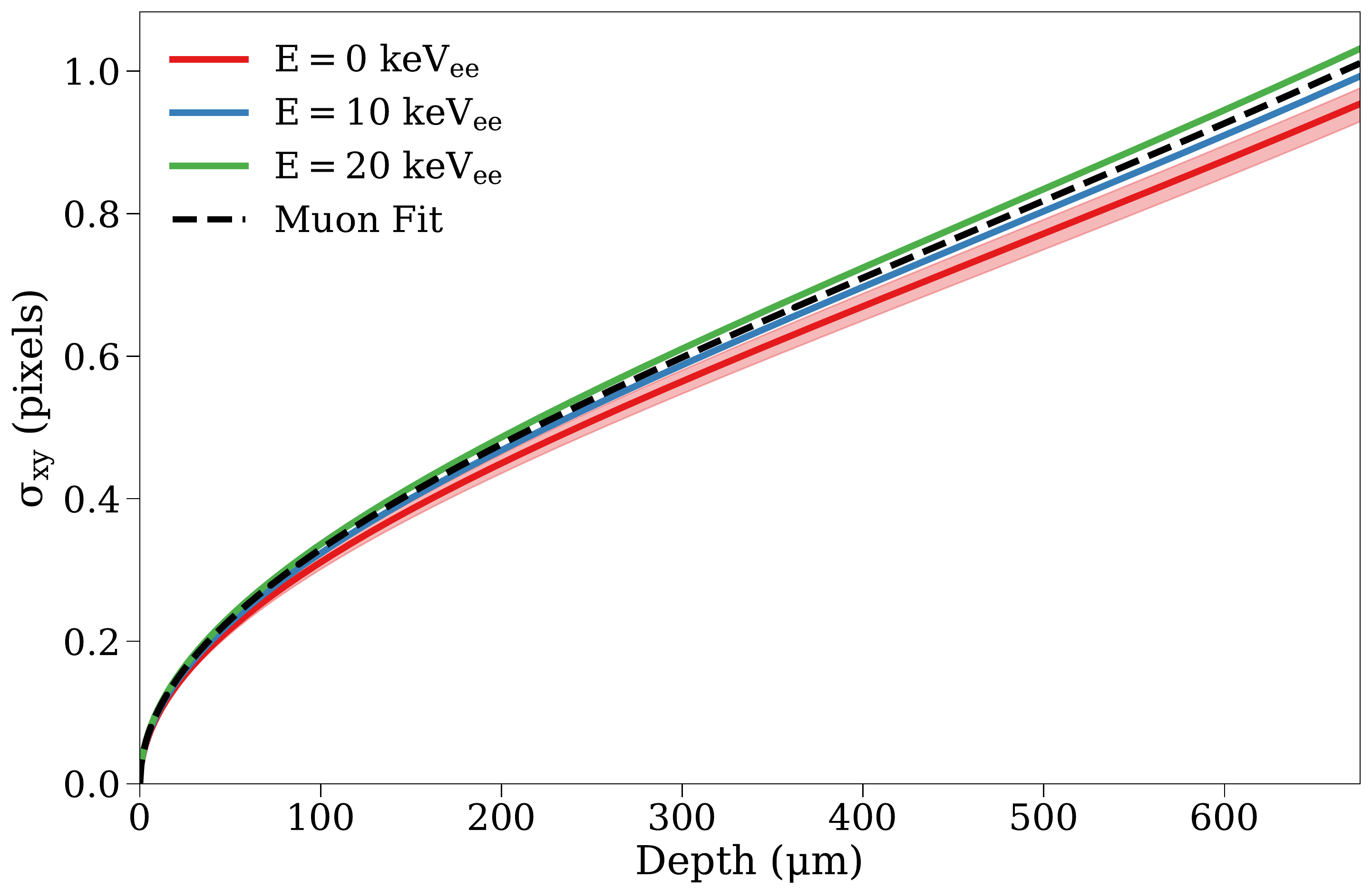}
	\caption{Diffusion model relating the reconstructed pixel width of a cluster ($\sigma_{xy}$ = \sx) to the true depth of the event in the CCD as calibrated from muon surface data (dashed). The baseline model with uncertainty band is provided for $E=0 \ \rm keV_{ee}$, along with overlaid solid lines accounting for linear energy corrections at $E=10,20 \ \rm keV_{ee}$. The lateral size of each pixel is $15~\mu$m.}
	\label{fig:diffusion}
\end{figure}

When ionization is first produced in the CCD bulk, it drifts in the electric field, and experiences stochastic thermal motion that leads to diffusion, which results in an increase in the lateral spread of charges on the pixel array. For a point-like interaction in the active region of the CCD, diffusion leads to a Gaussian distribution of charge with spatial variance $\sigma_{xy}^2=\sigma_{y}^2=\sigma_{x}^2$ that is proportional to the carrier transit time and, hence, positively correlated with the depth of the interaction. Because we use data from 1x100 readout mode in this analysis, the clusters are collapsed along the $y$ dimension and only the lateral spread in $x$ can be measured. Thus, we rely on $\sigma_x$ to reconstruct the depth $z$ of an event below the gates (defined as $z=0$). 
In general, diffusion can be modeled as~\cite{damic2016, steve}
\begin{equation}\label{eq:diffusion}
    \sigma_{xy}^2(z) = \sigma_x^2(z) = -a\ln|1-bz|,
\end{equation}
where the two parameters $a$ and $b$ are defined in the model as
\begin{equation}\label{eq:diffTheory}
\begin{split}
    a^{\mathrm{(th)}} &\equiv \frac{2 \epsilon_{\mathrm{Si}} k_B T}{\rho_n e} \approx 150 ~\mu\mathrm{m}^2 \\
    b^{\mathrm{(th)}} &\equiv \left( \frac{\epsilon_{\mathrm{Si}} V_b}{\rho_n z_D} + \frac{z_D}{2} \right)^{-1} \approx 10^{-3} ~\mu\mathrm{m}^{-1}.
\end{split}
\end{equation}
Here, $\epsilon_{\mathrm{Si}} = 636~e^2$~GeV$^{-1}$~fm$^{-1}$ is the permittivity of silicon at the operating temperature $T \approx 140$~K~\cite{permittivity}, $\rho_n~\approx 10^{11}~e$~cm$^{-3}$ is the nominal donor charge density~\cite{steve}, $k_B$ is the Boltzmann constant, $e$ is the charge of an electron, $V_b \approx 85$~V is the potential difference between the charge-collection well and the CCD backside\footnote{This differs from the substrate bias (70~V) since the potential minimum at the charge-collection well is $\sim 15$~V below ground~\cite{steve}.}, and $z_D \approx 669~\mu$m is the thickness of the CCD active region.

We use data taken with the DAMIC CCDs above ground at Fermilab (before deployment at SNOLAB) to extract the diffusion parameters.
These data were acquired with the same electronics as in SNOLAB and in 1x1 readout mode to maximize spatial resolution.
Surface laboratory images have a substantial background from muons, which deposit their energy in straight trajectories $\vec{[M]}$ through the silicon bulk.
From the ($x$,$y$) coordinates of the end points of an ionization track in an image, i.e., where the muon crosses the front ($z=0$) and the back ($z=z_D$) of the active region, we can reconstruct the $z$ coordinate at every point along the track.
We perform a fit to the charge distribution of the observed muon tracks by maximizing
\begin{equation}
    \log \mathcal{L_{\mathrm{muon}}} = \sum_{i=0}^N \log(f_q(q_i|a,b,\vec{[M]})),
\end{equation}
where $f_q(q_i)$ is the probability of measuring charge $q_i$ for the $i^{\mathrm{th}}$ pixel out of $N$ from a muon track constructed from trajectory $\vec{[M]}$ convolved with a Gaussian charge spread as a function of depth using Eq.~\eqref{eq:diffusion}.
From fits to a series of muon tracks, we obtain diffusion parameter values $a = 285 \pm 24~\mu\mathrm{m}^2$ and $b = (8.2 \pm 0.3) \times 10^{-4}~\mu\mathrm{m}^{-1}$, which are of the same order of magnitude as the theoretical calculations in Eq.~\eqref{eq:diffTheory}. 
In our 1x100 data taken at SNOLAB, we observe a percent-level deviation from this calibration that is proportional to the event energy $E$.
To correct for this dependence, we fit to a linear function the maximum observed spread $\sigma_{\rm max}$ of data clusters as a function of energy in 0.5~k\ev\ slices between 2--14~k\ev .
We find that $\sigma_x$ in the SNOLAB data is well-described after a multiplicative linear correction to Eq.~\eqref{eq:diffusion} of ($0.956+E/170 $~keV$_{\mathrm{ee}}$).
The final calibrated diffusion model is shown in Fig.~\ref{fig:diffusion}.

\subsection{Data Taking and Analysis}\label{S:Recon}

Since its initial installation in January 2017, the DAMIC detector acquired data almost continuously until September 2019 when the cryocooler failed.
From January to September 2017, data was taken in 1x1 readout mode dedicated to searches for $\alpha$ particles and decay sequences to determine the radioactive background sources in the detector (see Ref.~\cite{damicCoincidence2020}). Following a LED calibration campaign in the summer of 2017, from September 2017 to December 2018, we accumulated all data used for both the construction of the background model and the WIMP search; these data were taken in 1x100 readout mode with either 100 ks ($\approx 28$~hr) or 30 ks ($\approx 8$~hr) exposures, where each exposure results in seven images (one per CCD). 
The temperature of the CCDs (150K in the 1x1 data) was decreased with time to reduce dark current: down to 140K for the first two-thirds of the 1x100 data used in this analysis, and further to 135K for the remaining third. We were cautious with lowering the temperature because temperature-induced mechanical stresses are a known cause of amplifier luminescence.

The 1x100 images acquired at SNOLAB are 4116 columns $\times$ 42 rows. The first step in image processing is the removal of the image pedestal introduced by the digitizer baseline. The image pedestal is estimated to be $\sim 10^4$~ADU (analog-to-digital units) from the median value of every column and then subtracted from every pixel in the column.
The same procedure was repeated in row segments of 1029 pixels across the entire image to remove any residual pedestal trend in the $\hat{x}$-direction.
Noise picked up by the electronics chain of the system (e.g., by the cabling) is correlated between all CCD amplifier signals.
We suppress from each pixel the correlated noise by subtracting the weighted sum of the corresponding pixel in the mirror images from all CCDs with the weights evaluated to minimize the pixel variance. After this procedure, the per-pixel noise $\sigma_{pix}$ is $\approx$~1.6~$e^-$ ($\sim 6$~\ev{})~\cite{damicER}, where we calibrate the pixel value to collected charge using an LED installed inside the DAMIC cryostat by following the calibration procedure with optical photons detailed in Ref.~\cite{damic2016}. LED studies also demonstrate the linearity of the CCD energy scale to be within 5\% down to 10 $e^-$ ($\sim 40$ \ev{}) and confirm that the trailing charge after clocking the entire length of the serial register (4116 transfers) is $<$1\%. To translate between the collected charge and electron-equivalent energy we use 3.8\,\ev/$e^-$ from Ref.~\cite{senseiFANO} for the average kinetic energy deposited by a fast electron to ionize an e-h pair in silicon at the operating temperature of 140~K.
We refine the calibration of the electron-equivalent energy scale by fitting to known lines in the data from $^{210}$Pb X-rays (10.8, 12.7, and 13.0~keV) and copper $K_{\alpha}$ fluorescence (8.0~keV) for each individual CCD. We obtain an average gain of $0.257$~eV$_{\textrm{ee}}$~ADU$^{-1}$. Given the 16~bit dynamic range of the digitizer, this results in a (CCD-dependent) pixel saturation value of $\sim 14$~keV$_{\textrm{ee}}$. This calibration is in good agreement with similar CCDs calibrated on the surface with in-situ gamma and X-ray sources~\cite{damicCompton}.

Data selection based on the quality of the images is made upon experimental criteria.
The process starts with a visual inspection by the on-shift scientist during data taking, who flags any image with visible noise patterns for removal.
DAMIC data is divided into stable acquisition ``data runs" in between temperature changes or restart of the electronics, which were mostly caused by power outages at SNOLAB. 
Images at the beginning of each data run exhibit transients of high leakage current and are excluded. 
Following this removal of bad exposures, the ostensibly ``good" data set consists of 864 exposures that result in 6048 images.
We monitor radon levels around the DAMIC cryostat inside of the shield using a Rad7 monitor, and exclude periods when the average reading is above 5 Bq/m$^3$, excluding 63 exposures (441 images).
An additional set of 29 images across the 7 CCDs are removed as they show pixels with highly negative values ($<-5 \sigma_{\rm pix}$). Following image selection, the cumulative detector live time (per CCD) is 307~days.
Furthermore, we remove regions at the edge of the images by accepting the central region 128$<x\leq$3978  (5.78~cm in $\hat{x}$) and by excluding rows $y=1$ and 42, leaving 6.0~cm in $\hat{y}$. 
This edge cut removes $\approx$ 7$\%$ of pixels, which results in a reduced CCD target mass of 5.6~g.

Regions of spatially localized leakage current (as from lattice defects) that may mimic ionization events are excluded based on the procedure described in Ref.~\cite{damic2016}, in which maps of such areas are stored as masks that track individual pixels to be removed from consideration.
Masks are generated for every CCD in every data run based on the procedure described in Ref.~\cite{damic2016}. We calculate the median and median absolute deviation (MAD) of every pixel over all images in the data run and include in the mask pixels that either deviate more than three MAD from the median in at least 50$\%$ of the images or whose median or MAD is an outlier ($\gtrsim 5 \sigma$). Masks from 35 data runs are combined to generate a single ``iron mask'' per CCD.
The data runs considered consist of those used for this analysis and higher-temperature data runs acquired at 150~K and 170~K, which are more sensitive in identifying defects because of the strong dependence of leakage current on temperature~\cite{janesick2001scientific}. For the 150~K data runs, acquired in 1x1 readout mode, we rebin the masks into 1x100 format. The ``iron mask'' for each CCD is approximately the union of the masks from all data runs, except for isolated pixels that are only found in masks from a small number of data runs, consistent with statistical fluctuations in the pixel values. 
Averaging over all CCDs, the ``iron mask" removes 6.5$\%$ of pixels (keeping 93.5$\%$ of the average CCD mass).

We run clustering algorithms in unmasked regions of the images to identify charge spread over multiple pixels that originates from the same ionizing particle event. We apply two methods to cluster events, which we refer to as ``fast clustering'' and ``likelihood clustering'' and which are used in the background model construction (Section~\ref{S:Template}) and the WIMP search (Section~\ref{S:likelihood}), respectively.

\begin{figure}[t]
	\centering
	\includegraphics[width=0.49\textwidth]{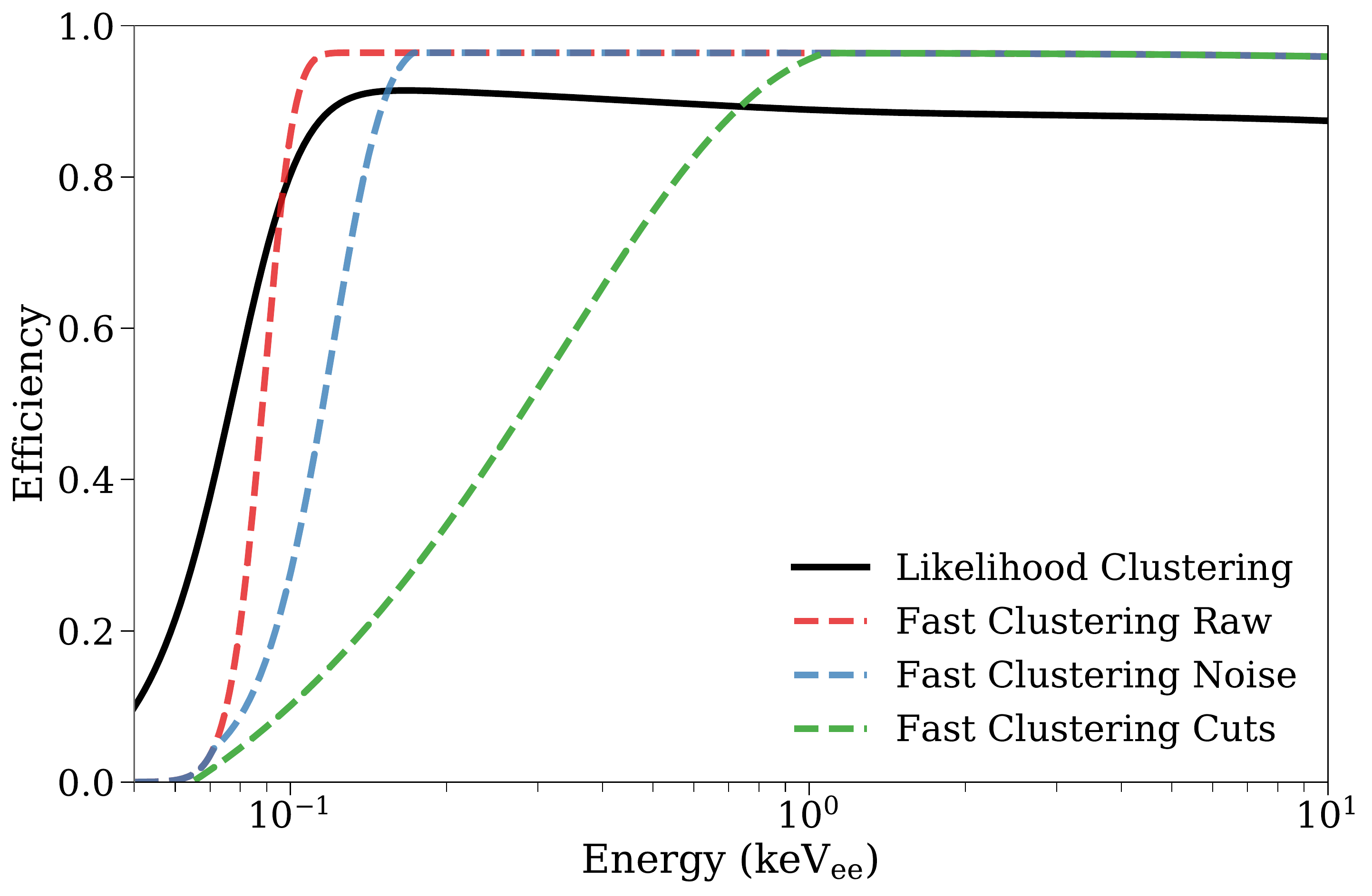}
	\caption{Comparison of fast vs. likelihood clustering efficiencies from our WIMP search threshold (50~\ev) to the maximum energy used in the likelihood clustering (10~k\ev). The fast clustering efficiency is determined using simulated bulk tritium decays, with curves shown for the ideal case without noise (red), with simulated pixel noise (blue), and after quality cuts (green). The likelihood clustering efficiency (black solid line) is determined from uniformly distributed events in energy and position added to zero-exposure ``blank'' images that contain only noise, with the efficiency drop-off below 120~\ev\, caused by the $\Delta LL$ selection and the plateau at higher energies dominated by the mask.}
	\label{fig:efficiencies}
\end{figure}

The ``fast clustering'' algorithm identifies contiguous groups of pixels each with signal larger than 4$\sigma_{\rm pix}$ ($\gtrsim$~24~\ev{}). This algorithm effectively clusters all types of ionization events (point-like low energy depositions, high-energy electrons, muon tracks, alpha particles, etc.) regardless of the pattern on the pixel array.
Because the fast clustering algorithm considers only pixels that contain charge, it can be used to efficiently translate many terabytes of simulated energy depositions into reconstructed coordinates without the need of a complete image simulation. 
If the sum of the pixel values in the cluster is $<$100~k\ev{}, we perform a 2D Gaussian fit to the pixel values to obtain the energy and $\sigma_x$ of the event.
We select events that do not contain and are not touching a masked pixel. 
We require that the integral of the best-fit Gaussian gives the same energy as the sum of pixel values to within 5$\%$, and that the fit converges with a minimum negative log-likelihood value that is consistent with other clusters of similar energy.
Figure~\ref{fig:efficiencies} shows the efficiency of the fast clustering algorithm for simulated events where the charge is distributed in an empty pixel array (raw), an array containing simulated readout noise, and
on the same array after applying the quality cuts mentioned here, which cause the efficiency to decrease rapidly for clusters with energies $<$1~k\ev.

Conversely, the ``likelihood clustering" algorithm is used to scan over a full image and calculate the likelihood of an energy deposition present in a window of predefined size. This algorithm only works for energies below 10~k\ev, where an energy deposition can reasonably be considered point-like (much smaller than the pixel size) prior to diffusion. Before running likelihood clustering on every image, in addition to the iron mask, we mask any pixels that are part of clusters found by the fast clustering algorithm with energies above $10$~k\ev{}, along with any pixels less than 200~pixels from the cluster in the $\hat{x}$-direction to avoid any low-energy events from charge transfer inefficiencies across the serial register. The algorithm is then used to iterate across the image row-by-row, calculating for each row segment (of variable but minimum width of 5 pixels) the likelihood that the pixel values can be described by white noise $\mathcal{L}_n$ or white noise plus a Gaussian distribution of charge $\mathcal{L}_G$. When the negative log-likelihood of their ratio, 
\begin{equation}
   \Delta LL \equiv - \log \frac{\mathcal{L}_G}{\mathcal{L}_n},
\end{equation}
is sufficiently negative, a cluster is identified. Most clusters identified in this way have a small negative $\Delta LL$ and can be attributed to noise.

\begin{figure}[t]
	\centering
	\includegraphics[width=0.49\textwidth]{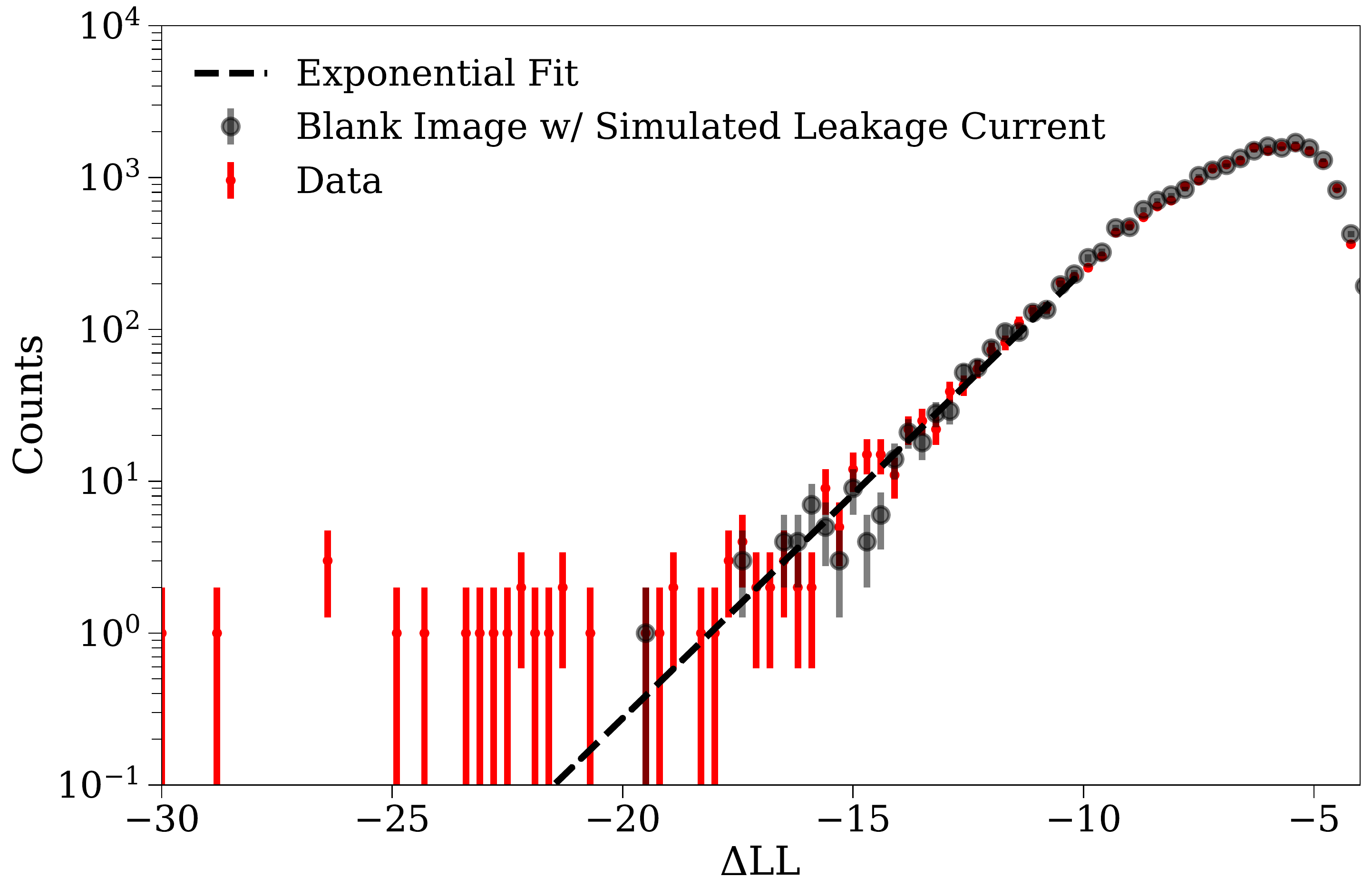}
	\caption{Distribution of the log-likelihood ratio of clusters from simulated noise images (black) compared to CCD data (red). The simulated distribution is fit (dashed line) to extract the $\Delta LL$ selection to reject noise clusters. The excess events in the tail of the data distribution above the expectation from noise correspond to low-energy ionization events.}
	\label{fig:dll}
\end{figure}

We produce simulated images starting from blank (zero exposure) images, which are acquired immediately after each exposed image so that they almost only contain the readout noise of the detector. We then introduce shot noise uniformly in space to simulate the effect of leakage current. The good agreement between the measured and simulated distributions at low negative $\Delta LL$ in Fig.~\ref{fig:dll} confirms that our modeling of noise clusters is accurate. To exclude such events from the WIMP search, we fit the simulated distribution with an exponential function and set a value of $\Delta LL < -22$ to ensure that there are fewer than 0.1 noise events expected in our dataset. 
This selection leads to a reconstruction efficiency $< 10\%$ below 50~\ev, which we set as our analysis threshold.
In addition, we apply the following quality criteria: we remove clusters containing or adjacent to a masked pixel, clusters that span more than one row, clusters too close to each other on an image such that their fitting windows overlap, and clusters where the maximum pixel contains less than 20$\%$ of the total charge (indicating a charge distribution that is too broad to arise from a point-like event). The efficiency of these cuts is shown in Fig.~\ref{fig:efficiencies} and is found to plateau around 90$\%$ above 120~\ev. 
The lower plateau value for the likelihood clustering compared to the fast clustering algorithm is because we consider clusters removed by the mask as an inefficiency in the likelihood clustering; since the simulation of events from fast clustering does not use blank CCD images, the CCD mask does not need to be applied.

\section{Background Sources and Detector Modeling}\label{S:Backgrounds}

As with all dark matter detectors, it is critical to examine all possible sources of background events in DAMIC at SNOLAB. These backgrounds are primarily radiogenic and come from radiochemical impurities in and activation of detector materials, contamination deposited on detector surfaces, and external sources. For an accurate estimate of the different background contributions in our detector, we simulate radioactive backgrounds with the \texttt{GEANT4} package, and include the effect of the partial charge collection region in the back of our sensors.

\subsection{Material Assays}
All materials used in the construction of DAMIC at SNOLAB were carefully catalogued and assayed for long-lived radioactivity. These assays focused primarily on the primordial $^{238}$U and $^{232}$Th chains, but also measured $^{40}$K in the detector materials. For the $^{238}$U chain, we consider the possibility that two different segments of this chain, delimited by $^{226}$Ra ($\tau_{1/2} = 1.6$~kyr) and $^{210}$Pb ($\tau_{1/2} = 22$~yr), may be out of secular equilibrium with $^{238}$U and with each other. A full list of the assay results of the materials used in the DAMIC at SNOLAB detector can be found in Table~\ref{tab:parttable}, and are detailed below.

\begin{table*}[!t]
\begin{center}
\resizebox{\textwidth}{!}{
\begin{tabular}{r|@{\hskip 0.1in} r@{\hskip 0.15in} r@{\hskip 0.15in} r@{\hskip 0.15in} r@{\hskip 0.15in} r@{\hskip 0.15in} r}
\hline \hline
\rule{0pt}{2.5ex} & \multicolumn{1}{c}{$^{238}$U} & \multicolumn{1}{c}{$^{226}$Ra} & \multicolumn{1}{c}{$^{210}$Pb} & \multicolumn{1}{c}{$^{232}$Th} & \multicolumn{1}{c}{$^{40}$K} & \multicolumn{1}{c}{$^{32}$Si}\\
\hline
\rule{0pt}{2.5ex}CCD / Si frame &$<$11~{\cite{damicCoincidence2020}} &$<$5.3~{\cite{damicCoincidence2020}} & $<$160*~{\cite{damicCoincidence2020}} &$<$7.3~{\cite{damicCoincidence2020}} &$<$0.5 $^{[M]}$ &$140 \pm 30$~{\cite{damicCoincidence2020}} \\
 Kapton Cable & 58000 $\pm$ 5000 $^{[M]}$ & 4900 $\pm$ 5700 $^{[G]}$ & not measured
& 3200 $\pm$ 500 $^{[M]}$ & 29000 $\pm$ 2000 $^{[M]}$ & N/A\\
OFHC Copper & $<$120 $^{[M]}$ &$<$130 $^{[G]}$ & 27000 $\pm$ 8000~\cite{xmass}
&$<$41 $^{[M]}$ &$<$31 $^{[M]}$ & N/A\\
Module Screws & 16000 $\pm$ 44000 $^{[G]}$ &$<$138 $^{[G]}$ &  27000 $\pm$ 8000$^\dagger$ 
& 2300 $\pm$ 1600 $^{[G]}$ & 28000 $\pm$ 15000 $^{[G]}$ & N/A\\
Ancient Lead &$<$23~\cite{spanishLead} & $<$260 $^{[G]}$ & $33000 \pm 10000^\ddagger$\cite{spanishLead}
& $2.3 \pm 0.1^\ddagger$\cite{spanishLead} &$<$5.8 $^{[M]}$ & N/A\\
Outer Lead &$<$13~{\cite{exo}} &  $<$200 $^{[G]}$ & ($19\pm5$)$\times 10^6$~\cite{exo} &$<$4.6~{\cite{exo}} &$<$220~{\cite{exo}} & N/A\\
\hline \hline
\end{tabular}}
\caption{\label{tab:parttable}Measured activities from material assay used to constrain the radioactivity in each simulated detector part, in units of $\mu$Bq/kg. A superscript $M$ ($G$) is used to indicate a value measured with mass spectrometry ($\gamma$-ray counting). The asterisk* on $^{210}$Pb in silicon is because this value is not used for the background model (but listed for completeness). The dagger$^\dagger$ on $^{210}$Pb in the brass screws indicates the crude assumption (in absence of direct assay results) that $^{210}$Pb activity in brass is comparable to copper, to show that this activity has minimal ($<$0.1 counts kg$^{-1}$ day$^{-1}$ k\ev{}$^{-1}$) effect on the final background model. The double-dagger$^\ddagger$ on $^{210}$Pb and $^{232}$Th in the ancient lead indicates a measurement which is fixed 
in the background-model fit to reduce degeneracy among the fit parameters.}
\end{center}
\end{table*}

For many of the materials, assays were performed using either Inductively-Coupled Mass Spectrometry (\text{ICP-MS}) or Glow Discharge Mass Spectrometry (GDMS). These techniques measure the elemental composition of a sample and estimate the activity of a specific isotope from its natural abundance. We adopt the standard values 1~Bq/kg $^{238}$U per 81 ppb U, 1~Bq/kg $^{232}$Th per 246 ppb Th, and 1~Bq/kg $^{40}$K per 32.3 ppm K~\cite{LOACH20166}. All materials constrained by mass spectrometry are indicated as such in Table~\ref{tab:parttable} with a superscript $M$.

Most materials used in the DAMIC at SNOLAB detector were screened by Germanium $\gamma$-ray spectrometry at the SNOLAB $\gamma$-ray counting facility~\cite{Lawson_2020}. This method non-destructively measures the isotopic abundance from the intensity of specific $\gamma$ lines from radioactive decays in the sample. All materials best-constrained by $\gamma$-ray counting are indicated in Table~\ref{tab:parttable} with a superscript $G$.
A special case is the Epotek 301-2 used to glue the CCD to its silicon frame, which was activated with neutrons from the nuclear research reactor at North Carolina State University before performing $\gamma$ ray spectroscopy of the activation products to estimate the abundance of the radiocontaminants initially present in the sample, a method known as Neutron Activation Analysis. The resulting measurement is below detectable limits and so the epoxy is omitted from this analysis; we place upper limits on Epotek 301-2 of $<220$, $<45$, and $<130$~$\mu$Bq/kg for $^{238}$U, $^{232}$Th, and $^{40}$K respectively, making this a suitable epoxy for future low radioactivity applications.

Some materials are best constrained by measurements published elsewhere. For example, the ancient lead used in the inner shield and around CCD~1 is from the same batch as the ``U. Chicago Spanish lead" sample in Ref.~\cite{spanishLead}. We thus use this measurement to constrain the bulk $^{210}$Pb content of our ancient lead. Similarly, the XMASS Collaboration has presented a robust analysis of the variation of bulk $^{210}$Pb content in commercially available OFHC copper~\cite{xmass}. While this is not a direct measurement of our copper, it provides an excellent starting point for our analysis. The PNNL electroformed copper (EFCu) used to house CCD~1 is far cleaner than the other materials of the detector~\cite{ARNQUIST2020163761} and as such is treated as perfectly radiopure. Finally, the low-activity lead used in the outer DAMIC shield is of the same origin as the lead assayed by the EXO Collaboration~\cite{exo}. We use the EXO measurements to confirm the negligible contribution this lead has on our background model.

While we have performed some direct assays of our CCD silicon, only the $^{40}$K estimate from Secondary Ion Mass Spectrometry (SIMS) is better than the constraints that we can place with the CCDs themselves. The CCD analysis, originally demonstrated in Ref.~\cite{damicBackgrounds2015} and recently updated in Ref.~\cite{damicCoincidence2020}, provides better numbers for many of the long-lived isotopes in our detector by leveraging spatially coincident events from the same decay chain occurring over long timescales.

\subsection{Material Activation}
In addition to these measured radioisotopes, we take into account the cosmogenic activation of the detector materials before they are taken underground to SNOLAB. Activation of a material happens when high-energy cosmogenic neutrons cause spallation of material nuclei~\cite{Cebrian:2017oft}. Here we consider only activation of the silicon CCDs and copper parts, including the Kapton readout cables, which are 70$\%$ copper by mass.

For copper activation, we consider the average activity $A$ of an isotope during the WIMP search exposure to be
\begin{equation}\label{eq:activation}
    A = \frac{S}{\lambda T_{\rm run}} \left( 1-e^{-\lambda T_{\rm act}} \right) \left( e^{-\lambda T_{\rm cool}} \right)  \left( 1-e^{-\lambda T_{\rm run}} \right)
\end{equation}
where the decay constant of the isotope is related to its half-life as $\lambda = \log(2)/\tau_{1/2}$, $S$ is the saturation activity of the isotope at sea level (from Ref.~\cite{laubenstein}), and $T_{\rm act}$, $T_{\rm cool}$, and $T_{\rm run}$ are the activation, cooldown, and run times, respectively~\cite{Baudis:2015kqa}. For copper activation, we consider seven activation isotopes, listed in Table~\ref{tab:chains}. The activation (cooldown) times of the copper modules, box, and vessel are 8 months, 16 months, and 1000 years (540 days, 300 days, and 6.6 years) respectively. The calendar run time is considered to be 441 days (September 27, 2017 -- December 18, 2018). We assume the initial activity of the copper vessel to be the saturation value by setting $T_{\rm act}$ to 1000 years since we do not know its exact history prior to manufacturing and being brought underground in 2012. This upper limit is used to determine that the activation of the copper vessel contributes less than 0.7~counts kg$^{-1}$ day$^{-1}$ k\ev{}$^{-1}$ to the detector background rate. 

\begin{table}[!t]
 \begin{center}
 \begin{tabular}{c@{\hskip 0.2in} c@{\hskip 0.2in} r }
    \hline \hline
    \rule{0pt}{2.5ex}Parent Chain & Isotope & Q value \\ \hline
       \rule{0pt}{2.5ex}$^{238}$U & $^{234}$Th & 274~keV  \\
       & $^{234m}$Pa & 2.27~MeV  \\ \hline
       \rule{0pt}{2.5ex}$^{226}$Ra & $^{214}$Pb & 1.02~MeV  \\
       & $^{214}$Bi & 3.27~MeV  \\ \hline
       \rule{0pt}{2.5ex}$^{210}$Pb & $^{210}$Pb & 63.5~keV  \\
       & $^{210}$Bi & 1.16~MeV  \\ \hline
       \rule{0pt}{2.5ex}$^{232}$Th & $^{228}$Ra & 45.5~keV  \\
       & $^{228}$Ac & 2.12~MeV  \\
       & $^{212}$Pb & 569~keV  \\
       & $^{212}$Bi & 2.25~MeV \\
       & $^{208}$Tl & 5.00~MeV \\ \hline
       \rule{0pt}{2.5ex}$^{40}$K & $^{40}$K & 1.31~MeV  \\ \hline
       \rule{0pt}{2.5ex}Copper & $^{60}$Co & 2.82~MeV  \\
       Activation & $^{59}$Fe & 1.56~MeV  \\
       & $^{58}$Co & 2.31~MeV  \\
       & $^{57}$Co & 836~keV  \\
       & $^{56}$Co & 4.57~MeV  \\
       & $^{54}$Mn & 1.38~MeV  \\
       & $^{46}$Sc & 2.37~MeV  \\
       \hline
       \rule{0pt}{2.5ex}$^{32}$Si & $^{32}$Si & 227~keV  \\ 
       & $^{32}$P & 1.71~MeV  \\ \hline
       \rule{0pt}{2.5ex}Silicon & $^{22}$Na & 2.84~MeV  \\ 
       Activation & $^{3}$H & 18.6~keV  \\ 
      \hline \hline
 \end{tabular}
 \caption{\label{tab:chains}Isotopes considered for the background model grouped by parent decay chain classification. Q values are provided for convenience from Ref.~\cite{Huang_2021,Wang_2021}.}
 \end{center}
 \end{table}

The cosmogenic activation of silicon at sea-level was only recently measured to be 124$\pm$24 atoms/kg-day for $^3$H and 49.6$\pm$7.3 atoms/kg-day for $^{22}$Na~\cite{damicActivation}\footnote{Ref.~\cite{damicActivation} also constrains the $^7$Be activation rate to be 9.4$\pm$2.0 atoms/kg-day. This isotope is not included in our background model since it should have decayed by the start of the WIMP search, nine months after the CCDs were brought underground, due to its short (53 day) half life.}. Previous constraints only provide approximate guidance on silicon activation rates~\cite{SuperCDMS:2018tqu}. Furthermore, the exact exposure history of the DAMIC CCDs is not well constrained. The original silicon ingot was pulled in September 2009, with CCD fabrication taking place in early 2016. The DAMIC CCDs were moved underground at SNOLAB on January 6, 2017. The total time spent on the surface (unshielded) prior to this date was 7.2 years at various altitudes, including roughly 53 hours of commercial air travel across 7 flights during the CCD manufacturing process. This is estimated at 10--11~km altitude minus take-off and landing to account for a surface equivalent exposure of ($2.0 \pm 0.2$) yr~\cite{flight}, resulting in an approximate sea-level-equivalent exposure time of roughly $9.2 \pm 0.2$~years. We assume an initial guess for the activity of bulk $^3$H in our CCDs of 0.3~mBq/kg (25 decays/kg-day) based on a preliminary analysis of 1x1 data, but choose to leave this as a free parameter in our analysis. $^{22}$Na decays into $^{22}$Ne, emitting a high-energy 1.27~MeV $\gamma$ ray that typically escapes the CCD. Positron emission occurs in 90.4\% of these decays, which leads to a $\beta$ track in the CCD. The remaining 9.6\% of decays occur by electron capture, resulting in  a K-shell line at 870~eV$_{\mathrm{ee}}$ of 8.9\% intensity from the deexcitation of the $^{22}$Ne atom~\cite{TabRad_v5}. This peak is directly measured in our CCDs and used to constrain the $^{22}$Na activity to 0.32$\pm$0.06~mBq/kg.
The corresponding L-shell line does not contribute significantly to our background since its mean energy is below the 50~eV$_{\mathrm{ee}}$ analysis threshold and its intensity is only 0.7\%.
    
\subsection{Surface Contamination}\label{SS:surface}
A dominant background and major uncertainty in the radioactive contamination of the DAMIC detector comes from the activity and location of surface $^{210}$Pb from ``radon plate-out'' on the detector surfaces. This happens when $^{222}$Rn ($\tau_{1/2} = 3.8$~days) in the air, emanating from materials following the decay of primordial $^{238}$U contamination, decays around detector parts during fabrication. The recoiling $^{218}$Po daughter ion strikes and sticks to nearby surfaces. The following sequence of $\alpha$ decays further embed the long-lived $^{210}$Pb ($\tau_{1/2} = 22.3$~years) daughter up to $100$~nm into a surface, where it will otherwise remain until it undergoes low energy $\beta$ decay. We assume that the profile of $^{210}$Pb embedded a distance $z'$ from a surface follows a complementary error function~\cite{RadonDiff} with characteristic maximum depth $z_M=50$~nm~\cite{SuperCDMSSoudan:2013ukv,ziegler}, i.e.,
\begin{equation} \label{eq:diffuse_depth}
    \mathrm{erfc}\left(\frac{z'}{z_M} \right) = \frac{2}{\sqrt{\pi}}\int_{z'/z_M}^{\infty}e^{-t^{2}}dt.
\end{equation}

The $\beta$ decays of $^{210}$Pb and its daughter $^{210}$Bi are a dominant background and will be discussed at length in the following sections. The daughter of $^{210}$Bi, $^{210}$Po emits $\alpha$ particles, which, while not a low-energy background for the WIMP search, are used for background studies.
For every $^{210}$Po $\alpha$ decay there is also a recoiling $^{206}$Pb nucleus with 103~keV of kinetic energy. 
These recoils would be a dangerous low-energy background if they were to deposit significant energy in the sensitive regions of the CCD. 
A simulation based on SRIM~\cite{SRIM} 
estimates their range to be $\sim$45\,nm, which results in most of their energy being deposited in inactive silicon, with an upper limit of the energy deposited in the sensitive region of 35~eV.
Since a 35\,eV nuclear recoil generates a signal $\ll50$\,\ev\ in silicon, this background is not considered in the analysis.

The activity of $^{210}$Pb is notoriously difficult to measure by standard assay techniques: its abundance is too low to be detectable by mass spectrometry for measurable activities in the detector because of its relatively short half-life, and its decay products are too low in energy to be efficiently detected by $\gamma$-ray counters~\cite{BUNKER2020163870}. Although we monitored the radon level in the air during CCD packaging and chemically cleaned the copper and lead parts to remove surface contamination, we have otherwise limited \emph{a priori} knowledge about which detector surfaces have experienced radon plate-out, or to what degree. As an initial guess, we use a surface activity of 70~decays~day$^{-1}$~m$^{-2}$ for $^{210}$Pb deposited on all surfaces~\cite{damicBackgrounds2015}, which is then left as a free parameter in our analysis. 
We allow for different surface activities of $^{210}$Pb on the front and back of the CCDs, although we require that all CCDs have the same contamination. 
The amount of deposition depends on the height of the air column above a surface~\cite{RadonDiff}. Since the CCDs are never placed face down to avoid damaging the wire bonds, it is likely that more $^{210}$Pb will be on the front surface, although we do not require so in the fit.
Additionally, the silicon wafers were stored for several years exposed to air in a vertical position before being manufactured into CCDs. 
Up to 10~$\mu$m of silicon is later removed from the front of the wafer in a polishing step during CCD fabrication, effectively eliminating any $^{210}$Pb deposited on the front side of the wafer. However, no significant thickness of silicon is removed from the backside before the deposition of the ISDP gettering layer, thus a layer of $^{210}$Pb contamination $\sim 3~\mu$m below the backside of the CCD is expected to be an important background.

We also consider $^{210}$Pb deposition on the detector surfaces around the CCDs. Our cleaning procedures remove a few $\mu$m of material from the copper and lead surfaces, which should effectively eliminate surface $^{210}$Pb contamination~\cite{BUNKER2020163870}. As a cross-check, we repeat the same analysis presented later in Section~\ref{S:Template}, allowing for surface $^{210}$Pb on the copper components, and find no substantive change to our result. Separately, we allow for surface $^{210}$Pb on the silicon frames that support the CCDs and find the magnitude of surface $^{210}$Pb activity needed to have an impact on the background model is roughly an order of magnitude higher than the amount placed on the CCD surfaces by the fit; since the history and treatment of all silicon surfaces is similar, we ignore this component.
In the background model construction, we do not consider any additional surface contamination in the form of dust particulates, for which the accumulation rates at SNOLAB have been measured~\cite{DIVACRI2021165051}.

In Section~\ref{S:Alphas}, we compare our fit result to an independent analysis on the activity of surface $^{210}$Pb in our detector from the rate of spatially coincident $^{210}$Pb-Bi decay sequences and the location of $^{210}$Po $\alpha$ decays.

\subsection{Muon and Neutron Backgrounds}

The cosmic muon flux at SNOLAB of 0.27\,m$^{-2}$d$^{-1}$~\cite{snolabhandbook} corresponds to 1 muon crossing the CCD array every $\sim 1000$ days. With a mean kinetic energy $>$300~GeV~\cite{meiandhime}, the muon would generate a prominent particle shower in the detector, with a large number of clusters across the CCD array within the same image exposure. No such muon events were observed in the WIMP search.

Neutrons can produce ionization signals in the detector either directly by nuclear recoils from the scattering of fast neutrons in the silicon or indirectly by the interactions of $\gamma$ rays emitted following the capture of thermal neutrons by detector materials.
The flux of external neutrons from the cavern walls (0.5~cm$^{-2}$\,d$^{-1}$~\cite{snolabhandbook}) is suppressed to a negligible level ($<10^{-3}$~cm$^{-2}$\,d$^{-1}$) by the polyethylene shield.
Neutrons may also be produced in the detector materials by i)~$\text X(\alpha,n) \text Y$ reactions involving light nuclei ($\sim 90\%$), ii)~spontaneous fission of uranium ($\sim 10\%$), and iii)~spallation induced by cosmic muons interacting in the polyethylene and lead shield ($<$1 d$^{-1}$).
Following a survey of all detector components, we determined that neutron production within the detector is dominated by $(\alpha,n)$ reactions in the VIB above the internal lead shield (30 d$^{-1}$).

Two sets of simulations were used to quantify the expected number of neutron-induced events in the WIMP search exposure.
The first simulation determined that the cavern neutron flux leads to only $5.6\pm0.5$ neutrons crossing the sensors throughout the WIMP search, corresponding to $< 1$ neutron scattering event. 
The second simulation estimated an expected number of 0.1 nuclear recoils in the WIMP search exposure from neutrons emitted by the VIB.
Since the number of ionization events induced by neutrons is orders of magnitude below the observed event rate, their contribution is not considered in the background model.

\subsection{GEANT4 Simulations}\label{S:Simulation}
\begin{figure}[t]
	\centering
	\includegraphics[width=0.5\textwidth, trim=0 0 0 0, clip=true]{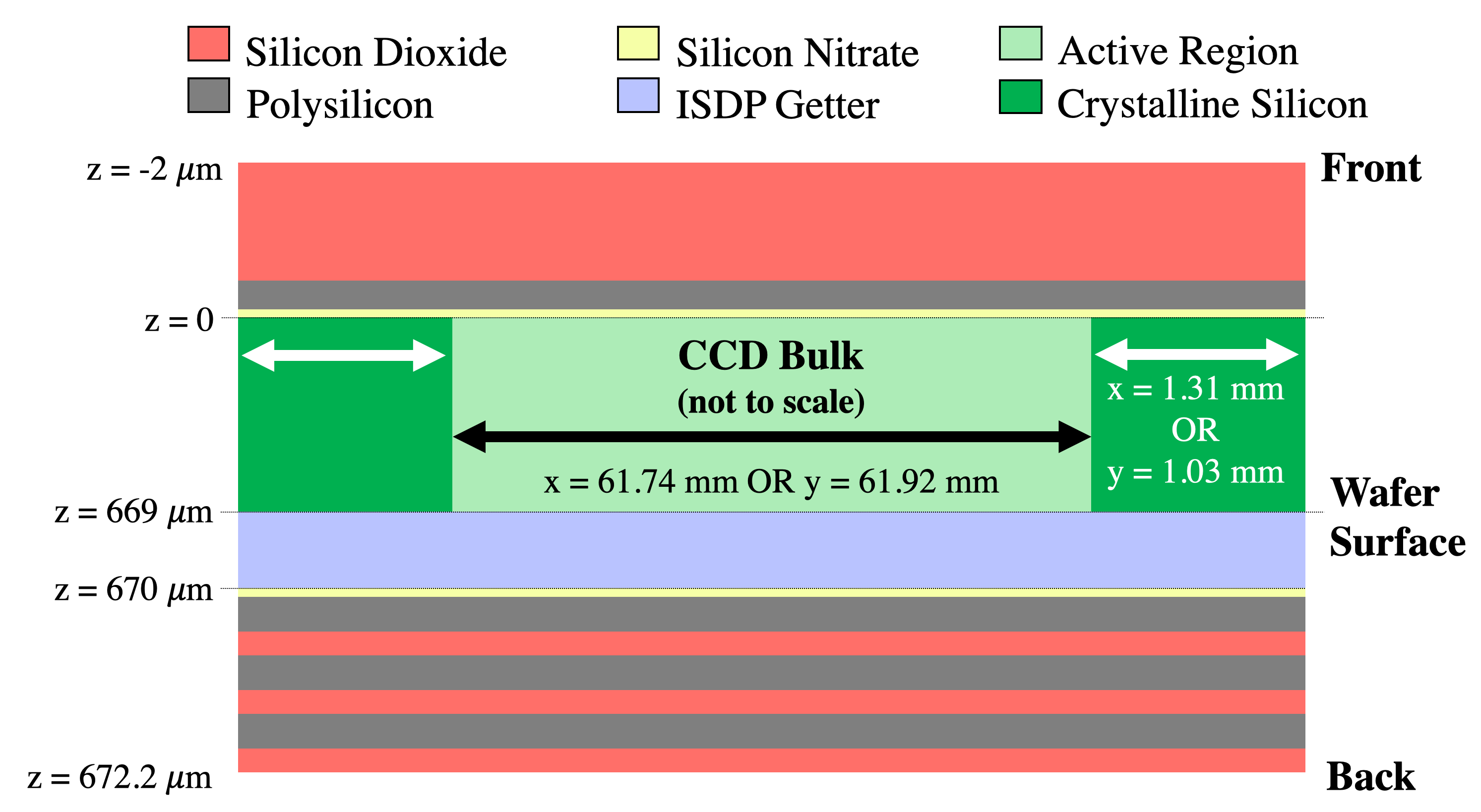}
	\caption{Cross-section of the \texttt{GEANT4} geometry of the DAMIC at SNOLAB CCD sensor, with each sub-layer colored according to the top legend. Key dimensions are indicated, with further detail in the text.}
	\label{fig:structure}
\end{figure}

To simulate the spectrum of ionization events from radioactive background sources, we utilize the \texttt{GEANT4} simulation framework~\cite{geant4} with the Livermore physics list~\cite{osti_295438,osti_5691165,osti_10121422}. 
The implemented physics models are evaluated for electron energy depositions down to 10~eV and photon cross sections down to 100~eV. 
The CCD geometry is broken down into a series of sub-regions to maximize accuracy, as shown in Figure~\ref{fig:structure}~\cite{joaothesis}.
The active region (where energy depositions are recorded) is a $61.74\rm{~mm} \times 61.92\rm{~mm} \times 669.0~\mu\rm{m}$ rectangular volume laterally centered on a rectangular silicon block of size $64.36\rm{~mm} \times 63.99\rm{~mm} \times 674.2~\mu\rm{m}$.
The frontside dead layer includes $2.0~\mu$m total of insulator ($1.6~\mu$m SiO$_2$), polysilicon gate electrodes ($0.3~\mu$m Si), and gate dielectric ($0.1~\mu$m Si$_3$N$_4$). Similarly, the backside dead layer consists of the ISDP gettering layer ($1.0~\mu$m Si), a dielectric layer ($0.1~\mu$m Si$_3$N$_4$), and three sets of alternating polysilicon ($0.4~\mu$m Si each) and silicon dioxide ($0.3~\mu$m SiO$_2$ each) layers, for a total simulated backside thickness of 3.2~$\mu$m.\footnote{There are slight differences in dimensions between the simulated backside layers and the SIMS results presented later in Section~\ref{S:PCC}. For the background model, these differences are largely absorbed into the partial charge collection uncertainty, addressed further in Section~\ref{S:PCC}.} In all cases above, the layers are listed according to increasing $z$ position (front-to-back). See Ref.~\cite{HOLLAND1989537} for more details on the CCD  structure.

We apply a different selection on the minimum range of secondary particles generated, i.e., a ``range cut,'' in different detector volumes to optimize computing time.
Components outside of and including the copper vessel are simulated with a 1~mm range cut. Components inside the vessel but outside the copper box (such as the cold finger) are simulated with a 100~$\mu$m range cut. The copper box itself and everything inside (lead bricks and copper modules, including screws and cables) are simulated with a 1.3~$\mu$m range cut. The CCDs are simulated with a 50~nm range cut, corresponding to the range of $\sim 20$~eV electrons~\cite{bethe1930,bethe1932}.

For the purpose of executing simulations, the detector is divided into 64 volumes, each belonging to one of the detector parts listed in Table~\ref{tab:parttable}. 
For each volume, we individually simulate up to $5\times10^8$ unique $\beta/\gamma$-decays from isotopes that can contribute to low energy backgrounds, optimizing the number of decays simulated to not introduce significant uncertainty from statistical fluctuations in the spectra. These isotopes come from the $^{238}$U, $^{232}$Th, and $^{40}$K decay chains, or material-specific isotopes (as from activation). 
In addition to isotopes uniformly distributed within detector volumes, we simulate $^{210}$Pb (and its daughter $^{210}$Bi) deposited on the surfaces of the detector in three groups: the front, the back of the CCDs, and the back of the silicon wafers prior to CCD fabrication. 
All isotopes considered are shown in Table~\ref{tab:chains}.

DAMIC simulation code developed within \texttt{GEANT4} outputs the precise energy and position coordinates of each interaction in our CCDs. We process the \texttt{GEANT4} raw outputs into a compressed format that retains all information for event reconstruction. This is done event-by-event for each simulated decay by taking each energy deposition and converting it into some number of electrons, assuming 3.8~\ev{}/$e^-$ and a Fano factor $F=0.129$~\cite{senseiFANO,karthikIonization}. The electrons are then distributed according to our diffusion model (Section~\ref{S:Recon}) on a grid where each cell is the size of a CCD pixel. For energy depositions in the partial charge collection region (Section~\ref{S:PCC}), we additionally assign some probability based on the depth of the interaction that a simulated electron will recombine. 

A custom software package is then used to convert this reduced, pixelated \texttt{GEANT4} output into an analogous output from a CCD image. To do this, we rebin to match our 1x100 image format, add white readout noise to each pixel, and simulate pixel saturation by setting a maximum energy per pixel based on the measured image pedestal values and CCD calibration constants.
Then, we run the ``fast clustering" algorithm, detailed in Section~\ref{S:Recon}, on the simulated pixels, which allows us to compress many terabytes of simulated data into a list of reconstructed cluster variables including $E$, $\sigma_x$, and mean $(x,y)$ for the simulated cluster.
We additionally carry through information about the simulated collected energy $E_{\rm sim}$, obtained by multiplying the simulated number of collected electrons by 3.8 \ev{}, and the mean depth $z$ of each event. We use only the reconstructed information in $E$ and $\sigma_x$ to construct the background model, leaving $(x,y)$ position information as a cross-check.

\subsection{Partial Charge Collection}\label{S:PCC}
The dominant uncertainty in the response of the DAMIC CCDs is the loss of ionization charge by recombination in the CCD backside.
The 1~$\mu$m thick ISDP layer that acts as the backside contact is heavily doped with phosphorous (P).
Phosphorous diffuses into the CCD during fabrication, which leads to a transition in P concentration from $10^{20}$ cm$^{-3}$ in the backside contact, where all free charge immediately recombines, to $10^{11}$ cm$^{-3}$ in the fully-depleted CCD active region, where there is negligible recombination over the free carrier transit time.
A fraction of the charge carriers produced by ionization events in the transition region may recombine before they reach the fully-depleted region and drift across to the CCD gates. This partial charge collection (PCC) causes a distortion in the observed energy spectrum because events occurring on the backside have a smaller ionization signal than they would if they were to occur in the fully-depleted region.

\begin{figure}[t]
	\centering
	\includegraphics[width=0.49\textwidth, trim=30 0 50 20, clip=true]{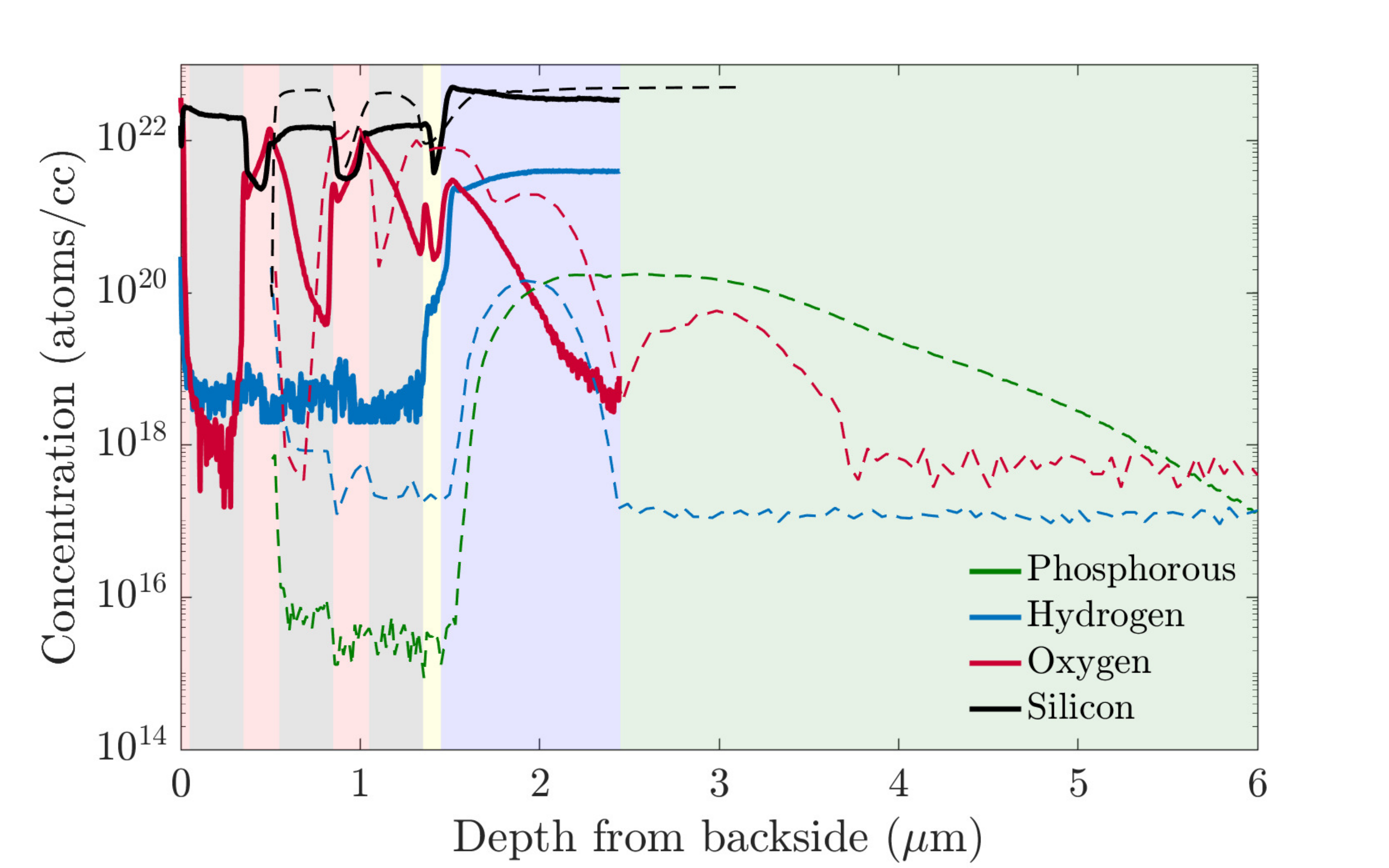}
	\caption{SIMS results of the CCD backside (lines): elemental concentrations of oxygen (red), phosphorous (green), and hydrogen (blue) as a function of depth, with silicon (black) for reference. Two different measurements using fine (solid) and coarse (dashed) beam widths are shown, which probe different depths into the CCD. The backside layer structure is clearly visible. Shaded areas from left to right: alternating SiO$_2$ (red) and polysilicon (grey) layers, Si$_3$N$_4$ (yellow), ISDP layer (blue), and high-resistivity $n$-type crystalline silicon (green).}
	\label{fig:sims}
\end{figure}

To construct an accurate model of the CCD backside, we performed measurements by secondary ion mass-spectrometry (SIMS) of the concentration of different elements from the backside of some wafer scraps from the same batch as the DAMIC at SNOLAB CCDs, with results shown in Fig.~\ref{fig:sims}. The first measurement had fine resolution and was used to measure silicon, hydrogen, and oxygen content $\sim 2.5~\mu$m into the backside of the wafer. The second measurement used a wider beam, reducing resolution, and measured silicon, hydrogen, oxygen, and phosphorous content up to $\sim 7.0~\mu$m deep (at which point all trace elements are below the measurement sensitivity). Both of these measurements clearly resolve the outermost 1.5~$\mu$m polysilicon and silicon-dioxide layers.

\begin{figure*}[t]
	\centering
	\includegraphics[width=0.49\textwidth]{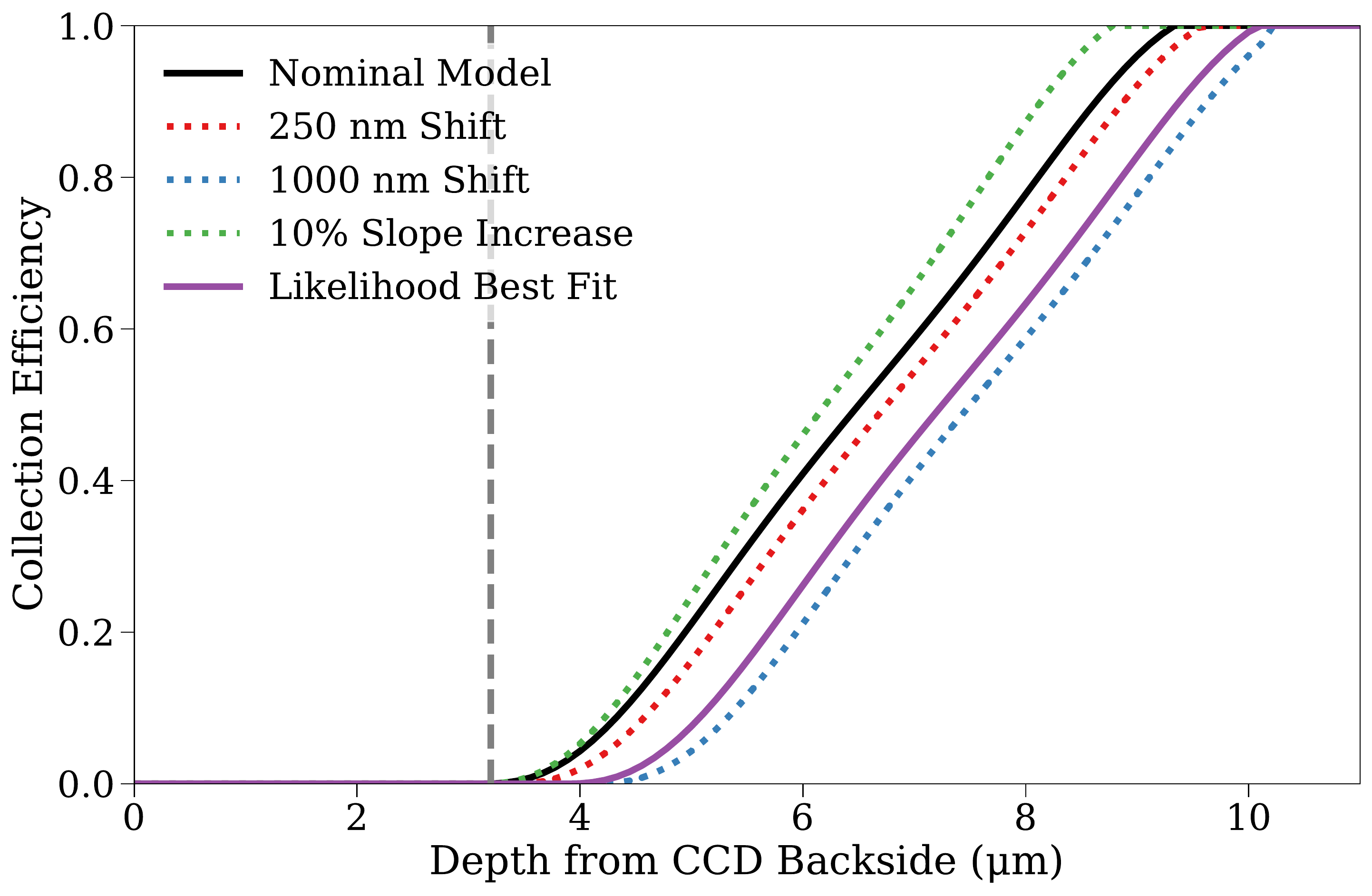}
	\includegraphics[width=0.49\textwidth]{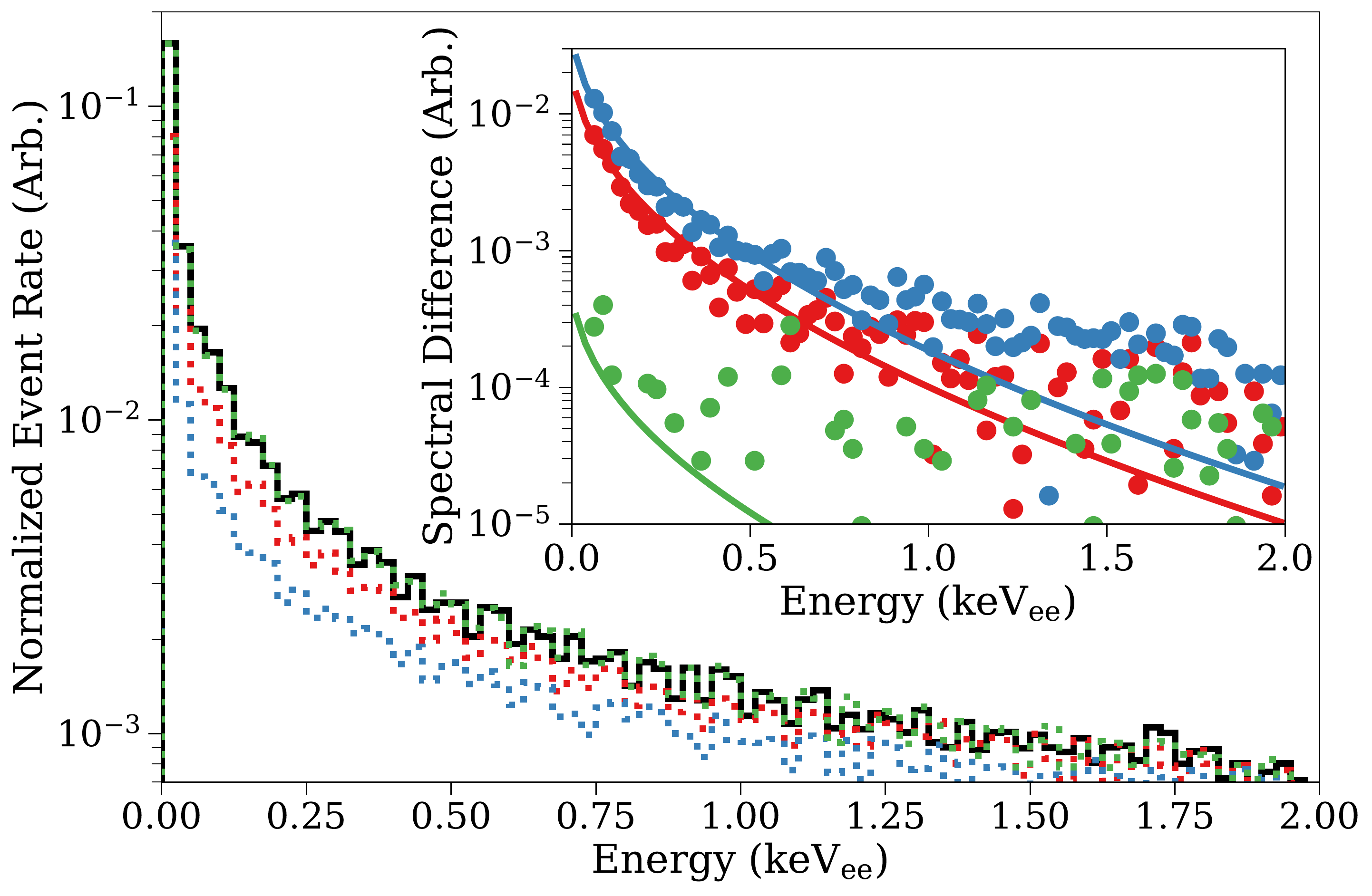}
	\caption{{\bf Left:} Nominal (black line) and variations (dotted lines) in the PCC model within systematic uncertainty. The simulated backside of the wafer is marked by the dashed-gray vertical line. The dotted-red (blue) line represents a shift of the nominal PCC model 0.25 (1.0) $\mu$m into the CCD and dotted-green represents and shrinking of the PCC region by $\sim10\%$. The solid purple line represents the best-fit result presented in Section~\ref{S:likelihood}. {\bf Right:} Spectra from $\rm {}^{210}Pb$ and $\rm {}^{210}Bi$ events on the backside of the silicon wafer for the different PCC models (same color legend as left), normalized to the simulated number of  $\rm {}^{210}Pb$ decays. The inset shows the spectral difference between the nominal and the varied PCC-model examples, with the result of the fit with Eq.~\ref{eq:pcc} overlaid. The position of the ``turn on'' point of the PCC region dominates the difference in the relative spectra. 
	}
	\label{fig:pcc_sims}
\end{figure*}

An extrapolation of the P concentration from the SIMS result with an exponential function suggests that the transition from the ISDP layer to the bulk value of $10^{11}$~cm$^{-3}$ occurs over a distance of $\approx 8$~$\mu$m.
The P concentration determines the transport properties (mobility, lifetime) of the free charge carriers and the electric field profile, which control the fraction of the free charge carriers that survive recombination.
We perform a numerical simulation to estimate the charge collection efficiency as a function of depth from the P concentration profile, with details provided in Appendix \ref{A:pcc}.
The nominal result, which shows a transition from zero to full charge collection over $\approx 5$~$\mu$m (from 4 to 9 $\mu$m from the backside surface), is presented by the black line in Fig.~\ref{fig:pcc_sims}.

The exact location and profile of the PCC region has a significant impact on the spectral shape of radioactive decays originating on the backside of the CCDs, specifically $^{210}$Pb and $^{210}$Bi decays just below the ISDP layer, which were originally on the backside of the wafer (see Section~\ref{SS:surface}).
To quantify the systematic uncertainty that PCC introduces in our background model, we compare simulated events on the back of the CCDs under different PCC-model assumptions.
We vary the slope and position of the PCC curve and analyze the resulting spectra from backside ${}^{210}$Pb and ${}^{210}$Bi decays. Three examples from this ensemble are shown by the dotted lines in Fig.~\ref{fig:pcc_sims}. The spectral difference between the nominal and varied models can be generally parametrized by the function:
\begin{equation} \label{eq:pcc}
    f(E | \beta) = N\exp ( -\sqrt{E} / \beta),
\end{equation}
where the amplitude $N$ depends on the specific PCC model and is primarily sensitive to the distance between the backside and the depth at which the collection efficiency ``turns on.'' We find a single $\beta = 0.18 \ \rm keV_{ee}^{-1/2}$ to describe well all models in the ensemble, reducing Eq.~\eqref{eq:pcc} to depend on a single parameter $N$.

As shown in the inset of Fig.~\ref{fig:pcc_sims}, the parametrization describes the spectral differences well at low energies where they are most significant. The accuracy of the parametrization worsens above $\sim 1$ k\ev , which only accounts for $<2\%$ of the spectral difference and is negligible compared to the statistical uncertainty of the measured spectrum for the WIMP search.

The SIMS profile also shows an unexpectedly high concentration of hydrogen in the 1~$\mu$m thick ISDP region. While the origin of the hydrogen is unknown, it likely is captured during the deposition of the ISDP layer, which occurs at $600^{\circ}$C in the presence of SiH$_4$ and PH$_3$ gases~\cite{hydrogenDespotion}. The hydrogen itself is not a problem but likely contains trace levels of radioactive tritium. Given the hydrogen concentration of $10^{20.2-21.6}$~H/cm$^3$ measured by SIMS and the approximation that the tritium content in hydrogen in both of these gases is comparable to water ($\approx 10^{-17}$~$^3$H/H~\cite{PLASTINO200768,plastino2011}), we expect tritium activities in the ISDP layer on the order of 1--30~mBq~kg$^{-1}$. 
This is higher than the final best-fit activity of bulk $^3$H that we find from direct silicon activation, but due to the low mass of the ISDP layer, it only amounts to 10--300~nBq per CCD (less than one decay per month).
Moreover, our simulations suggest that only a small fraction of the low energy $\beta$'s originating in the ISDP layer penetrate into the charge collection region and contribute to the observed event rate.
Thus, we exclude this background from our model.

\section{Background Model}\label{S:BackModel}

For comparison with data, our understanding of radioactive background sources and detector response must be converted into a background model: the expected distribution of ionization events in our CCDs in energy and depth.
We construct the background model from a fit to data from CCDs 2--7 above 6\,k\ev\ with templates generated from our \texttt{GEANT4} simulation output.
We validate the fit result with CCD~1 above 6~k\ev\ and demonstrate consistency with independent estimates of $^{210}$Pb surface activity from individual event identification.
The background model is extrapolated below 6\,k\ev\ for our WIMP search.

\subsection{Template Fitting}\label{S:Template}
We perform a binned-likelihood Poisson template fit to data in $(E,\sigma_x)$ space starting from the over $1000$ combinations of \texttt{GEANT4} simulations for each isotope-volume. We consider events reconstructed by our fast clustering algorithm with energies between 6--20~k\ev, where we do not expect a WIMP signal since other silicon-based experiments have placed strong bounds on WIMP masses that populate this high-energy range ($m_\chi>10$~GeV)~\cite{CDMS:2013juh}. Although reconstructed events begin to saturate pixels above 14~k\ev, we consider clusters up to 20~k\ev\ that do not have a saturated pixel to accept the full tritium $\beta$ spectrum, which is expected to be a dominant bulk component.  
We develop the background-model construction on a subset (64$\%$) of the data, adding the remaining data for the final background model once the methodology has been fixed. We reserve all data from CCD~1 as a further cross-check, since the background environment in this CCD is different from CCDs 2--7. 

We group like-simulations together according to common materials and decay chains.
The grouping for materials is according to the detector parts in Table~\ref{tab:parttable}, with further subdivision of the copper into the OFHC copper vessel, box, and modules (because of their different activation histories), and the EFCu CCD 1 module.
All EFCu is assumed to be perfectly radiopure, as it contributes $\ll 1$~count kg$^{-1}$ day$^{-1}$ k\ev{}$^{-1}$. For each detector part, we further group simulated isotopes by decay chain, according to Table~\ref{tab:chains}. To avoid introducing uncertainty from the limited statistics of subdominant components, we discard any simulations that result in fewer than 0.1 events kg$^{-1}$ day$^{-1}$ and less than 1000 simulated events reaching the CCDs, leaving 289 simulations. 
The result of this grouping is 49 distinct sets of simulations, from which we construct templates (one per set).

For template $l$, the number of expected events $\nu_{ijl}$ in bin $(i,j)$ (of width 0.25~k\ev\ in $E$ and 0.025~pixels in \sx ) is calculated by summing over the integer number of events $n_{ijm}$ in each binned simulation output $m$ normalized by the corresponding template activity $A_l$ in Table~\ref{tab:parttable}, the mass of the material simulated $M_m$, the efficiency-corrected exposure of the data ($\epsilon_{data}~t_{\rm run}$), and the efficiency-corrected number of decays simulated $(\epsilon_{sim}~N_m)$ according to
\begin{equation} \label{eq:normalization}
\nu_{ijl} = \sum_m n_{ijm} \times \frac{A_l~M_m~(\epsilon_{data}~t_{\rm run})}{(\epsilon_{sim}~N_m)}.
\end{equation} 
The efficiencies of the data $\epsilon_{\rm data}$ and simulation $\epsilon_{\rm sim}$ are different since we find that the cluster selection on simulation to be 97.0$\%$, higher than the 93.3$\%$ observed for the data (Fig.~\ref{fig:efficiencies}). Additional study determines that the selection efficiency for simulated events varies slightly with depth, being nearly fully efficient at the front of the CCD and closer to 96$\%$ efficient at the back; this percent-level effect is noted but not taken into account in this analysis. No energy dependence in these efficiencies was observed.
Furthermore, we include in $\epsilon_{\rm sim}$ a correction for the fraction of the CCDs that are masked (6.5\%) rather than applying the iron mask (see Section~\ref{S:Recon}) to maximize the simulation statistics. 

In total, the data is divided into 2464 bins, 56 in energy between 6--20~k\ev\ and 44 bins in $\sigma_x$ between 0.1--1.2~pixels. We implement a custom binned likelihood fit using \texttt{TMinuit} to compare the observed number of events $k_{ij}$ in each bin against the expected number of events from simulation $\nu_{ij}$, obtained from the addition of the simulated templates $l$ with scaled amplitudes:
\begin{equation}
    \nu_{ij} = \sum_l C_l \nu_{ijl}.
\end{equation}
Under the assumption that the probability of observing $k_{ij}$ events in a bin from an expectation of $\nu_{ij}$ is determined by a Poisson distribution, the two-dimensional log-likelihood is then calculated to be the sum over all energy and \sx\, bins as
\begin{equation}
    LL_{2D} = \sum_i \sum_j \left(k_{ij} \log(\nu_{ij}) - \nu_{ij} - \log(k_{ij}!) \right).  
\end{equation}

We introduce Gaussian constraints as additional terms in the total log-likelihood for the subset of templates for which there is an independent estimate of the corresponding radioactivity:
\begin{equation}\label{eq:nuis}
    LL_{total} = LL_{2D} - \sum_n \frac{(C^0_n - C_n)^2}{2 \sigma_n^2},
\end{equation}
where $C^0_n$ is the expected value of the scale factor of the $n^{th}$ template with uncertainty $\sigma_n$ based on the measurement of the activity. 
For species for which the constraint comes from a measurement, the uncertainty is taken from Table~\ref{tab:parttable}.
For upper limits, the Gaussian constraint is only included when $C_l$ exceeds the upper limit with an uncertainty of 10\%.
For the activation of the copper, which is not measured but calculated according to Eq.~\eqref{eq:activation}, we assume a 10$\%$ uncertainty. 
The tritium activity in the silicon bulk and all surface $^{210}$Pb templates are unconstrained in the fit.

\begin{figure*}[t]
	\centering
	\includegraphics[width=\textwidth]{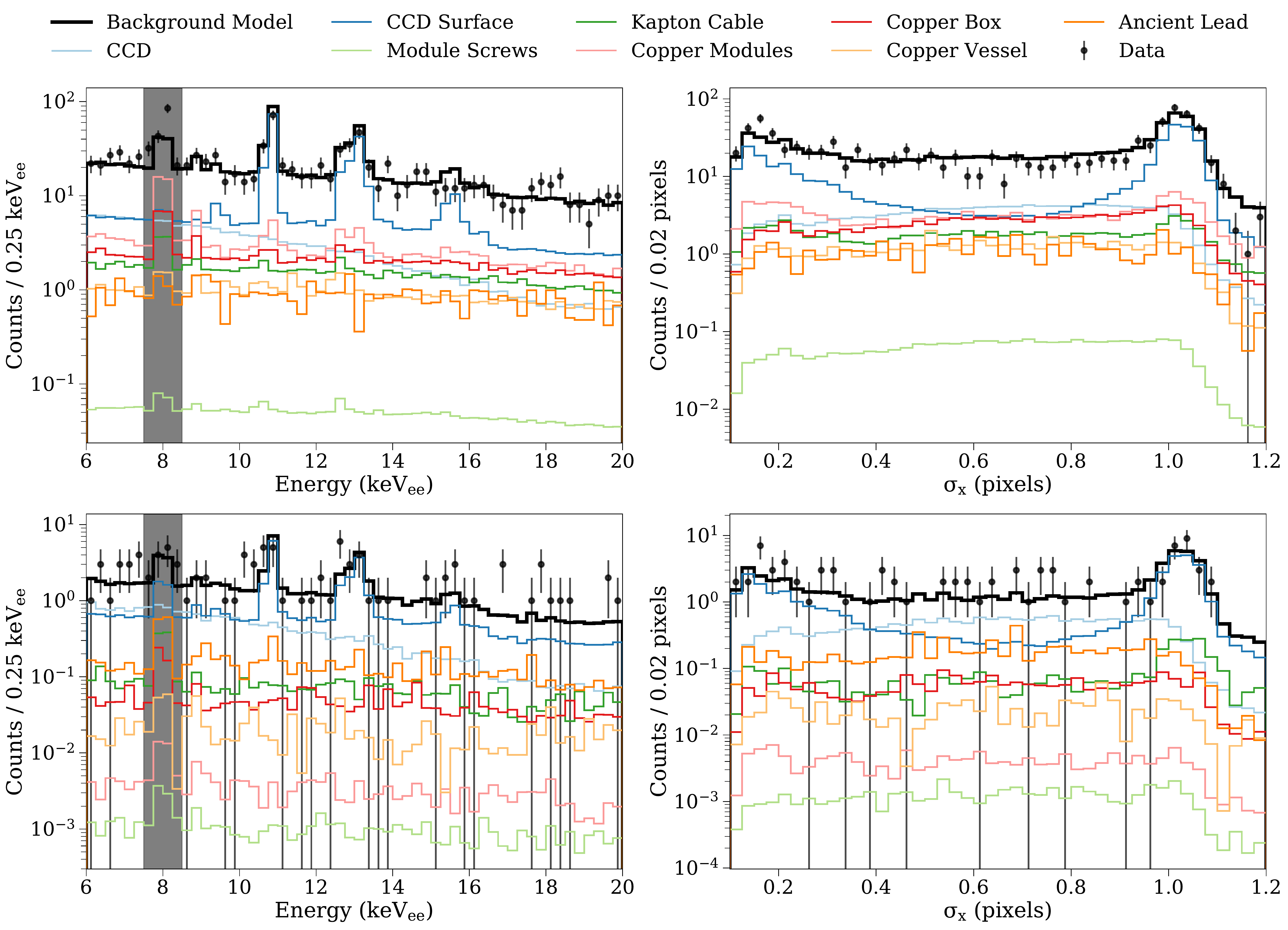}
	\caption{Output of the template fit between energies of 6--20~k\ev, and $\sigma_x$ of 0.1--1.2 pixels for CCDs 2--7 (top) and CCD 1 (bottom) projected onto the energy (left) and $\sigma_x$ (right) axes. Rather than show projections for all 49 templates, they have been grouped by detector part for clearer demonstration. The shaded region in the energy projections (left) indicates the bins containing copper fluorescence that are excluded from the fit and are not included in the \sx~projection (right).}
	\label{fig:fit_output}
\end{figure*}

We minimize $-LL_{\rm total}$ for the combined 2D spectra for CCDs 2--7. 
We exclude from the fit the 7.5--8.5~k\ev\ region (removing 176 bins) because we are not confident in our prediction of the amplitude of the copper fluorescence peak with \texttt{GEANT4}. The result of the fit for CCDs 2--7, projected onto the fast-clustered energy and $\sigma_x$ axes, is shown in Fig.~\ref{fig:fit_output}, along with the cross-check using CCD~1. We also compared the background model prediction of the $(x,y)$ distribution of clusters to the data and find both to be statistically consistent with a uniform distribution. 
We provide a full list of best-fit scaling factors $C_l$ for each template in Table~\ref{tab:fit_result}, along with calculated differential rates in different energy ranges. Note that the ``fast clustering" algorithm used in the construction of the background model does not efficiently reconstruct events below 1~k\ev~(see Fig.~\ref{fig:efficiencies}), so the background model construction is effectively blind to the energy range most relevant for the WIMP search. 

For the fit to CCDs 2--7, we find a goodness of fit $p$-value of 0.004, as determined from the distribution of Monte Carlo trials drawn from the best fit PDF. 
In Appendix~\ref{A:template}, we detail this procedure and show how this poor goodness of fit is dominated by outlier bins above 14~k\ev, where saturation is likely not being perfectly modeled.
Moreover, we show that removing outlier bins in the saturation region and performing the same procedure results in an improved $p$-value of 0.049 and less than $2\%$ change in the background model for any region of $E_{\rm sim}$ and depth $z$ below 6~k\ev. 
Thus, we conclude that the background model constructed from data between 6--20~k\ev, is statistically consistent with that same data below 14~k\ev. 

\begin{table*}[htbp]
  \centering
    \begin{tabular}{r @{\hskip 0.1in} lr @{\hskip 0.1in}|@{\hskip 0.1in} c @{\hskip 0.1in}|@{\hskip 0.1in}r@{\hskip 0.1in}| @{\hskip 0.2in}c @{\hskip 0.1in} c @{\hskip 0.2in}|@{\hskip 0.2in}c @{\hskip 0.1in}c@{\hskip 0.1in}}
    \hline \hline
        \rule{0pt}{2.5ex} & \multirow{2}{*}{Detector Part} & \multirow{2}{*}{Chain} & \multirow{2}{*}{$C_l$}  & \multirow{2}{*}{Best-Fit Activity} & \multicolumn{2}{@{\hskip -0.35in}c}{Rate (dru): CCDs 2--7} & \multicolumn{2}{@{\hskip -0.2in}|@{\hskip 0.1in}c}{Rate (dru): CCD 1} \\
         & & & & & 1--6 k\ev & 6--20 k\ev & 1--6 k\ev & 6--20 k\ev \\ \hline
    \rule{0pt}{2.5ex}1     & CCD & $^{238}$U & 0.897 & $\lesssim 9.86~\mu$Bq/kg & 0.01  & 0.01  & $<0.01$  & $<0.01$ \\
    2     & CCD & $^{226}$Ra & 0.900 & $\lesssim 4.79~\mu$Bq/kg & 0.01  & 0.01  & $<0.01$  & $<0.01$ \\
    3     & CCD & $^{232}$Th & 0.900 & $\lesssim 6.56~\mu$Bq/kg & 0.01  & 0.03  & 0.01  & 0.02 \\
    4     & CCD & $^{40}$K & 0.910 & $\lesssim 0.42~\mu$Bq/kg & $<0.01$  & $<0.01$  & $<0.01$  & $<0.01$ \\
    5     & CCD & $^{22}$Na & 1.066 & $340 \pm 60~\mu$Bq/kg & 0.17  & 0.16  & 0.10  & 0.09 \\
    6     & CCD & $^{32}$Si & 1.042 & $150 \pm 30~\mu$Bq/kg & 0.19  & 0.17  & 0.15  & 0.13 \\
    7     & CCD & $^{3}$H & 1.131 & $330 \pm 90~\mu$Bq/kg & 2.86  & 0.78  & 2.40  & 0.66 \\
    \rule{0pt}{3ex}8     & CCD (front surf.) & $^{210}$Pb & 1.658 & $69 \pm 12~$nBq/cm$^2$ & 1.45  & 1.67  & 0.53  & 0.88 \\
    9     & CCD (back surf.) & $^{210}$Pb & $<10^{-4}$ & $<0.1~$nBq/cm$^2$ & $<0.01$  & $<0.01$  & $<0.01$  & $<0.01$ \\
    10    & CCD (wafer surf.) & $^{210}$Pb & 1.343 & $56 \pm 8~$nBq/cm$^2$ & 2.43  & 1.84  & 1.98  & 1.18 \\
    \rule{0pt}{3ex}11    & Copper Box & $^{238}$U & 0.900 & $\lesssim 110~\mu$Bq/kg & 0.01  & 0.01  & $<0.01$  & $<0.01$ \\
    12    & Copper Box & $^{226}$Ra & 0.900 & $\lesssim 120~\mu$Bq/kg & 0.19  & 0.15  & 0.03  & 0.02 \\
    13    & Copper Box & $^{210}$Pb & 0.380 & $10 \pm 6~$mBq/kg & 0.33  & 0.20  & 0.02  & 0.01 \\
    14    & Copper Box & $^{232}$Th & 0.900 & $\lesssim 36~\mu$Bq/kg & 0.08  & 0.06  & 0.01  & 0.01 \\
    15    & Copper Box & $^{40}$K & 0.900 & $\lesssim 28~\mu$Bq/kg & $<0.01$  & $<0.01$  & $<0.01$  & $<0.01$ \\
    16    & Copper Box & Act. & 1.015 & $280 \pm 30~\mu$Bq/kg & 0.63  & 0.49  & 0.10  & 0.08 \\
    \rule{0pt}{3ex}17    & Copper Modules & $^{238}$U & 0.900 & $\lesssim 110~\mu$Bq/kg & 0.05  & 0.03  & $<0.01$  & $<0.01$ \\
    18    & Copper Modules & $^{226}$Ra & 0.900 & $\lesssim 120~\mu$Bq/kg & 0.21  & 0.17  & $<0.01$  & $<0.01$ \\
    19    & Copper Modules & $^{210}$Pb & 0.557 & $15 \pm 4~$mBq/kg & 1.18  & 0.71  & $<0.01$  & $<0.01$ \\
    20    & Copper Modules & $^{232}$Th & 0.900 & $\lesssim 36~\mu$Bq/kg & 0.10  & 0.08  & $<0.01$  & $<0.01$ \\
    21    & Copper Modules & $^{40}$K & 0.900 & $\lesssim 28~\mu$Bq/kg & $<0.01$  & $<0.01$  & $<0.01$  & $<0.01$ \\
    22    & Copper Modules & Act. & 1.006 & $130 \pm 10~\mu$Bq/kg & 0.30  & 0.23  & 0.01  & 0.01 \\
    \rule{0pt}{3ex}23    & Kapton Cable & $^{238}$U & 1.016 & $59 \pm 5~$mBq/kg & 0.51  & 0.30  & 0.23  & 0.11 \\
    24    & Kapton Cable & $^{226}$Ra & 1.362 & $7 \pm 5~$mBq/kg & 0.24  & 0.18  & 0.05  & 0.03 \\
    25    & Kapton Cable & $^{232}$Th & 1.010 & $32 \pm 0.5~$mBq/kg & 0.17  & 0.13  & 0.04  & 0.02 \\
    26    & Kapton Cable & $^{40}$K & 1.003 & $29 \pm 2~$mBq/kg & 0.09  & 0.05  & 0.04  & 0.02 \\
    27    & Kapton Cable & Act. & 1.000 & $140 \pm 10~\mu$Bq/kg & 0.01  & 0.01  & $<0.01$  & $<0.01$ \\
    \rule{0pt}{3ex}28    & Ancient Lead & $^{238}$U & 0.911 & $\lesssim 21~\mu$Bq/kg & $<0.01$  & $<0.01$  & $<0.01$  & $<0.01$ \\
    29    & Ancient Lead & $^{226}$Ra & 0.900 & $\lesssim 230~\mu$Bq/kg & 0.44  & 0.36  & 0.21  & 0.18 \\
    30    & Ancient Lead & $^{210}$Pb & 1.000 & $\sim 33~$mBq/kg & 0.04  & 0.03  & 0.24  & 0.18 \\
    31    & Ancient Lead & $^{232}$Th & 1.000 &  $\sim 2.3~\mu$Bq/kg & $<0.01$  & $<0.01$  & $<0.01$  & $<0.01$ \\
    32    & Ancient Lead & $^{40}$K & 0.916 & $\lesssim 5.3~\mu$Bq/kg & $<0.01$  & $<0.01$  & $<0.01$  & $<0.01$ \\
    \rule{0pt}{3ex}33    & Outer Lead & $^{238}$U & 0.916 & $\lesssim 12~\mu$Bq/kg & $<0.01$  & $<0.01$  & $<0.01$  & $<0.01$ \\
    34    & Outer Lead & $^{226}$Ra & 0.909 & $\lesssim 190~\mu$Bq/kg & $<0.01$  & $<0.01$  & $<0.01$  & $<0.01$ \\
    35    & Outer Lead & $^{210}$Pb & 1.000 & $18 \pm 5~$Bq/kg & $<0.01$  & $<0.01$  & $<0.01$  & $<0.01$ \\
    36    & Outer Lead & $^{232}$Th & 0.907 & $\lesssim 4.2~\mu$Bq/kg & $<0.01$  & $<0.01$  & $<0.01$  & $<0.01$ \\
    37    & Outer Lead & $^{40}$K & 0.906 & $\lesssim 200~\mu$Bq/kg & $<0.01$  & $<0.01$  & $<0.01$  & $<0.01$ \\
    \rule{0pt}{3ex}38    & Module Screws & $^{238}$U & 1.000 & $20 \pm 40~$mBq/kg & $<0.01$  & $<0.01$  & $<0.01$  & $<0.01$ \\
    39    & Module Screws & $^{226}$Ra & 0.900 & $\lesssim 1.4~$mBq/kg & 0.01  & 0.01  & $<0.01$  & $<0.01$ \\
    40    & Module Screws & $^{210}$Pb & 1.000 & $27 \pm 8~$mBq/kg & $<0.01$  & $<0.01$  & $<0.01$  & $<0.01$ \\
    41    & Module Screws & $^{232}$Th & 1.024 & $2.4 \pm 1.6~$mBq/kg & 0.02  & 0.01  & $<0.01$  & $<0.01$ \\
    42    & Module Screws & $^{40}$K & 1.000 & $28 \pm 15~$mBq/kg & $<0.01$  & $<0.01$  & $<0.01$  & $<0.01$ \\
    43    & Module Screws & Act. & 1.000 & $89 \pm 9~\mu$Bq/kg & $<0.01$  & $<0.01$  & $<0.01$  & $<0.01$ \\
    \rule{0pt}{3ex}44    & Copper Vessel & $^{238}$U & 0.903 & $\lesssim 110~\mu$Bq/kg & $<0.01$  & $<0.01$  & $<0.01$  & $<0.01$ \\
    45    & Copper Vessel & $^{226}$Ra & 0.900 & $\lesssim 120~\mu$Bq/kg & 0.10  & 0.09  & 0.01  & 0.01 \\
    46    & Copper Vessel & $^{210}$Pb & 0.731 & $20 \pm 8~$mBq/kg & 0.06  & 0.03  & $<0.01$  & $<0.01$ \\
    47    & Copper Vessel & $^{232}$Th & 0.900 & $\lesssim 36~\mu$Bq/kg & 0.04  & 0.03  & $<0.01$  & $<0.01$ \\
    48    & Copper Vessel & $^{40}$K & 0.901 & $\lesssim 28~\mu$Bq/kg & $<0.01$  & $<0.01$  & $<0.01$  & $<0.01$ \\
    49    & Copper Vessel & Act. & 0.486 & $400 \pm 440~\mu$Bq/kg & 0.33  & 0.27  & 0.05  & 0.04 \\ \hline
    \rule{0pt}{2.5ex} & Total & \multicolumn{3}{c}{} & \multicolumn{1}{@{\hskip -0.2in}|c}{12.28} & 8.29 & 6.22 & 3.70  \\
    \hline \hline
    \end{tabular}%
    \caption{\label{tab:fit_result}Results from the template fit used to construct the background model. 
    The parameter $C_l$ indicates the best-fit fraction of the initial guess for each template's normalization, with the fit uncertainty propagated into the best-fit activity. 
    For activation (Act.) templates, all activities listed are for $^{60}$Co, the dominant isotope contributing to the template.
    The mean differential rate in counts kg$^{-1}$ day$^{-1}$ k\ev$^{-1}$ (abbreviated dru) is calculated separately for CCDs 2--7 and CCD 1 in the energy ranges of 1--6~k\ev\, and 6--20~k\ev, excluding Si and Cu fluorescence lines respectively. See text for more details.}
\end{table*}%

\begin{figure*}[t]
	\centering
	\includegraphics[width=\textwidth]{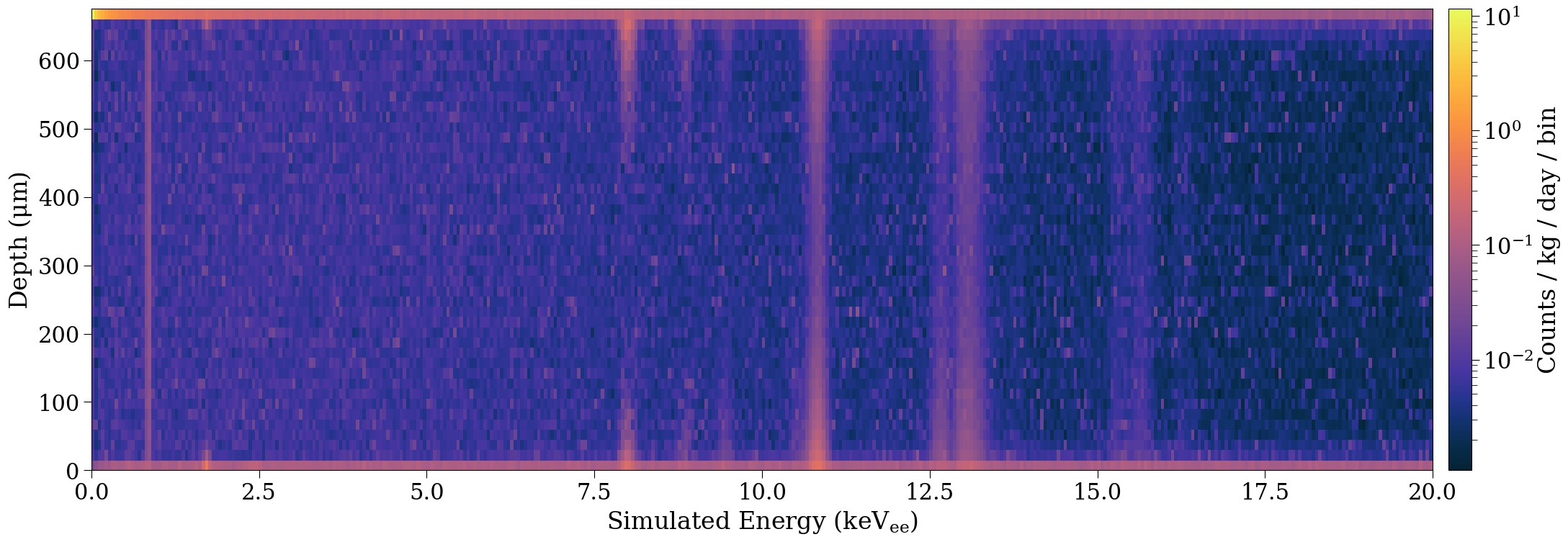}
	\caption{The background model template (for CCDs 2--7) in raw simulated energy $E_{\rm sim}$ and depth ($z=0$ corresponds to the front of the CCD) according to the best-fit combination of the 49 templates used in the binned likelihood fit above 6~k\ev. The color bar indicates the rate of events expected per kg-day per 50~\ev~$\times$~15~$\mu$m bin.}
	\label{fig:ezmodel}
\end{figure*}

\subsection{Surface \text{$^{210}$P\lowercase{b}} Analysis}\label{S:Alphas}
One of the unique features of the CCD technology as a particle detector is the ability to identify events coming from the same decay chain over long time periods. Recently, the DAMIC Collaboration published an analysis of spatially coincident $\beta$ and $\alpha$ events with time separation up to weeks to measure $^{32}$Si bulk contamination and set upper limits on bulk $^{210}$Pb, $^{238}$U, and $^{232}$Th~\cite{damicCoincidence2020}. The $^{210}$Pb limits were set under the assumption that all measured coincident $^{210}$Pb-Bi decays occurred in the bulk silicon. In reality, these events are far more likely to originate from $^{210}$Pb deposited on the external surfaces of the CCD and silicon wafer by radon plate-out, as in our background model. Here, we compare the results from the background model fit with the number of $^{210}$Pb-Bi event pairs from Ref.~\cite{damicCoincidence2020}, under the assumption that these events originate from the surfaces of the CCDs.

We briefly summarize the analysis from Ref.~\cite{damicCoincidence2020}. We used 181.3 days of data in 1x1 format instead of the data in 1x100 format used for most of the analysis presented in this paper. The detector setup was the same but the data was acquired earlier, with data acquisition starting before CCD~2 became operational so it was not included in the analysis. Classification of $\alpha$'s and $\beta$'s was based on the variable $f_{pix}$, the fraction of pixels over threshold in the smallest rectangle containing the clustered pixels. This selection was tuned on simulation to correctly classify $>99.9\%$ of all simulated $\beta$'s and 100\% of $\alpha$'s with energies $>2$~MeV. For two events to qualify as a coincident \text{$^{210}$Pb-Bi} candidate, they must both be classified as $\beta$'s, occur within the same CCD in images less than 25~days apart ($\approx5 \tau_{1/2}$ of $^{210}$Bi), include a minimum of one pixel overlap, and the energy of the first $\beta$ must be reconstructed in the range 0.5--70~k\ev. In total, 69 event pairs were found in the six CCDs with an exposure of 6.5~kg-days.

As presented in Table~\ref{tab:fit_result}, we find a best-fit $^{210}$Pb activity of $69 \pm 12$ ($56 \pm 8$)~nBq/cm$^2$ for the front (wafer back) surface in our background model fit. To translate the fit result to the expected number of $^{210}$Pb-Bi event pairs observed in Ref.~\cite{damicCoincidence2020}, we first determine the selection efficiencies for front (wafer back) surface $^{210}$Pb-Bi decay sequences from 50,000 \texttt{GEANT4} simulated decays to be $\epsilon_{sel} = 0.138~(0.315)$. The wafer back surface has a higher selection efficiency because it is closer to the active bulk, meaning that, despite the PCC region, it has a higher probability for both decays in a sequence to be reconstructed and tagged. As in Ref.~\cite{damicCoincidence2020}, the time efficiency to select decay sequences is $\epsilon_{t} = 0.748$ because of the readout time between images and the detector downtime. Given the surface area of 38~cm$^2$ of each CCD, $25.6 \pm 4.5$ ($47.4 \pm 6.8$) event pairs are expected from front (wafer back) surface $^{210}$Pb. Additionally, $2.5 \pm 0.1$ accidental event pairs are expected from random events overlapping spatially, and $19.4 \pm 5.1$ event pairs are expected from $^{32}$Si-P decays. The resulting $94.9 \pm 9.6$ expected $^{210}$Pb-Bi pairs is slightly larger than the 69 candidate pairs observed.

We also compare the background model fit result to the observed number of $\alpha$'s in the front and the back of the CCDs.
As shown in Ref.~\cite{damicCoincidence2020}, front and back $\alpha$'s can be distinguished by the aspect ratio of the clusters through a selection on $\sigma_x / \sigma_y$.
Furthermore, the energy of most of the observed $\alpha$'s is consistent with the decay of surface $^{210}$Po, daughter of $^{210}$Bi, with an energy distribution peaked at 5.3 MeV and a long tail toward lower energies caused by energy losses in the CCD dead layers.
Assuming secular equilibrium along the $^{210}$Pb-Bi-Po chain, the fit result corresponds to $41\pm7$ ($33\pm5$) $^{210}$Po decays in the front (wafer back) of each CCD in the data set used in Ref.~\cite{damicCoincidence2020}.
The total observed number of $\alpha$'s cannot be directly compared to these values since they include the contribution from $^{210}$Po decays on the copper surrounding the CCDs\footnote{The observation of $^{210}$Po $\alpha$'s from the copper does not imply a presence of $^{210}$Pb on the copper surface since the copper cleaning procedure is known to be more efficient in removing $^{210}$Pb than $^{210}$Po~\cite{BUNKER2020163870}.}.
However, assuming that the contribution from the copper is the same to the front and back CCD surfaces (treating CCD~1 separately, and excluding the top and bottom surfaces of the stack of CCDs 2--7), the front-back difference in the number of $^{210}$Po decays per CCD, $\Delta_{F-B}$, can be compared.
We estimate the number of surface $^{210}$Po decays from the number of $\alpha$'s with energy $<5.4$~MeV. This selection excludes higher energy $\alpha$'s from other decays in the U and Th chains, while retaining $>$95\% of surface $^{210}$Po decays that deposit energy in the CCDs.
After applying a simple geometrical simulation to estimate the event acceptance, we obtain $\Delta_{F-B} = 40\pm11$ from $\alpha$ counting, compared to $8\pm9$ from the background model fit.

Overall, we regard the $\sim 2 \sigma$ variation (considering only statistical uncertainties) between the background model fit result and the $^{210}$Pb-Bi and $^{210}$Po counting exercises as satisfactory.
These cross checks generally confirm the magnitude of $^{210}$Pb contamination on the CCD surfaces, and the relative contamination between the front and the back of the CCDs. 
We remind the reader that the $^{210}$Pb surface activity was left unconstrained in the template fit (Sec.~\ref{S:Template}).

\subsection{Extrapolation}\label{S:Model}
Following this fit, we use the simulation output to construct a single template for our background model in simulated variables $E_{\rm sim}$ vs. $z$ according to the best-fit scale factors $C_l$. Before adding the contributions from each individual template $l$, we remove any remaining statistical anomalies. For any template that contributes a total integral of less than 0.1 events/kg-day between 0--20~k\ev\ (less than 1 event in the WIMP search), we average over $E_{\rm sim}$ by overwriting the content of all bins for the same $z$ with their mean value. This smooths templates (2), 4, 9, (11), 15, (17--20), 21, (27), 28, 31--38, (39), 40, (41), 42--44, (47), and 48, where templates in parentheses are smoothed only for CCD 1. Refer to Table~\ref{tab:fit_result} for template identification. Second, after adding the templates, we scan for any anomalously high bins, defined as being 20 standard deviations away from the mean of the bulk background and not adjacent to another high bin (as in an energy peak). In total, 3 (4) bins are removed for CCDs~2--7 (CCD~1), which account for a total rate of 0.07 (0.16) events/kg-day, or less than 1 event in the combined exposure.
The resulting background model template for CCDs~2--7 in simulated coordinates is shown in Fig.~\ref{fig:ezmodel}. In order to maximize our resolution at this stage, we choose fine binning of 10~\ev{} in $E_{\rm sim}$ and 15~$\mu$m in $z$. The background model has a bulk background rate of $3.1 \pm 0.6$ ($6.5 \pm 0.4$) counts kg$^{-1}$ day$^{-1}$ k\ev{}$^{-1}$ between 2.5--7.5~k\ev{} in CCD~1 (CCDs 2--7), where the error bars provided are statistical, and an implicit energy threshold set by \texttt{GEANT4} of $\sim 10$~\ev{}\footnote{The Livermore physics list is validated down to 250~eV~\cite{Cirrone:2010zz,Pandola:2014uea}, and we independently extend this validation down to 50~\ev{} using our data~\cite{joaothesis}.}.

In Section~\ref{S:Template}, we constructed the background model using a fit to only data from CCDs~2--7, excluding data from CCD~1. 
CCD~1 is treated separately since the overall background rates are roughly $50\%$ lower than in CCDs~2--7, primarily because of the EFCu module and additional ancient lead shield. 
Using simulated events for CCD~1 and the best-fit parameters $C_l$, we construct an analogous background model for CCD~1. 
Any event will be compared against the relevant background model according to the CCD in which it was measured. 

\begin{figure}[t]
	\centering
     \includegraphics[width=0.5\textwidth]{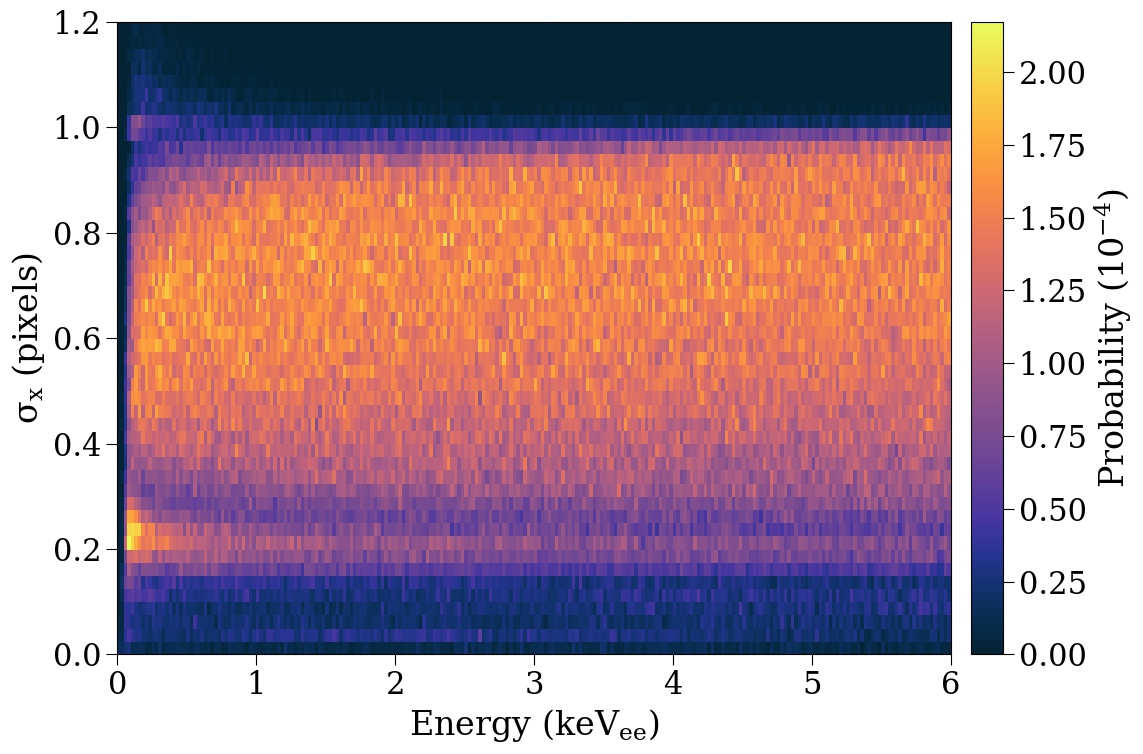}
	\caption{Response PDF of the CCD, shown in $E$ ($25 \ \rm eV_{ee}$ bin width) vs. $\sigma_x$ ($0.025$ pixels bin width) space, using the likelihood clustering algorithm employed for the WIMP search. The PDF was generated from simulated events, uniformly distributed in energy and depth, added onto blank images that were then processed through our analysis pipeline. 
	}
	\label{fig:response}
\end{figure}

\begin{figure*}[ht]
    \centering
    \includegraphics[width=0.95\textwidth]{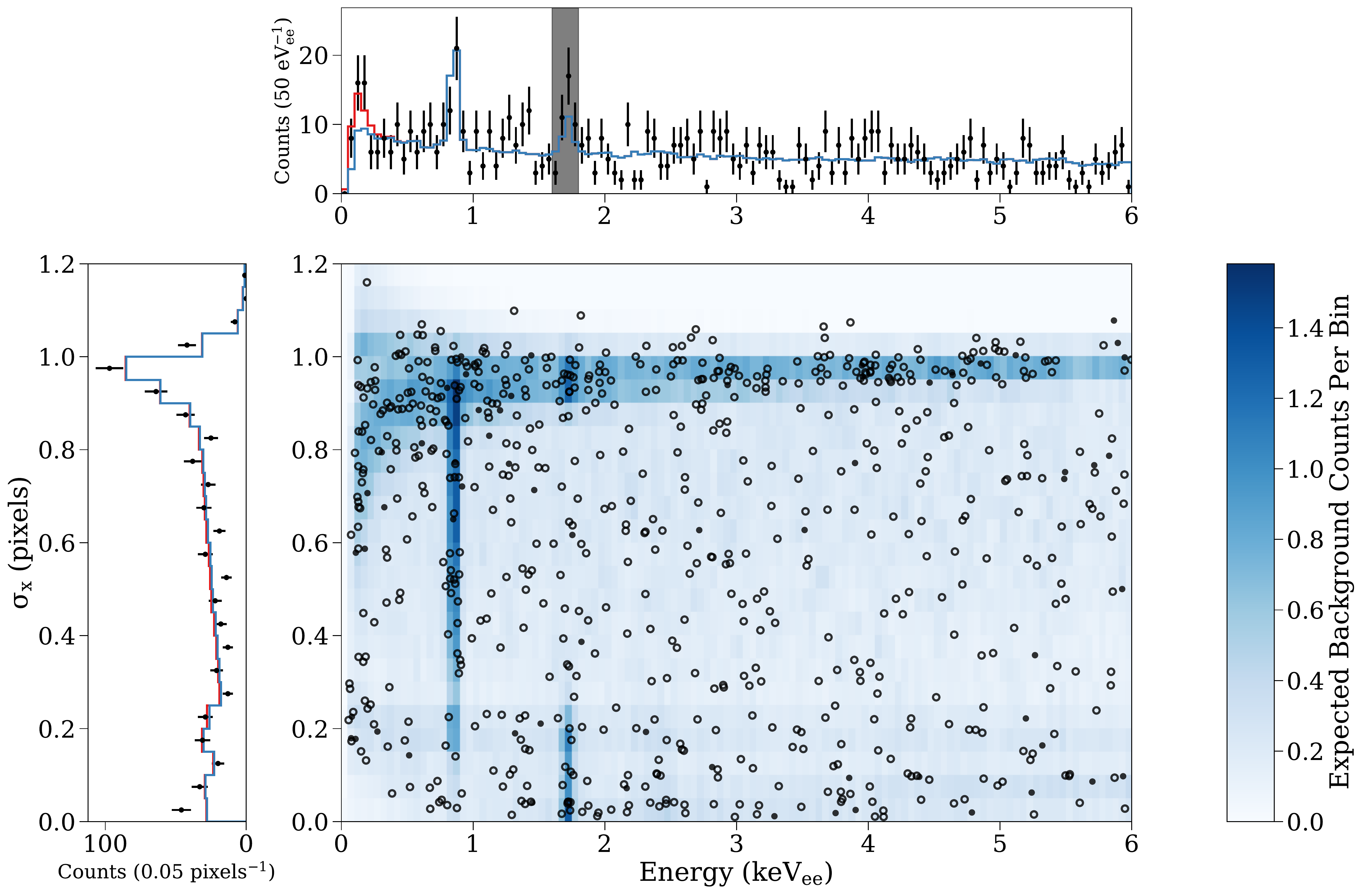}
    \caption{Result from the fit to the data with our background model plus a generic signal component as described in the text. This result was previously reported in Ref.~\cite{damic2020}. The scatter plot shows the likelihood-clustered data from CCDs 2--7 (CCD~1) with open (filled) circles overlaid on the combined background model for all CCDs. The central plot shows the background model component of the likelihood fit, where the color axis has units of expected counts per bin in the 11 kg-day exposure. The top (left) plots are the projections on the energy (\sx) axes. In each projection, the red line is the best-fit global minimum (background + generic signal) and blue is the background component of the global fit. The region between $1.6$ and $1.8$ $\rm keV_{ee}$ that includes the silicon fluorescence is excluded from the fit.}
    \label{fig:likelihood-fit-results}
\end{figure*}

While the background model represents our best fit above 6~k\ev{}, there is still substantial uncertainty in our PCC model at lower energies (Section~\ref{S:PCC}). This uncertainty is restricted to the highest $z$-bin (back of the CCD) in Fig.~\ref{fig:ezmodel}, and can be approximated as an additive correction proportional to $\exp ( -\sqrt{E_{\rm sim}} / 0.18$~keV$^{-1/2}$). Furthermore, there is an underlying, approximately constant spectrum in the highest $z$-bin from ionization events in the fully-depleted region of the CCDs, thus we split the background model shown in Fig.~\ref{fig:ezmodel} into two pieces: a ``lower-bound'' background model with the exponential PCC correction subtracted, such that the background rate in the highest $z$ bin is approximately flat in energy, and a positive, additive component proportional to $\exp ( -\sqrt{E_{\rm sim}} / 0.18$~keV$^{-1/2}$). 
We treat the positively valued amplitude of the additive exponential as a free parameter in our WIMP-search fit.
We perform this procedure for the CCD 1 and CCDs 2--7 background models separately and find that the exponential component in CCD 1 corresponds to the same fraction of events in the highest $z$ bin as in CCDs 2--7, but its amplitude is 25\% smaller due to the lower background rates in CCD~1 external to the sensor.
Thus, to reduce the number of free parameters in our WIMP-search fit, we fix the ratio of the additive exponential component between CCDs 2--7 and CCD 1 to 75\%. 

We perform the WIMP-search fit on events identified by the ``likelihood clustering'' algorithm presented in Section~\ref{S:Recon}, which is designed to efficiently identify low-energy clusters on full CCD images.
Thus, we need to translate the final background model in simulated coordinates into reconstructed coordinates by the likelihood clustering algorithm, correctly accounting for all reconstruction efficiencies in the data down to our analysis threshold of 50~\ev.

\begin{table*}[!t]
 \begin{center}
 \begin{tabular}{r@{\hskip 0.1in} | @{\hskip 0.1in}c@{\hskip 0.1in} c@{\hskip 0.1in} c@{\hskip 0.1in} c@{\hskip 0.1in} c@{\hskip 0.1in} c }
    \hline \hline
    \rule{0pt}{2.5ex}Parameter & Null hypothesis & All events & CCD 1 only & CCDs 2--7 only & $>$200~\ev{} & $n_{\rm pix} > 1$ \\ \hline
       \rule{0pt}{2.5ex}$s$~[events] & 0 & $17.1 \pm 7.6$ & $6.4\pm3.0$ & $8.9\pm7.2$ & 0 & $13.9\pm6.8$  \\
       $\epsilon$ [\ev{}] & - & $67 \pm 37$ & $89\pm50$ & $51\pm39$ & - & $78\pm33$  \\
       $b_1$~[events] & 56.2 & $57.6 \pm 3.3$ & $56.0\pm3.1$ & - & $54.8 \pm 3.0$ & $43.6\pm2.5$  \\
       $b_{2\text{--}7}$~[events] & 625 & $609 \pm 21$ & - & $613 \pm 21$ & $591 \pm 21$ & $535 \pm19$  \\
       $c_1$~[events] & 5.4 & $0.9 \pm 1.1$  & $0 \pm 0.9$ & - & $0.40 \pm 0.87$ & $1 \pm 1.1$ \\
       $c_{2\text{--}7}$~[events] & 41.6 & $6.6 \pm 8.9$ & - & $5.0 \pm 7.0$ & $3.0 \pm 6.5$ & $8 \pm 8.7$  \\  \hline
       \rule{0pt}{2.5ex}exposure [kg-day] & - & 10.9 & 1.6 & 9.3 & 10.9 & 10.3 \\
       no-signal $p$-value & - & $2.2\times10^{-4}$ & $5.8\times10^{-4}$ & 0.039 & 1 & $5.1 \times 10^{-3}$  \\
              g.o.f. $p$-value & - & 0.10 & 0.94 & 0.21 & 0.32 & 0.69  \\
      \hline \hline
 \end{tabular}
 \caption{\label{tab:dmresult}Results from the unbinned extended likelihood fit between the background model and data below 6~k\ev{} and down to the 50~\ev{} analysis threshold (unless otherwise listed). Uncertainties are those returned by {\tt MINUIT}. A no-signal $p$-value of $1$ indicates no difference between the global minimum and background-only ($s=0$) fit results. The bottom row provides the goodness of fit (g.o.f.) $p$-value for the global minimum. The ``Null hypothesis'' column provides the expected values of the parameters from the extrapolation below 6~k\ev{} of the background model as constructed in Section~\ref{S:BackModel}. }
 \end{center}
 \end{table*}

We sample from the background model (both the lower bound and the additive exponential component) separately for CCDs~2--7 and CCD~1 to obtain $E_{\rm sim}$ and $z$ values for simulated events. We then apply the CCD-response simulation (diffusion, binning, saturation, etc.) and add the pixel values from the events randomly onto blank images. 
To best represent the data, we also add shot noise from leakage current to the simulated images at the levels measured for the exposed images. A total of 830 simulated images with 250 randomly sampled events each (for a total of 200,000 simulated events per CCD) is then run through the same analysis pipeline as the data (including masking, likelihood clustering and cluster selection) to obtain the final simulated clusters and their reconstructed coordinates. Unlike the analysis that is performed to translate simulated \texttt{GEANT4} events into reconstructed clusters in Section~\ref{S:Template}, this procedure accurately reflects data all the way down to our analysis threshold of 50~\ev , and implicitly accounts for all reconstruction efficiencies, solely at the cost of processing time. The only efficiency correction that must be made is for the removal of clusters that overlap or contain more than one ionization event (see Section \ref{S:Recon}), which is higher in the simulated images because of the higher spatial density of events.

We generate the response PDF of our detector (Fig.~\ref{fig:response}) by repeating the procedure above with events sampled uniformly in energy and depth.
For a model of the hypothetical WIMP signal, we scale the response PDF as a function of energy according to the signal spectrum.

\section{Low Energy Analysis}\label{S:likelihood}

We perform a WIMP search by comparing the data below 6 k\ev{} to the background model constructed in Section~\ref{S:Model}.
The results, reported in Ref.~\cite{damic2020}, are summarized and expanded below.

Images correlated with periods of higher leakage current due to LED flashing and temperature cycling show a large increase in the total number of clusters and were excluded from the WIMP search.
These 443 images (8\% of the total) were identified as those that have an average charge per pixel larger than $\sim 0.47 e^-$ (6.76 ADU average pixel value).
Their removal results in a total exposure for the WIMP search of 10.93~kg-days. 
These images were used in the background model construction, since the higher leakage current has no effect above 6~k\ev.

A two-dimensional $(E, \sigma_x)$ unbinned likelihood fit was performed to the final sample of events identified by the likelihood clustering following the formalism in Ref.~\cite{damic2016}. 
The fit was performed jointly to the two data sets $k$ from CCD~1 and CCDs~2--7, each containing $N_k$ events in a fractional exposure $\gamma_k$.
The fit function was constructed from the addition of probability density functions (PDFs) of the lower-bound background model $f_{b_k}$, the PCC correction $f_c$, and a ``generic signal'' of a population of events distributed uniformly in space with an exponentially decreasing spectrum $f_s$:
\begin{equation}\label{eq:signal}
    f_s(E, \sigma_x | \epsilon) = \frac{1}{\epsilon}\exp( -E / \epsilon ) \times R(E, \sigma_x),
\end{equation}
where the exponential decay is multiplied with our detector response function $R(E, \sigma_x)$ shown in Fig.~\ref{fig:response}.
We choose this functional form to capture the most general DM models, such as the quenched~\cite{chavarria2016measurement} WIMP nuclear recoil spectrum. 
The decay energy $\epsilon$ and amplitude $s$ of a generic signal, the amplitudes of the lower-bound background models $b_{k}$, and the PCC correction\footnote{We note that $c_{k}$ is a single free parameter because $c_{2\text{--}7} \equiv 6 ~c_1/0.75$, as described in Section~\ref{S:Model}.} $c_{k}$ were free parameters in the fit.
We constrained $b_{k}$ to $b_k'$, the predicted normalization from the background model fit above $6 \ \rm keV_{ee}$, with fractional uncertainty $\sigma_{b_k}/b_k = 6\%$.
We perform the fit in the region $E\in[0.05, 6] \ \rm keV_{ee}$ (excluding Si $K$ fluorescence $E\in[1.6, 1.8] \ \rm keV_{ee}$) and $\sigma_x$ $ \in [0, 1.2]$ pixel by minimizing with {\tt MINUIT} the extended negative log-likelihood function:
\begin{widetext}
\begin{equation}\label{eq:extendedlikelihood}
\begin{split}
     \log \mathcal{L}(s, \epsilon , & \hspace{2.5pt} b_{1,2\text{--}7}, c_{1,2\text{--}7}) = \\ & \sum_{k}^2 \left( -(\gamma_k s + b_k + c_k) + \sum_i^{N_k} \Big( s \gamma_k f_s(E_i, \sigma_{x_i} | \epsilon) + b_k f_{b_k}(E_i, \sigma_{x_i}) + c_k f_{c}(E_i, \sigma_{x_i}) \Big)  + \frac{(b_k-b_k')^2}{2\sigma_{b_k}^2} \right).
\end{split}
\end{equation}
\end{widetext}

Figure~\ref{fig:likelihood-fit-results} shows the comparison between the best-fit background model (in blue) and data (markers). 
The best-fit values for the lower-bound number of background events are $b_1 = 57.6 \pm 3.3$ and $b_{2\text{-}7} = 609 \pm 21$.
The best-fit value for $c_1 = 0.9 \pm 1.1$ and $c_{2\text{--}7} = 6.6 \pm 8.9$ corresponds to a distance between $^{210}$Pb contamination on the back side of the original wafer and the start of charge collection of $0.75^{+0.50}_{-0.35}$\,$\mu$m (see purple line in Fig.~\ref{fig:pcc_sims}). 
Our best fit exhibits a preference for an exponential bulk component with $s=17.1\pm7.6$ events and $\epsilon=67\pm37$ \ev{}.
We estimate a goodness-of-fit $p$-value of 0.10 from the minimum negative log-likelihood distribution obtained from running the fit procedure on Monte Carlo samples drawn from the best-fit PDF. 
Note that the outlier first bin in the \sx\ projection arises from front-side events above 1~k\ev.

\begin{figure}
    \centering
    \includegraphics[width=\linewidth]{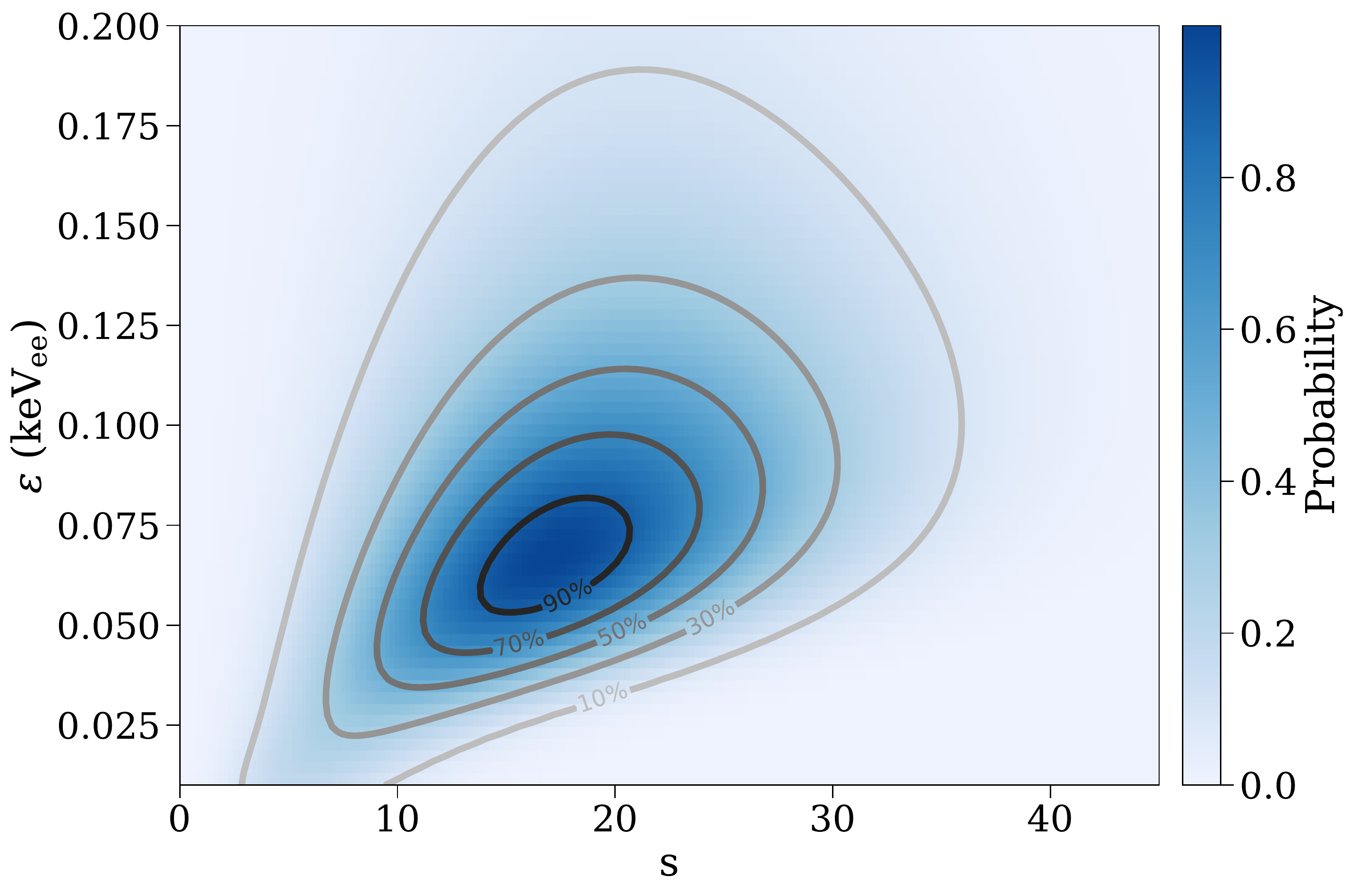}
    \caption{Fit uncertainty in the number of excess signal events over the background model ($s$) and characteristic decay energy ($\epsilon$) of the generic signal spectrum. The color axis represents the $p$-value from likelihood-ratio tests to fit results with constrained $s$ and $\epsilon$. Contours are drawn for $p$-values of 0.9, 0.7, 0.5, 0.3, and 0.1.}
    \label{fig:s-epsilon-contour}
\end{figure}

\begin{figure}
    \centering
    \includegraphics[width=\linewidth]{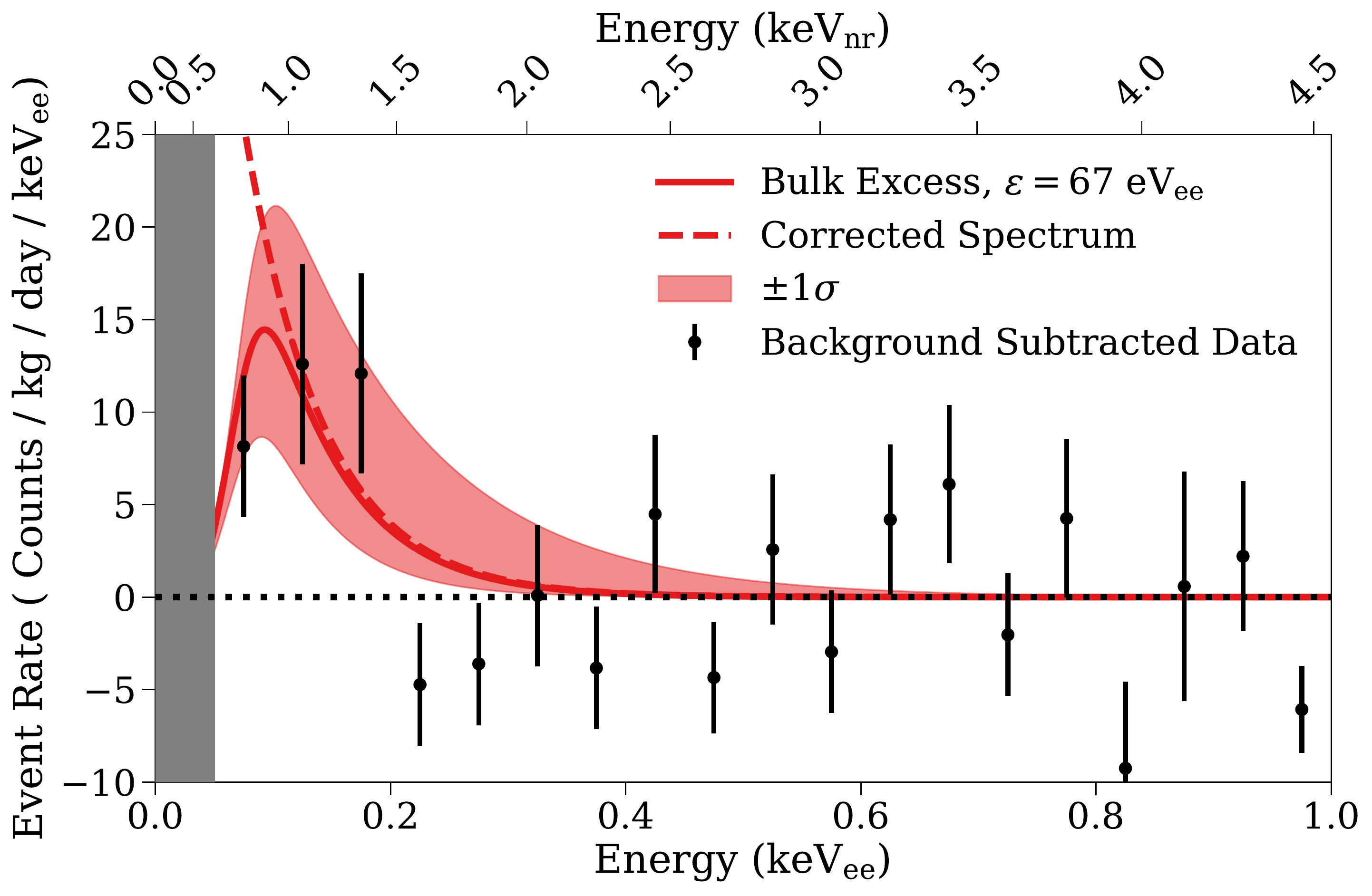}
    \caption{Energy spectrum of the best-fit generic signal (red lines) overlaid on the background-subtracted data (markers). Both the fit spectrum that includes the detector response (solid line) and the spectrum corrected for the detection efficiency (dashed line) are provided. The red shaded region represents the 1-$\sigma$ uncertainty from the likelihood-ratio tests. For reference, the equivalent nuclear recoil energy ($\rm keV_{nr}$) is shown on the top axis; the ionization efficiency is taken from the direct calibration performed in Ref \cite{chavarria2016measurement}.}
    \label{fig:excess-energy-projection}
\end{figure}

We computed the uncertainty about the best fit by performing likelihood ratio tests between the global best fit and fit results with constrained values of $s$ and $\epsilon$. Figure \ref{fig:s-epsilon-contour} shows the probability of the $s\text{-}\epsilon$ parameter space derived from the likelihood-ratio test statistic assuming that its distribution follows a $\chi^2$ distribution with two degrees of freedom (Wilks' theorem).
We confirmed by Monte Carlo that Wilks' theorem is an appropriate approximation for our data.
The background-only hypothesis $s=0$ is disfavored with a $p$-value of $2.2\times10^{-4}$ ($3.7\sigma$).
Figure~\ref{fig:excess-energy-projection} shows the bulk excess spectrum obtained by projecting the best-fit result onto the energy axis and subtracting the best-fit background model amplitude, which includes the partial charge collection component. The $\pm1\sigma$ band was derived from the contour in Fig.~\ref{fig:s-epsilon-contour}. 
The displayed error bars are the Poisson uncertainties in the bin contents since the uncertainty in the background model amplitude is small in comparison.
The energy bins below 200~\ev{} show a clear excess above background with a spectrum that is well described by the best-fit generic signal model. 
Furthermore, we confirm that events with energies below 200~\ev{} are distributed uniformly in $(x,y)$ and time, as expected for a dark matter signal.

\begin{figure}[t]
	\centering
	\includegraphics[width=\linewidth]{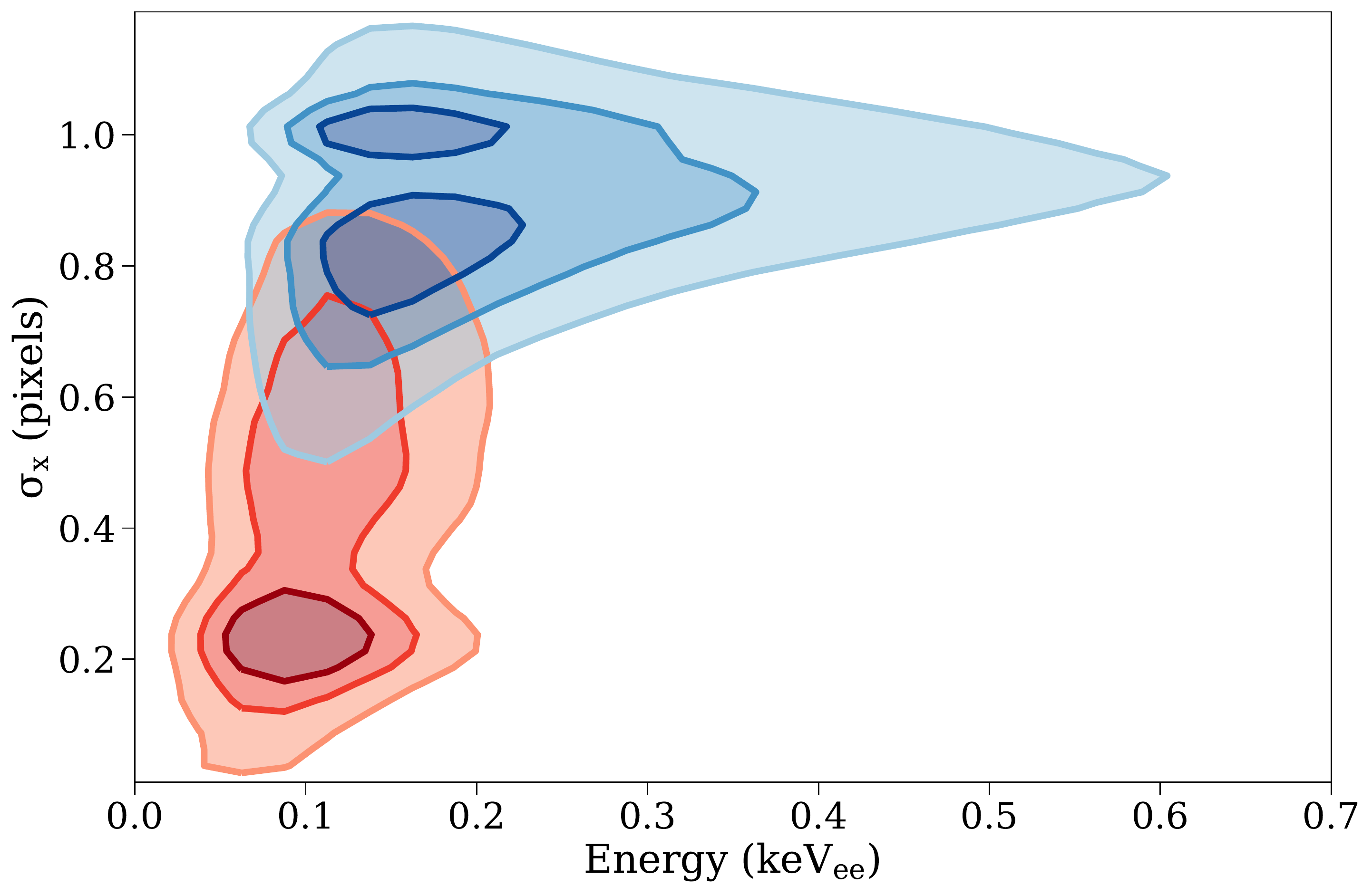}
	\caption{Comparison in reconstructed energy-\sx~space between the best-fit generic signal (red) and the additive backside PCC (blue) components of the fit. Contours are drawn to contain 10$\%$, 40$\%$, and 70$\%$ of events expected from each distribution.}
	\label{fig:back_uncertainty}
\end{figure}

Several systematic tests were performed to confirm the consistency of our fit result, with results summarized in Table~\ref{tab:dmresult}:

\begin{itemize}
    \item If we perform the fit to events with energies $>$200~\ev{} only, the result is consistent with the background-only hypothesis and shows no preference for a generic signal. 
    \item If we perform the fit to the data from CCD~1 or CCDs~2--7 separately, the resulting generic signal is statistically consistent between the two datasets with a higher statistical significance in CCD~1, which has the lowest background.
    \item A comparison between the two-dimensional profile of the best-fit generic signal and the additive PCC component (Fig.~\ref{fig:back_uncertainty}) shows that the excess of events with \sx$\sim 0.2$\,pixel that drive the statistical significance of the generic signal are in a region of parameter space that is not populated by events from the CCD backside.
    \item Although front-surface ($z\sim0$) events can populate the region of parameter space with \sx$\sim 0.2$\,pixel, a fit to the data after removing clusters where only one pixel has a value greater than 1.6\,$\sigma_{\rm pix}$\textemdash a selection that removes 56\% of front-surface events but only 6.5\% of bulk events\textemdash still returns a statistically significant generic signal with values for $s$ and $\epsilon$ consistent with the original result~\cite{damic2020}.
    \item If we introduce in our fit an artificial energy offset in the generic signal spectrum~\cite{dessert2020systematics}, we only observe a statistically significant excess for offsets below $\sim$100~\ev , which further confirms that our background model describes the data well above 200~\ev{}, and that the reported signal excess does not originate from unexpectedly large statistical fluctuations in the data.
\end{itemize}

Events occurring in the proximity of the serial register~\cite{senseiSR} were excluded as a possible source for the observed excess. 
These events generate ionization charge in the field-free regions peripheral to the CCD active region that can diffuse and reach the serial register. 
Clusters from these events appear as horizontal streaks in 1x1 images. 
In 1x100 images, where clusters appear as horizontal sequences of contiguous pixels, it is not possible to identify serial register events from their topology. 
However, since any charge in the serial register is discarded before the image is read out, serial register events that occur during an exposure are excluded from the analysis. 
Serial register events are only a problem if they occur during CCD readout. 
However, CCD readout accounts for only 0.01$\%$ of the overall target exposure and contributes a negligible number of clusters, as verified with blank images.


Our analysis suggests the presence of bulk ionization events at a rate of a few per kg-day in the range $E\in[0.05, 0.20] \ \rm keV_{ee}$ with a uniform distribution in the bulk silicon whose source is not included in our background model. 
The origin of these events (e.g., whether they are electronic or nuclear recoils, or some unexpected instrumental effect) remains unknown. 
In Ref.~\cite{damic2020}, we employ the background model presented here to set the most stringent limits with a silicon target on the WIMP-nucleon spin-independent scattering cross section for WIMP masses in the range $1\text{-}9 \ \rm GeV / c^2$.

\section{Conclusion}\label{S:Discussion}
In this paper, we present the first comprehensive background model for a CCD detector down to an unprecedentedly low energy of 50~\ev . The background model shows excellent agreement with the data down to 200~\ev, with an unexpected and statistically significant excess of events at the lowest energies discussed in Section~\ref{S:likelihood}.

In the process of constructing this background model, we compiled and determined the relative contributions of various radioactive background sources.
We note that our dominant background (see Table~\ref{tab:fit_result}) comes from cosmogenic $^3$H in the bulk of our CCDs and $^{210}$Pb deposited on the surfaces of the CCDs, including the surface of the wafer prior to CCD fabrication, now embedded $\sim 2.5~\mu$m beneath the CCD back surface.

The measured activity of bulk $^3$H from the template fit allows us to determine (according to the exposure history of our CCDs) the activation rate of cosmogenic $^3$H in silicon.
Under the assumption that activation of the pure silicon starts at $t_0$ with the formation of the original ingot used to produce the wafers (total sea-level equivalent exposure time $9.2 \pm 0.2$ years), we obtain a sea-level equivalent activation rate of $76 \pm 23$~atoms/kg-day.
Alternatively, if we assume that $t_0$ coincides with CCD fabrication (total sea-level equivalent exposure time $2 \pm 1$ years), during which the wafers were briefly processed at $900^{\circ}$C, potentially ``baking-out'' any $^3$H in the silicon, we obtain an activation rate of $205 \pm 90$ ~atoms/kg-day\footnote{The larger uncertainty in this case comes from not knowing at what dates in the fabrication process the baking took place and which flights come after this step.}.
Both of these numbers are consistent with recent measurements~\cite{damicActivation}, which leaves the possibility that the activation clock for tritium may be ``reset'' during CCD fabrication.
Under the assumption that $^{22}$Na cannot be baked out from the wafers\footnote{In fact, Na does have significant mobility in silicon at high temperatures~\cite{na22mobility1,na22mobility2}, but this warrants further study.}, we constrain the sea-level activation rate of $^{22}$Na in silicon to be $48 \pm 10$~atoms/kg-day, in excellent agreement with Ref.~\cite{damicActivation}.

The fit chooses to place $^{210}$Pb surface contamination on the front of the CCD and the back of what was originally the silicon wafer, the two surfaces most likely to contain such contamination (despite no input bias towards these surfaces in the fit).
Furthermore, the fit returns slightly more contamination on the front of the CCD than on the back of the wafer.
The measured activities and locations of $^{210}$Pb surface contamination from the template fit were confirmed by an independent analysis that leverages on the high spatial resolution of CCDs (Section~\ref{S:Alphas}), used previously to constrain radiocontaminants in the bulk silicon~\cite{damicCoincidence2020}.
These results demonstrate that our fitting methodology can effectively discern between different spectral components that originate from different locations in the detector.

In our modeling of the response of the detector to surface backgrounds, we discovered that diffusion of phosphorous from the backside ISDP layer into the high-resistivity bulk silicon causes a region of significant charge recombination that extends a few $\mu$m into the CCDs (Section~\ref{S:PCC}).
Partial charge collection (PCC) in this region causes a significant distortion of the spectrum from $^{210}$Pb decays on the backside of the CCDs, which is the dominant systematic uncertainty in our WIMP-search fit.
Our PCC profile, obtained from the measured properties of minority-carrier transport in silicon (details in Appendix~\ref{A:pcc}), was experimentally confirmed with direct measurements using a $^{55}$Fe x-ray source in Ref.~\cite{fnalPCC}.
High levels of hydrogen were also measured in the ISDP layer. This hydrogen is expected to contain minute levels of tritium, which due to the proximity to the PCC region, are expected to be preferentially reconstructed at the low energies where a WIMP signal is expected.
Although the background rate from $^3$H decays in the ISDP region is too small to be observable with this analysis, it could become a problem for future CCD dark matter experiments with greater sensitivity.

This analysis lays the foundation for the background modeling for the upcoming DAMIC-M experiment~\cite{damicm} and will inform the analysis and design of other CCD dark matter experiments (e.g., SENSEI~\cite{sensei} and Oscura~\cite{brn}).
It demonstrates the importance of limiting cosmogenic activation of the silicon and exposure of the detector surfaces to radon, including on the wafers prior to CCD fabrication.
It identifies important low-energy backgrounds associated with the CCD backside that must be addressed for future CCD dark matter detectors, e.g., by removing $\sim 10$~$\mu$m from the backside of the CCDs after fabrication~\cite{steve}.
Our background-model fit demonstrates the limitation of using ancient lead for shielding since the dominant external background in CCD~1 already comes from $^{210}$Pb decays in the ancient lead (Table~\ref{tab:fit_result}), which motivates the development of new methods of radiation shielding for future large-scale solid-state dark matter detectors.
Finally, successful mitigation of some of the identified backgrounds may require underground radon-free facilities for critical fabrication activities (e.g., sensor packaging, copper growth and machining, etc.) and further improvements in the design of dark matter detectors.

\begin{acknowledgments}
We are grateful to SNOLAB and its staff for support through underground space, logistical and technical services. SNOLAB operations are supported by the Canada Foundation for Innovation and the Province of Ontario Ministry of Research and Innovation, with underground access provided by Vale at the Creighton mine site.
We acknowledge financial support from the following agencies and organizations:
National Science Foundation through Grants No.\ NSF PHY-1912766 and NSF PHY-1806974; 
Kavli Institute for Cosmological Physics at The University of Chicago through an endowment from the Kavli Foundation; 
Gordon and Betty Moore Foundation through Grant GBMF6210 to the University of Washington; 
Fermi National Accelerator Laboratory (Contract No. DE-AC02-07CH11359); 
Institut Lagrange de Paris Laboratoire d'Excellence (under Reference No. ANR-10-LABX-63) supported by French state funds managed by the Agence Nationale de la Recherche within the Investissements d'Avenir program under Reference No. ANR-11-IDEX-0004-02; 
Swiss National Science Foundation through Grant No. 200021\_153654 and via the Swiss Canton of Zurich; 
Project PID2019-109829GB-I00 funded by MCIN/ AEI /10.13039/501100011033; 
Mexico's Consejo Nacional de Ciencia y Tecnolog\'{i}a (Grant No. 240666) and  Direcci\'{o}n General de Asuntos del Personal Acad\'{e}mico--Universidad Nacional Aut\'{o}noma de M\'{e}xico (Programa de Apoyo a Proyectos de Investigaci\'{o}n e Innovaci\'{o}n Tecnol\'{o}gica Grants No. IB100413 and No. IN112213);
STFC Global Challenges Research Fund (Foundation Awards Grant ST/R002908/1).
\end{acknowledgments}

\appendix

\section{Partial Charge Collection (PCC) Efficiency Model}\label{A:pcc}
\begin{figure}
	\includegraphics[width=\linewidth]{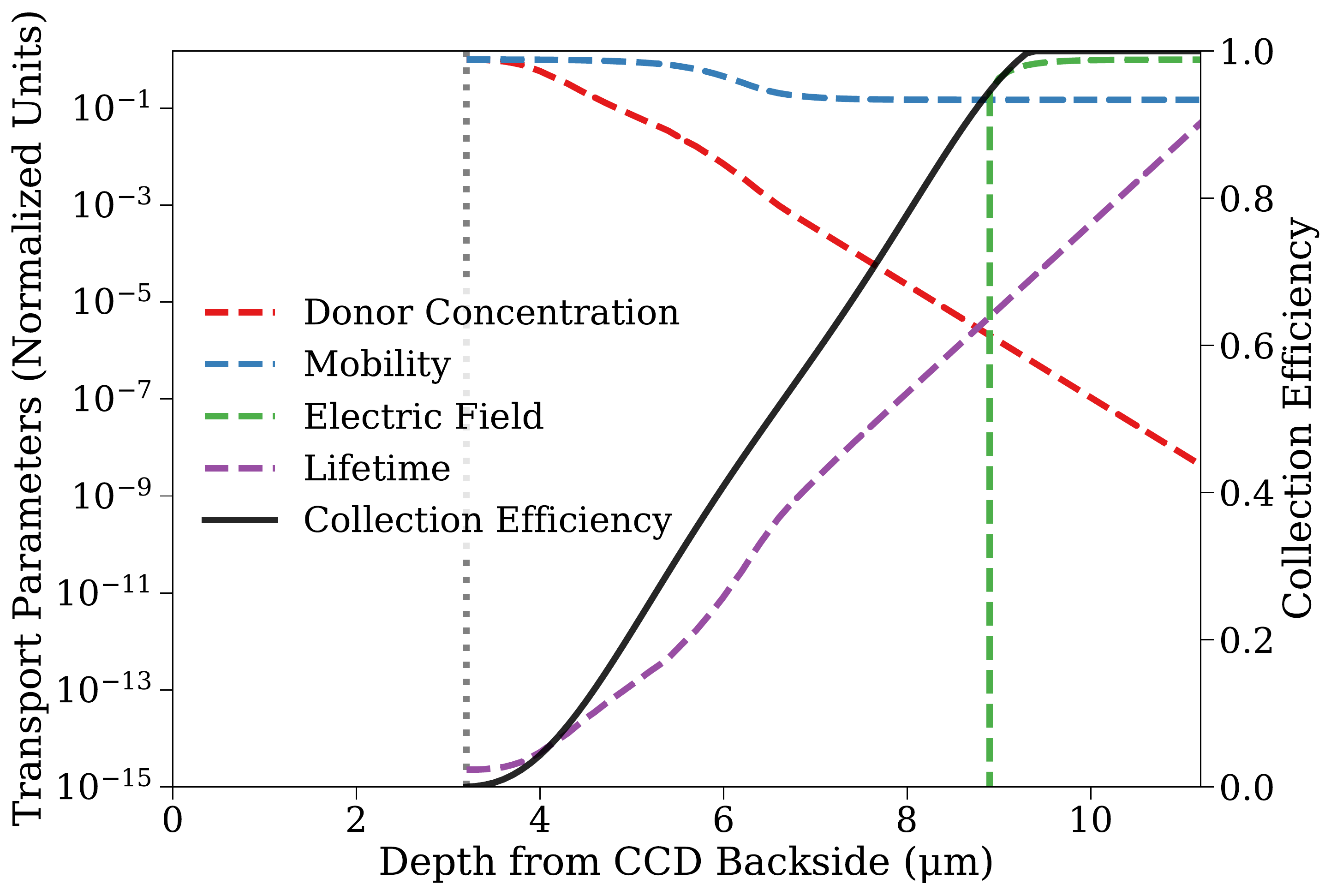}
	\caption{Properties that determine charge transport and recombination in silicon (dashed lines) in normalized units of $10^{20}$ cm$^{-3}$ (donor P concentration), $55 \times 10^{9} \ \rm \mu m^2 / V / s$ (mobility), $0.1 \ \rm V / \mu m$ (electric field), and $1.2\times10^{5} \ \rm s$ (lifetime) as a function of depth from the CCD backside ($z''$). The calculated charge collection efficiency is shown by the solid black line. The position of the original wafer surface is denoted by the vertical dotted grey line.}
	\label{fig:pcc_model}
\end{figure}

\begin{figure*}[!t]
	\centering
	\includegraphics[width=0.99\textwidth, trim=150 0 150 20, clip=true]{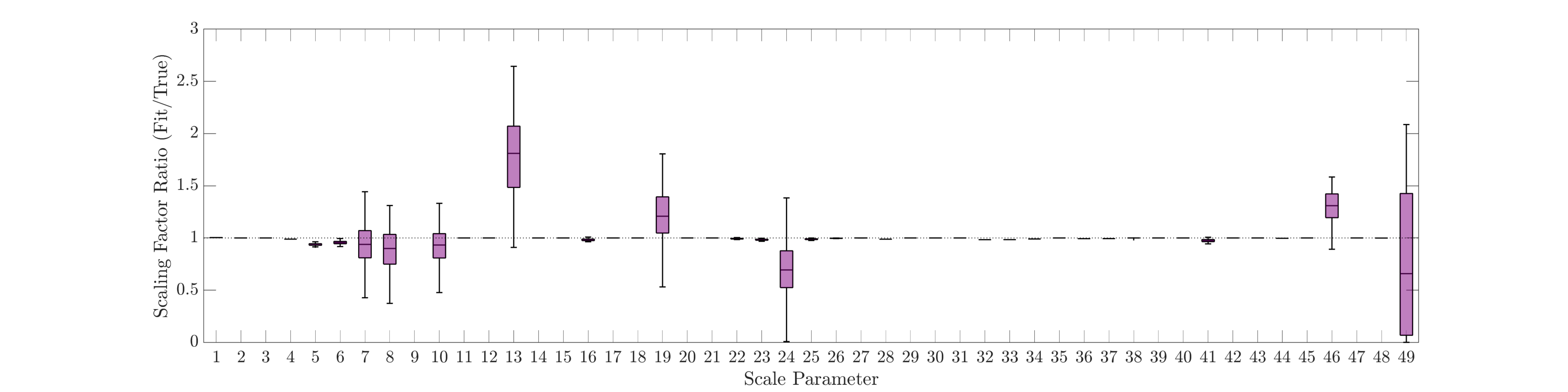}
	\caption{Comparison of the ratio of the fit parameters $C_l$ from 100 fake data samples pulled from the best-fit background model over the true best-fit values in Table~\ref{tab:fit_result}.
	For each $C_l$, the central horizontal mark indicates the median of the 100 samples with the upper and lower ranges of the purple box indicating the interquartile range (25--75$\%$ of the samples).
	A distribution centered around 1 indicates no bias in the fit methodology and the spread of the fake trials gives a best attempt to quantify the systematic uncertainty of the fit methodology on that parameter.
	For many fit parameters $C_l$, the variation relative to the best-fit value is so small that the ratio appears on this axis as a single horizontal line at 1.}
	\label{fig:LL_trials}
\end{figure*}

\begin{figure}[!t]
	\centering
	\includegraphics[width=0.49\textwidth, trim=30 0 50 20, clip=true]{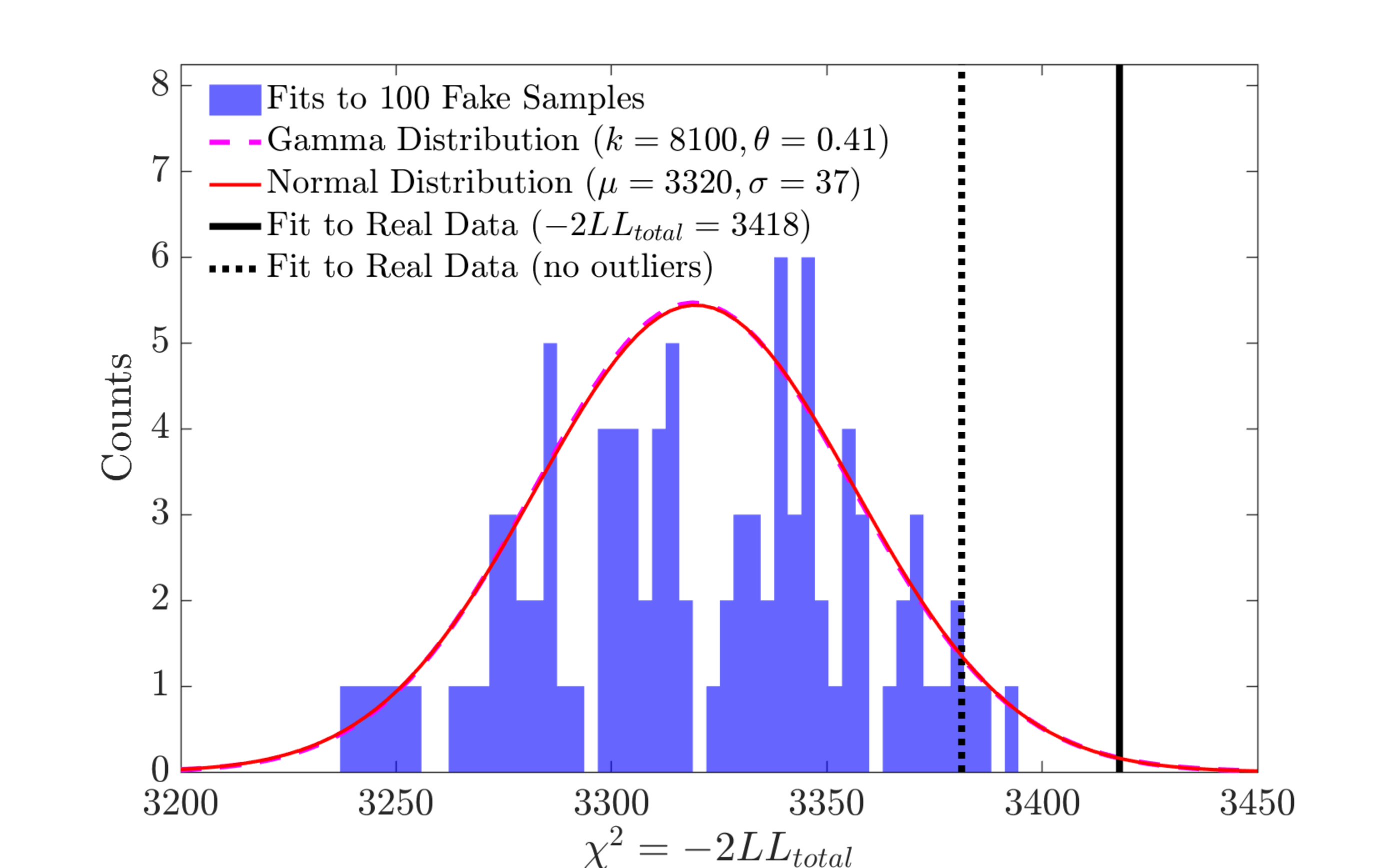}
	\caption{Distribution of the minimum log-likelihood from running the background model fit separately on 100 fake samples drawn from the background model PDF (blue histogram).
	Fits of this distribution to Gamma (magenta dashed) and Normal (red solid) distributions are overlaid.
	The minimum log-likelihood from the best-fit to the data is given with (solid black) and without (dashed black) the three outlier bins above 14~k\ev\ mentioned in the text.
	}
	\label{fig:LL_dist}
\end{figure}

As discussed in Section~\ref{S:PCC}, the presence of the ISDP layer on the back of the CCD during fabrication leads to the diffusion of phosphorous into the bulk silicon. The SIMS measurements show that the donor concentration decreases from $\sim 10^{20}$~cm$^{-3}$ at the backside contact to $\sim 10^{11}$~cm$^{-3}$ in the CCD active region over a distance of $\sim 8 \rm \ \mu m$. The concentration of donors changes the mobility and lifetime of the minority charge carriers (holes)~\cite{del1987modelling}\cite{wang1990temperature}, and the electric field profile near the back of the CCD. These variations can be many orders of magnitude; Fig.~\ref{fig:pcc_model} shows the P concentration and the resulting transport properties as a function of the distance from the CCD backside $z''$.

For $z''<8.8 \ \rm \mu m$, there is a field-free region where diffusion, a relatively slow process, is the only mechanism for charge to reach the fully-depleted active region, where it will be drifted and eventually collected at the CCD gates. At high P concentration close to the CCD backside, the carrier lifetime $\tau(z'')$ is very short and all charge recombines before diffusing significantly. However, at intermediate P concentration where $\tau(z'')$ is sufficiently long, some of the charge diffuses into the active region before it recombines, which creates the region of PCC.
We calculated the collection efficiency as a function of $z''$ by performing a numerical simulation of the evolution of charge packets. We assumed that $|\vec{E}_x| = |\vec{E}_y| = 0$ and therefore the problem is invariant in the $(x,y)$ coordinates. We considered the region from the back surface of the original wafer to a point in the active region where any charge can be assumed to be efficiently drifted and collected at the CCD gates, i.e., $3.2 < z'' < 11.2 \ \rm \mu m$. We divided the region into $50$ nm bins and assumed all charge in a given bin evolves in the same manner.

\begin{figure*}[!t]
	\centering
	\includegraphics[width=0.49\textwidth, trim=30 0 50 20, clip=true]{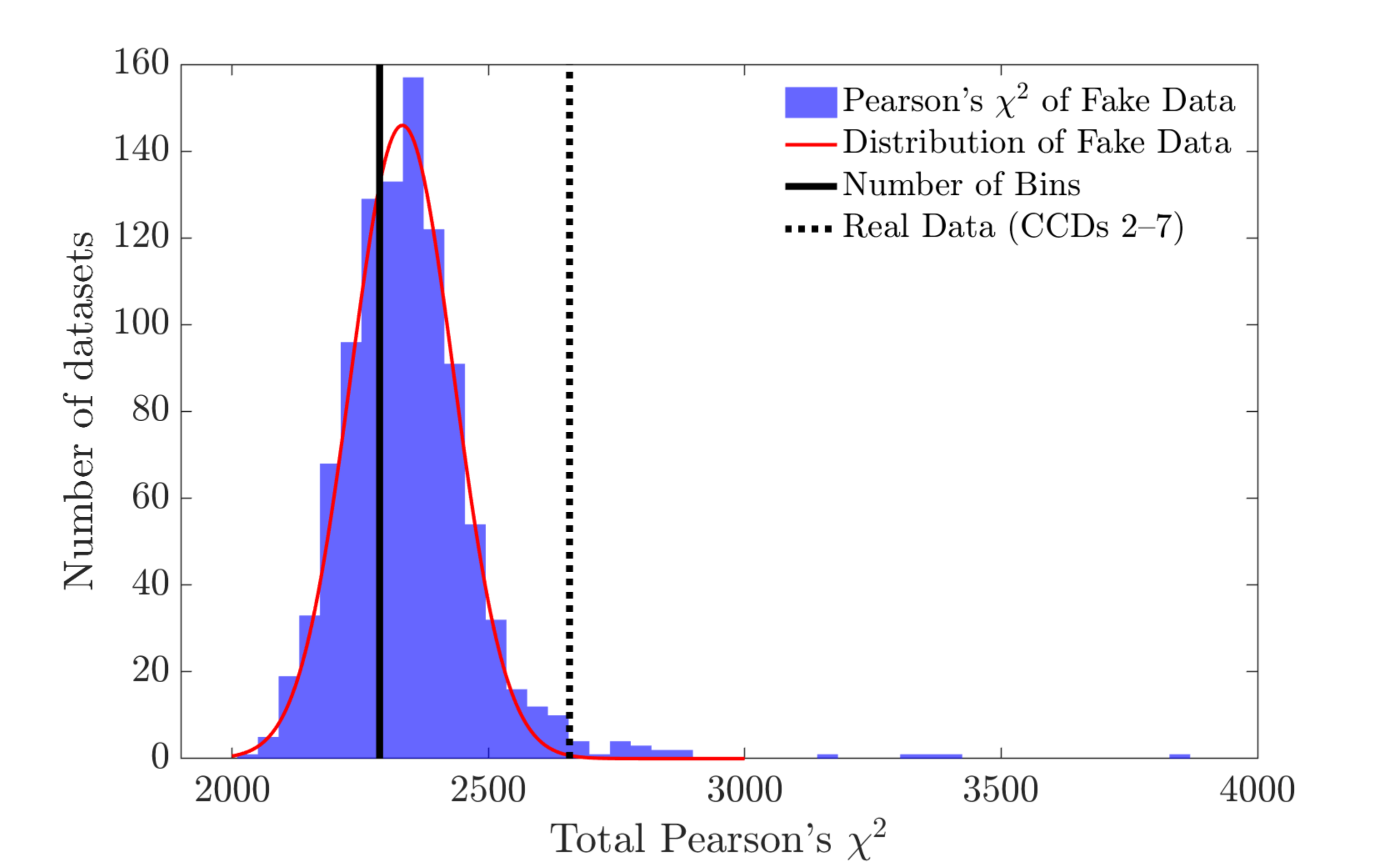}~
	\includegraphics[width=0.49\textwidth, trim=20 0 50 20, clip=true]{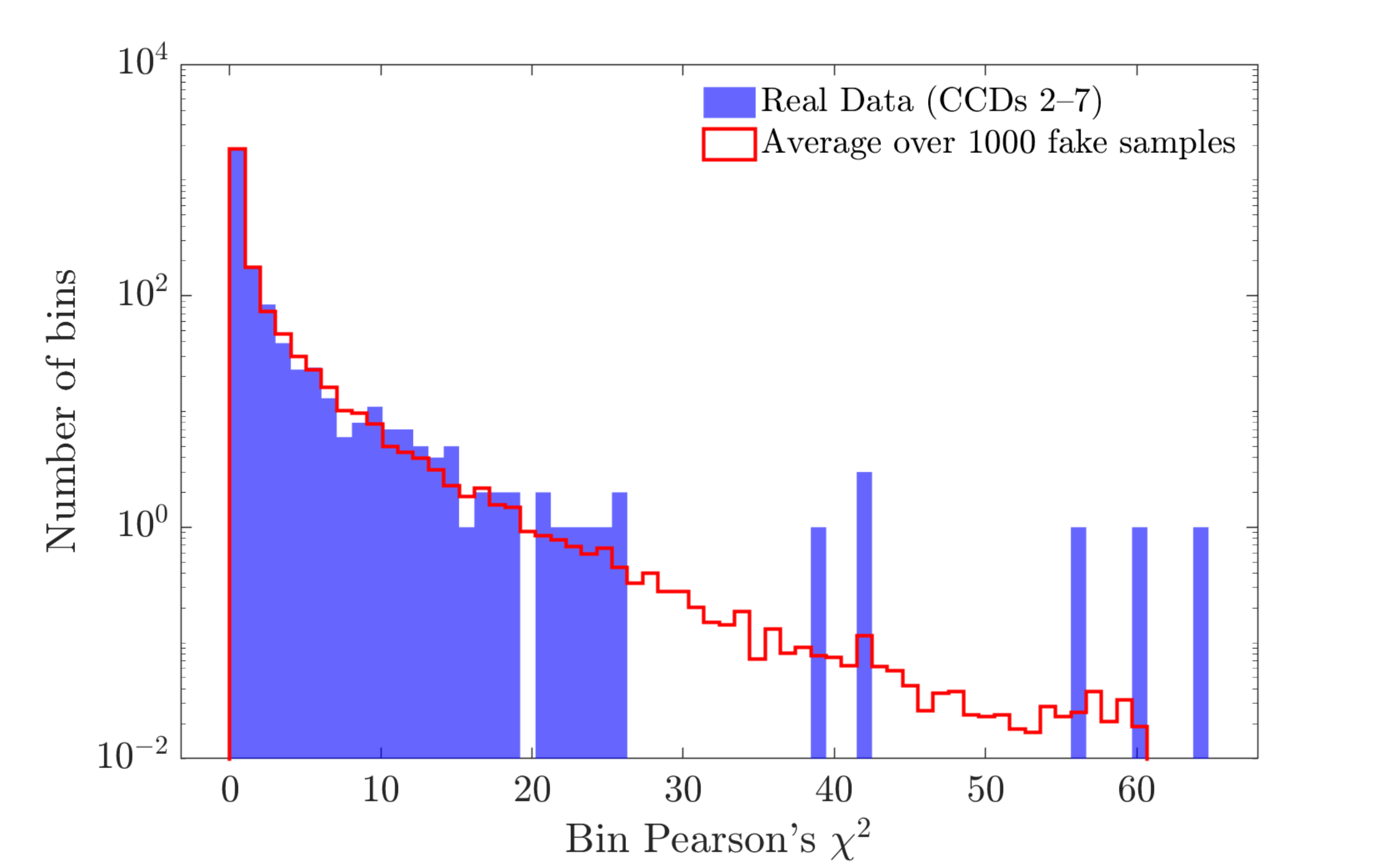} \\
	\includegraphics[width=0.49\textwidth, trim=30 0 50 20, clip=true]{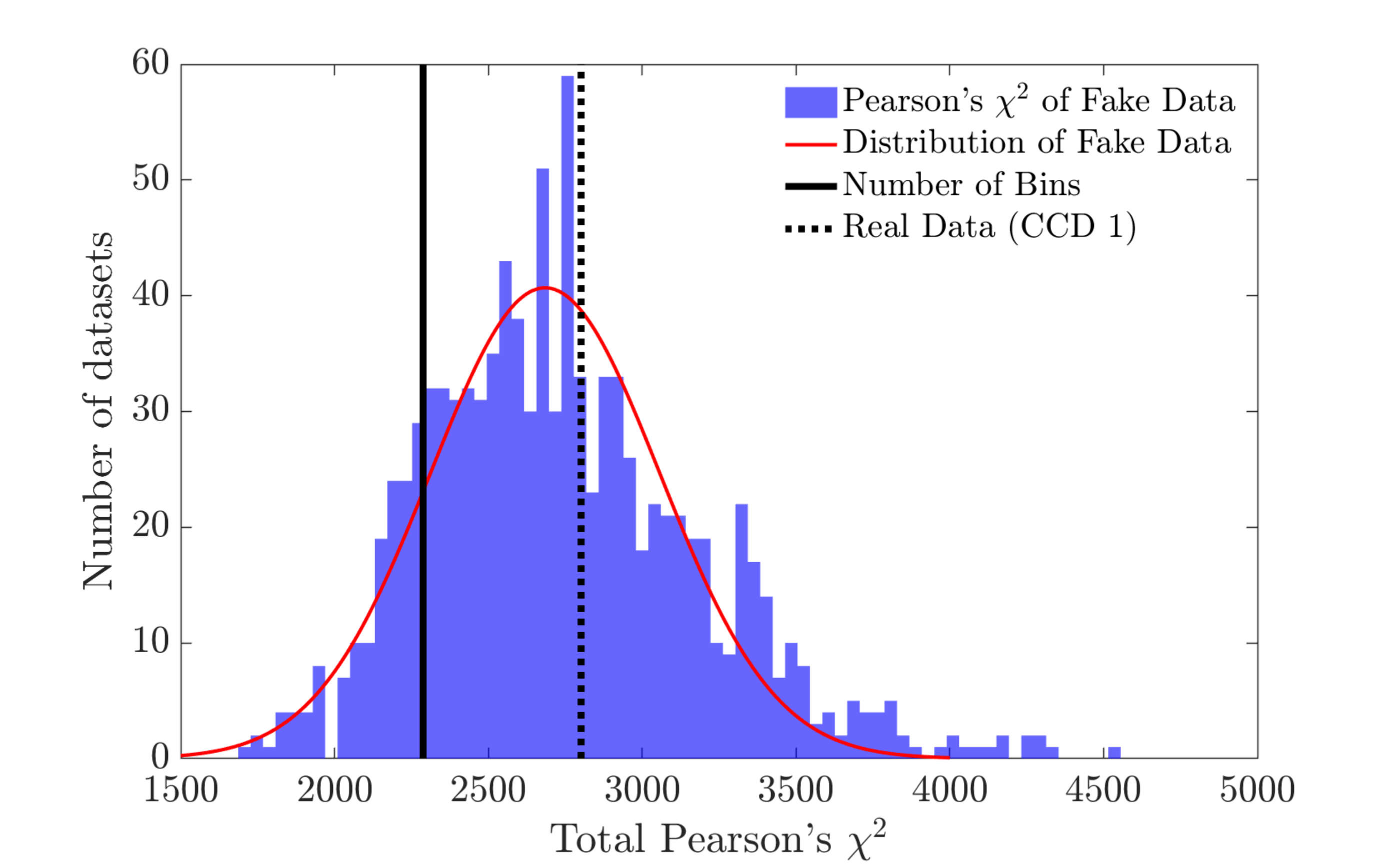}~
	\includegraphics[width=0.49\textwidth, trim=20 0 50 20, clip=true]{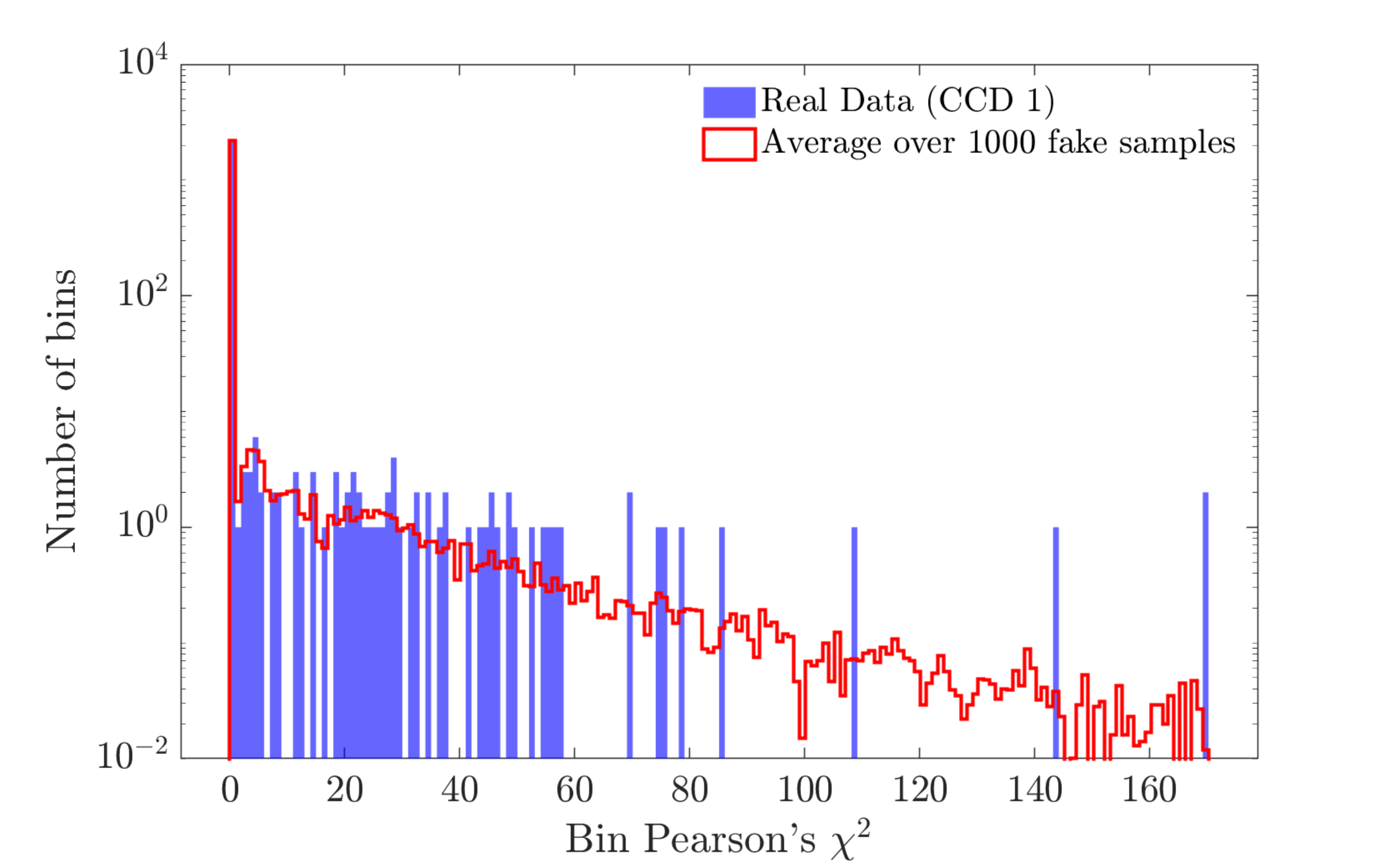}
	\caption{\textbf{Left:} Distribution of the total Pearson's $\chi^2$ for 1000 fake samples drawn from the best-fit background model (blue histogram) compared against the value from the background-model fit to real data (dashed vertical).
	The red line is a Gaussian fit to the fake samples, while the black line marks the number of bins considered.
	The vertical dashed lines shows the corresponding result from the fit to data.
	\textbf{Right:} Distribution of the contribution of individual bins to Pearson's $\chi^2$.
	Blue histogram are the results from the best-fit to data.
	The red histogram shows the average distribution from the best-fit to 1000 fake data samples.
	Top panels are for CCDs 2--7 while the bottom panels are for CCD 1.
	}
	\label{fig:stats_chisq}
\end{figure*}

We placed a $\delta$ function of charge at a specific $z''$ position and evolved the charge distribution in $\Delta t = 30$~ps temporal steps. The spatial distribution evolves via two channels: diffusion and drift in the electric field. To simulate diffusion, the charge in each spatial bin is redistributed every time step following a Gaussian with variance~\cite{gordondiff}
\begin{equation}
	\sigma^2(z'') = \frac{2kT\mu(z'') \Delta t}{e},	
\end{equation}
where $k$ is the Boltzmann constant, $T$ is the temperature, $\mu(z'')$ is the hole mobility, and $e$ is the fundamental charge. Additionally, the position of the charge drifts as
\begin{equation}
	z''_{t+1} = z''_t + \mu(z'') E_z(z'') \Delta t,
\end{equation}
where $z''_t$ and $z''_{t+1}$ are the spatial bin positions of the charge before and after the time step, and $E_z(z'')$ is the $z$ component of the electric field. 
Finally, at each time step the charge carriers have some probability to recombine, thus the charge $q$ in each bin decreases as
\begin{equation}
	q_{t+1}(z'') =  q_t(z'') \left ( 1 - \frac{\Delta t}{\tau(z'')} \right ),
\end{equation}
which is a good approximation since $\Delta t \ll \tau(z'')$ in the PCC region.

After each time step, the charge that is collected by the CCD is defined as any charge that enters in the active region $z''>11.2~\mu$m. We ran the simulation for $500$~ns and repeated over a range of depths. For every initial $z''$, the charge left in the partial charge region at the end of the simulation was $<10^{-4}$ of the initial value. The charge collection efficiency is the ratio of the collected charge to the initial charge deposited.

Our numerical simulations suggests that below $z''\sim 3.5 \ \rm \mu m$ we collect no ionization charge. The collection efficiency turns on for $z''>3.5 \ \rm \mu m$ and rises nearly linearly until full collection efficiency at $z''\sim 10 \ \rm \mu m$.

\section{Systematic Checks on the Background Model}\label{A:template}


In this Appendix, we test whether the data used to construct the background model is statistically consistent with Monte Carlo samples drawn from that background model. 
In order to check for any biases in our two-dimensional binned likelihood fit to construct the background model, 
we draw 100 fake samples (of the same size as the data) from the best-fit background model PDF and re-run the fitting algorithm on them (Section~\ref{S:Template}).
We compare the ratio of the template fit parameters $C_l$ between the 100 trials and the input ``true" values (Table~\ref{tab:fit_result}).
Figure~\ref{fig:LL_trials} uses a box-and-whisker plot to show the distribution of each fit parameter $C_l$ across all 100 trials relative to the input value.
The only significant bias is to add extra $^{210}$Pb in the OFHC copper volumes (parameters 13, 19, and 46 in Table~\ref{tab:fit_result}). This is not an issue for our WIMP analysis, and indicates that the best fit to data overestimates external $^{210}$Pb components. Notably, the dominant of these components (parameter 19---the copper modules) is the closest to its input value. The variation of the other dominant background components (parameters 7, 8, and 10) in the trial fits is within the interquartile range. The other noticeable variation (parameters 24 and 49) comes from poorly constrained templates which contribute $\sim 5\%$ to the overall background combined. While not visible given the y-scale, the largest remaining bias is to include $\sim 5\%$ less intrinsic $^{22}$Na and $^{32}$Si than in the background model (parameters 5 and 6), which is consistent with the fact that the background-model fit result (see Table~\ref{tab:fit_result}) increases these templates by roughly the same fraction over the mean value of the Gaussian constraint. Thus, we conclude that the fit methodology is not significantly biased. 

To estimate the goodness of the fit, we compare the minimum log-likelihood test statistic  $-2 LL_{total}$ of the background-model fit to data to the distribution of the results from the fits to the 100 fake samples. 
We find that the data gives a worse fit than the fake samples (Fig.~\ref{fig:LL_dist}), which indicates some systematic inconsistency between the data and the best-fit background model.
We obtain a $p$-value of 0.0039 (2.7$\sigma$ discrepancy) from a one-sided $Z$-test based on a Gaussian fit to the $-2 LL_{total}$ distribution of the fake samples. 
A fit with a gamma function returns a scale parameter $\theta=0.41$, significantly different than $\theta=2$ expected for a $\chi^2$ distribution. 
If we scale the $-2 LL_{total}$ values by a factor of 5 ($2/0.41=4.88$), we can mimic a $\chi^2$ distribution with 16200 (2 times the gamma distribution shape parameter $k$) degrees of freedom. 
Thus, we can consider the goodness of fit to be $-2 LL_{total}^{\rm data} (\times 5) = 3418(\times 5) / 16200 = 1.03$ ($p$-value of 0.0041), similar to the result from the Gaussian fit. 
Both of these fits are shown in Fig.~\ref{fig:LL_dist}.

To understand the origin of this inconsistency, we draw an additional 1000 samples (of the same size as the data) from the best-fit background model PDF.
For every sample, we compute the total Pearson's $\chi^2$ and the corresponding contribution from each bin relative to the best-fit background model PDF.
The left-hand side panels of Fig.~\ref{fig:stats_chisq} show the distribution of total Pearson's $\chi^2$ for the fake data samples with results for CCDs 2--7 (CCD 1) in the top (bottom) panel.
The corresponding result from the fit to the data is marked by the vertical dashed line.
The right-hand side panels show the distribution of the contribution of bins to the  Pearson's $\chi^2$ averaged over all fake data samples compared to the distribution from the best-fit to data.
Overall, we find excellent agreement between the generated samples and the data in CCD 1.
For CCDs 2--7, three statistically outlier data bins result in a mismatch between data and the fake samples: 1 event with $E \in [18.00,18.25]$~\ev~ and $\sigma_x \in [0.150,0.175]$~pixels, 3 events with $E \in [18.25,18.50]$~\ev~ and $\sigma_x \in [0.550,0.575]$~pixels, and 2 events with $E \in [19.50,19.75]$~\ev~ and $\sigma_x \in [0.325,0.350]$~pixels. Notably, all of these bins are at high energies where saturation occurs. 
Removing these outlier bins from the data does not substantially change the fit result ($\sim 2\%$ effect below 6~k\ev,) but does significantly improve the goodness-of-fit. Specifically, using the gamma distribution from before, the updated best fit to data results in $-2 LL_{total}^{\rm data} (\times 5) = 3381(\times 5) / 16200 = 1.02$ ($p$-value of 0.049). Therefore, we conclude that the discrepancy between the maximum likelihood of the fit samples compared to the data in Fig.~\ref{fig:LL_dist} is not caused by any mismodeling that affects the extrapolation of our background model to the energy region for the WIMP search, and is primarily restricted to the saturation region above 14~k\ev.

\bibliography{ref.bib}

\end{document}